\documentclass[aps,reprint,superscriptaddress,amsmath,amssymb,pra,floatfix]{revtex4-2}
\usepackage{physics}
\usepackage{siunitx}
\usepackage[colorlinks,linkcolor=blue,citecolor=blue,anchorcolor=blue,urlcolor=blue]{hyperref}
\usepackage{bm}
\usepackage{graphicx}
\usepackage{float}
\usepackage{xcolor}

\newcommand{\me}{{e}}

\newcommand{\vo}[1]{{\vb{#1}}}

\begin{document}
    \title{Spinor matterwave control with nanosecond spin-dependent kicks}
    \date{today}
    \author{Liyang Qiu}
    \email{lyqiu16@fudan.edu.cn}
    \author{Lingjing Ji}
    \author{Jiangyong Hu}
    \author{Yizun He}
    \author{Yuzhuo Wang}
    \author{Saijun Wu}
    \email{saijunwu@fudan.edu.cn}
    \affiliation{
        Department of Physics, State Key Laboratory of Surface Physics and Key Laboratory of Micro and Nano Photonic Structures (Ministry of Education), Fudan University, Shanghai 200433, China.}

    \begin{abstract}
    Significant aspects of advanced quantum technology today rely on rapid control of atomic matterwaves with hyperfine Raman transitions. Unfortunately, efficient Raman excitations are usually accompanied by uncompensated dynamic phases and coherent spin-leakages, preventing accurate and repetitive transfer of recoil momentum to large samples. We provide systematic study to demonstrate that the limitations can be substantially overcame by dynamically programming an adiabatic pulse sequence. Experimentally, counter-propagating frequency-chirped pulses are programmed on an optical delay line to parallelly drive five $\Delta m=0$ hyperfine Raman transitions of $^{85}$Rb atoms for spin-dependent kick (SDK) within $\tau=40$~nanoseconds, with an $f_{\rm SDK}\approx 97.6\%$ inferred fidelity. 
    Aided by numerical modeling, we demonstrate that by alternating the chirps of successive pulses in a balanced fashion, accumulation of non-adiabatic errors including the spin-leakages can be managed, while the dynamic phases can be robustly cancelled. Operating on a phase-stable delay line, the method supports precise, fast, and flexible control of spinor matterwave with efficient Raman excitations. 
    \end{abstract}


\maketitle

\section{Introduction}

Precise control of effective 2-level systems is instrumental to modern quantum technology. In atomic physics, such controllable two-level systems are naturally defined on a pair of long-lived atomic internal states and are often referred to as atomic spins. To control the external motion, quantized photon recoil momentum can be transferred to the atom during spin-flips driven by Raman excitations~\cite{Kasevich.1991,Featonby.1996, McGuirk2000}. By laser cooling~\cite{MetcalfBook}, the atomic motion can be sufficiently slowed that the momentum transfers by the rapid optical pulses are effectively instantaneous ``spin-dependent kicks'' (SDK)~\cite{Mizrahi2013, Jaffe.2018}, with the excitation efficiency insensitive to the atomic motion. SDKs are therefore a class of broadband spinor matterwave control techniques with an achievable accuracy similar to those for 2-level internal atomic spin controls~\cite{Harty2014, Ballance2016, Wang2016}. Beyond traditional applications such as to enhance the enclosed area of light pulse atom interferometers~\cite{McGuirk2000,Kotru2015,Jaffe.2018}, SDKs emerge as an important technique to control spin-motion entanglement and to improve the scalability of ion-based quantum information processing~\cite{Garcia-Ripoll2003,
Duan2004, Mizrahi2013,Lo2015}.


For precise spinor matterwave control, it is essential to operate SDK at high enough speed to suppress low-frequency noises and to compose multiple operations within a limited duration. Furthermore, except for working with microscopically confined samples~\cite{Mizrahi2013,Lo2015,Ballance2016}, high quality optical control needs to be highly resilient to intensity errors associated with illumination inhomogeneity~\cite{Ma2020,He2020b}. To meet the fidelity requirements by the next generation quantum technology~\cite{Schafer2018, Fluhmann2019,Szigeti2020, Wu2020,Anders2021,Greve2021}, particularly for large samples, high speed implementation of error-resilient composite techniques~\cite{Genov2014,Low2016,Saywell2018} are likely required. However, the SDK speed is practically limited by intricate requirements associated with precise Raman control, including the following two categories.

\begin{figure}[t]
    \centering
    \includegraphics[width=\linewidth]{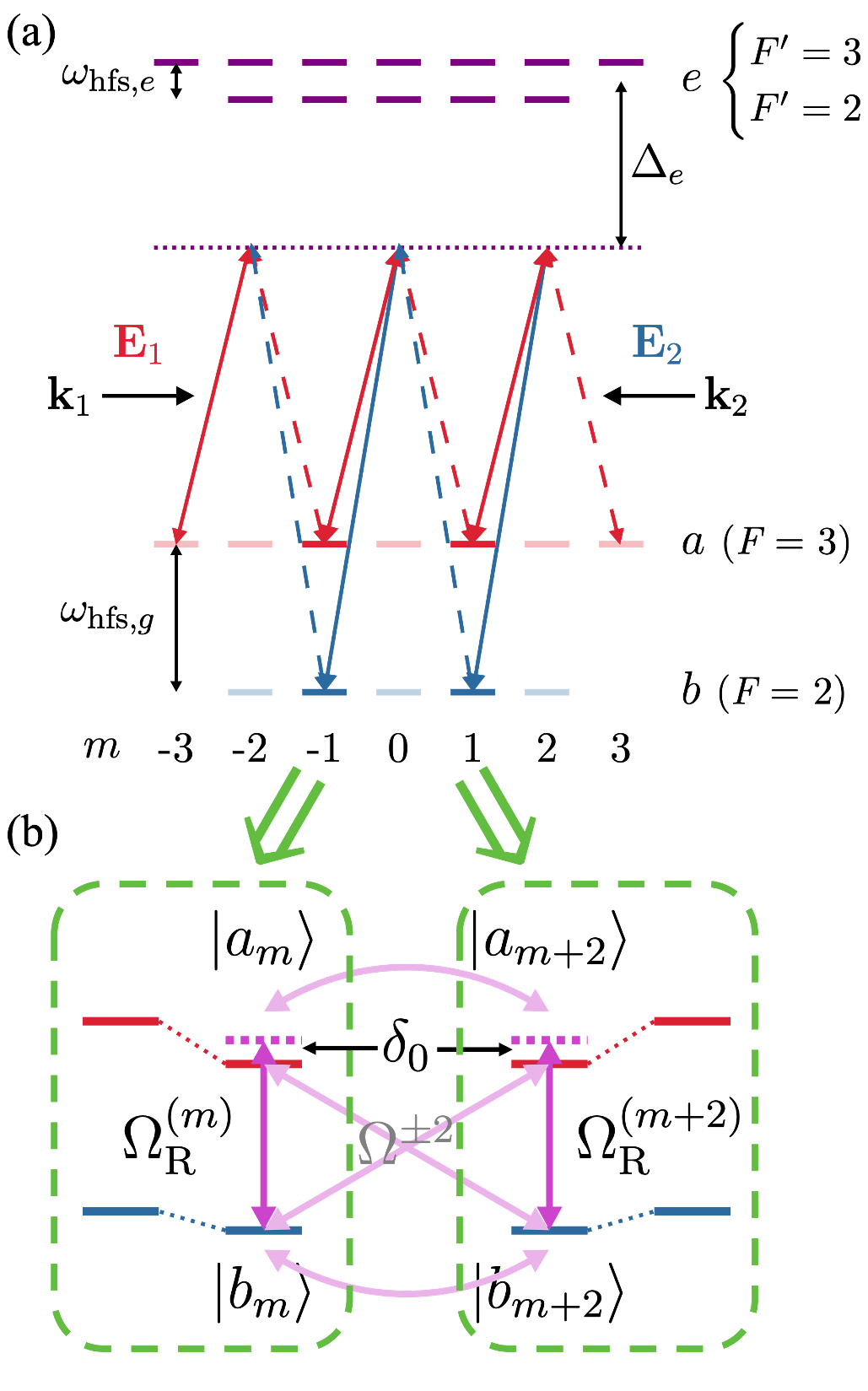}
    \caption{Schematic of SDK on a hyperfine manifold, with the $^{85}$Rb D1 line as the example in this work. (a) Cross-linearly polarized, counter-propagating ${\bf E}_{1,2}$ resonantly drive the $\Delta m=0$ and $\Delta m=\pm 2$ Raman transitions. The optical couplings are illustrated with solid and dashed lines, with $m=\pm 1,\pm 3$ ground states as examples. When the single-photon detuning $\Delta_e$ is much larger than the excited state hyperfine splitting $\omega_{{\rm hfs},e}$, the $\Delta m=\pm 2$ couplings through the $F'=2,3$ intermediate states largely cancel each other, leading to decoupled $\{|a_m\rangle,|b_m\rangle\}$ sub-spin systems. (b) The reduced level diagrams for the weakly coupled $\{|a_m\rangle,|b_m\rangle\}$ systems ($m=\pm 1$). The bare levels are relatively shifted and Raman coupled. The effective Hamiltonian parameters $\Omega^{\pm 2}$, $\Omega_{\rm R}^{(m)}$, $\delta_0$ for Eqs.~(\ref{eq:leakrabi})(\ref{eq:RamanRabi})(\ref{eq:h0}) are marked. The differential Stark shift $\delta_0$ is largely responsible for the SDK dynamic phase (Eq.~(\ref{eq:sdkevo})). Similarly coupled sub-spin systems composed of $\{|a_m\rangle,|b_m\rangle\}$ with $m=0,\pm 2$ are not shown. }
    \label{fig:1}
\end{figure}

The first type of SDK speed constraints are associated with multi-level dynamics accompanying the Raman control of ``real'' atoms. To avoid  spontaneous emission, the Raman excitation bandwidth has to be much smaller than the single-photon detuning $\Delta_e$ (Fig.~\ref{fig:1}(a)) to avoid populating the excited states. But a moderate $\Delta_e$ is typically required for efficient control with limited laser power, as well as for managing the differential Stark shifts (Fig.~\ref{fig:1}(b))~\cite{Weiss1994}. In addition, as to be clarified further,  to avoid the $m$-changing transitions during Raman excitation~\cite{Happer1972}, one usually lifts the Zeeman degeneracy with a strong enough quantization field~\cite{Kasevich1992}. The typically moderate field strength also limits the speed of the Raman control.



The second type of SDK speed constraints are associated with the requirement of controllable direction of the recoil momentum transfer. As in Fig.~\ref{fig:1}, when counter-propagating ${\bf E}_{1,2}$ pulses are applied to drive a Raman transition, to ensure the preference for the atom to absorb a photon from the ${\bf E}_1$ field followed by a stimulated emission into the ${\bf E}_2$ field, certain mechanism is needed to prevent the time-reversed Raman process from occurring. For hyperfine spin control of alkaline-like atoms, the directionality is ultimately supported by the ground-state hyperfine splitting $\omega_{{\rm hfs},g}$~\cite{Mizrahi2013}. Practically, however, when the Raman interferometry is operated in the retro-reflection geometry with established advantages~\cite{Durfee2006,Canuel2006,Malossi2009,Zhou2015,Perrin2019,Hartmann2020}, the directionality is typically protected by the moderate 2-photon frequency differences introduced either by additional frequency modulations~\cite{McGuirk2000,Durfee2006} or Doppler shifts of moving atoms~\cite{Canuel2006}, which in turn limit the SDK speed.

\begin{figure}[t]
    \centering
    \includegraphics[width=\linewidth]{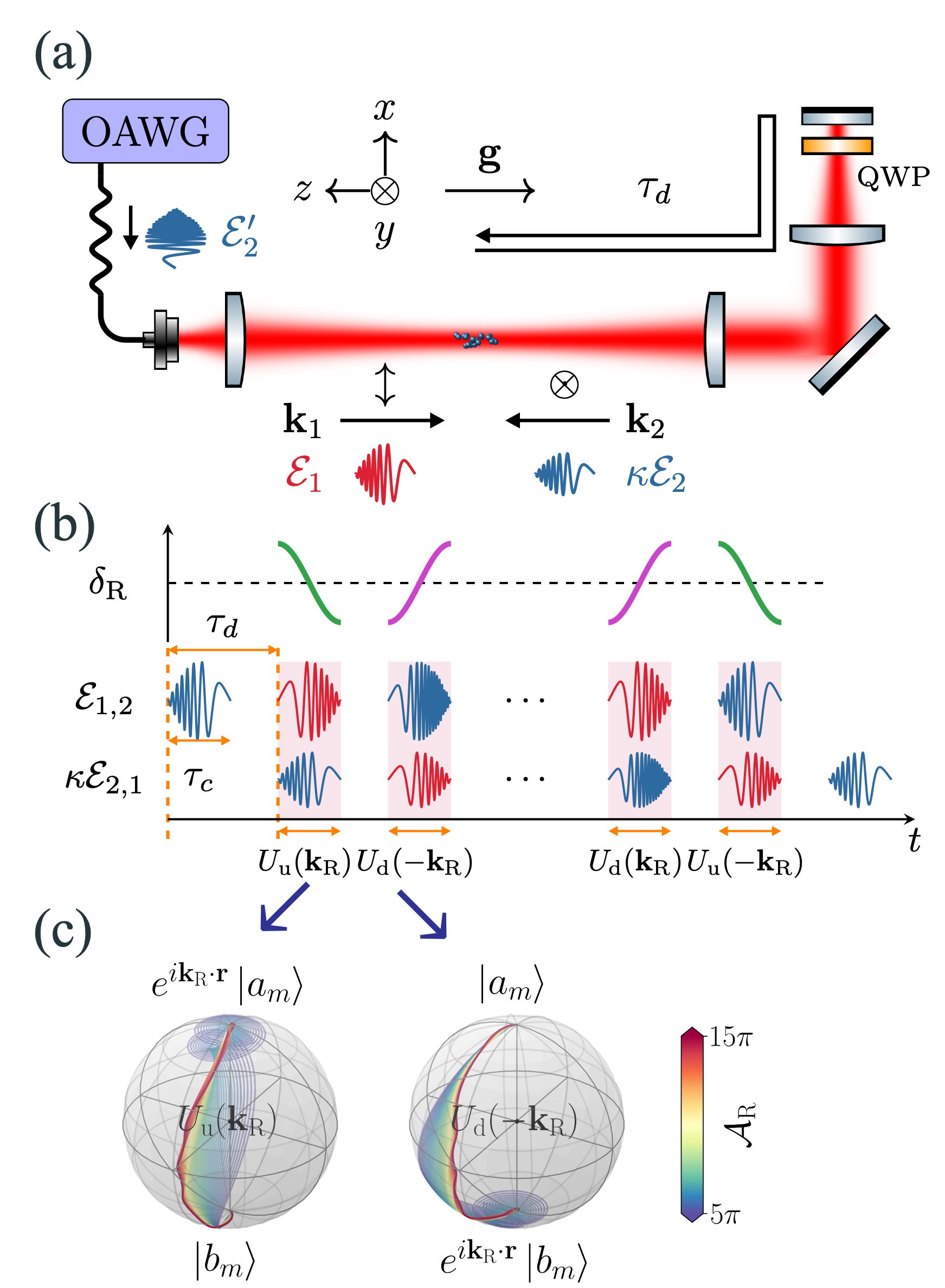}
    \caption{(a) The schematic setup. Frequency-chirped optical waveforms (illustrated with the real part of the complex envelop function $\mathcal{E}_{1,2}$, sharing a same color-coding as those for ${\bf E}_{1,2}$ in Fig.~\ref{fig:1}(a)) are programmed by an optical arbitrary waveform generator (OAWG) and sent through a single mode fiber to the delay line. An ${\bf E}_1$ pulse collides with a retro-reflected ${\bf E}_2$ pulse at the atomic sample to drive the $|a_m\rangle\leftrightarrow |b_m\rangle$ Raman transition while imparting $\pm \hbar \vb{k}_{\rm R}$ kicks to atoms reaching $|a_m\rangle$ and $|b_m\rangle$ respectively. Next, after a delay time $\tau_{\rm d}$, the ${\bf E}_1$ pulse is retro-reflected to meet ${\bf E}_2'$ at the same location to further boost the $|a_m\rangle$, $|b_m\rangle$ separation. QWP: quarter-wave plate. (b) The timing sequence for the ``balanced chirp-alternating scheme'' (Eq.~(\ref{eq:nSDKuddu})). The two-photon detunings $\delta_{\rm R}$ are plotted vs time. The adiabatic controls with ``up-chirp'' and ``down-chirp'' 2-photon detuning $\delta_{\rm R}$ are referred to as $U_{\rm u}$ and $U_{\rm d}$ respectively.
    Here the amplitudes for the retro-reflected pulses are reduced by a $\kappa$ factor. 
    Typical Bloch sphere dynamics of an $\{|a_m\rangle,|b_m\rangle\}$ spinor ($m=0$) are given in (c) at various laser intensities parametrized by the Raman pulse area $\mathcal{A}_{\rm R}$. The weak $\Delta m=\pm2$ couplings (Fig.~\ref{fig:1}(b), Eq.~(\ref{eq:leakrabi})) are ignored. } 
    \label{fig:1b}
\end{figure}

The purpose of this work is to improve the speed and precision of spinor matterwave control in presence of the above mentioned technical barriers. In particular, we study Raman adiabatic SDK~\cite{Kotru2015,Jaffe.2018} within nanoseconds in a retro-reflection optical setup~\cite{Peters1997, Perrin2019}, for alkaline atoms with Zeeman degeneracy prone to coherent spin leakage. As schematically summarized in Fig.~\ref{fig:1} (level diagram) and Fig.~\ref{fig:1b} (experimental implementation),  counter-propagating frequency-chirped pulses are programmed on the optical delay line to adiabatically drive the $F_b=2\leftrightarrow F_a=3$ Raman transitions of $^{85}$Rb through the D1 line, while imparting $\pm \hbar \vb{k}_{\rm R}$ momentum ``kicks'' to atoms. Here $\vb{k}_{\rm R}=\vb{k}_2-\vb{k}_1$ is the difference of k-vectors between the pulse pair. The adiabatic Raman transfer technique~\cite{Kral2007,Kotru2015,Jaffe.2018} (Fig.~\ref{fig:1b}(b)(c)) provides the intensity-error resilience for the focused laser to address a mesoscopic sample.
To enforce the directionality in the standard retro-reflection geometry~\cite{Peters1997, Perrin2019},  the regular method of resolving the 2-photon frequency differences~\cite{McGuirk2000,Durfee2006,Canuel2006} are abandoned. Instead, we exploit the fact that the nanosecond pulses are short enough to be spatially separated (Fig.~\ref{fig:1b}(a)), allowing controllable collision of specific Raman pulses within the atomic sample, at specific instances, to drive the Raman transitions. 

Experimentally, equipped with a wideband optical arbitrary waveform generation system (OAWG)~\cite{He2020a}, we drive adiabatic SDKs within tens of nanoseconds, the fastest realization to date~\cite{Kotru2015,Jaffe.2018}, by shaping counter-propagating pulses  (Fig.~\ref{fig:1b}) with merely tens of milliwatts laser power. The spin population and momentum transfer are measured and compared with full-level numerical simulations, with which we infer an SDK fidelity 
of $f_{\rm SDK}\approx 97.6(3)\%$. The $\delta f\sim 2.5\%$ infidelity, as unveiled by the full-level numerical simulation, is primarily limited by spontaneous emission and the $\Delta m=\pm 2$ leakage among Zeeman sublevels within the ground state hyperfine manifold (Fig.~\ref{fig:1}(b)) at the moderate $\Delta_e=2\pi\times 10$~GHz detuning.

Aided by numerical modeling, we further demonstrate that beyond the incoherent momentum and population transfer, high-fidelity spinor matterwave ``phase gates''  capable of coherently transferring many recoil momenta can be synthesized by nanosecond adiabatic SDKs on the delay line. This is despite the coherent spin leakage and intensity-dependent diffraction phase broadening~\cite{Weiss1994, Estey2015,Jaffe.2018,Kristensen2021} by individual ``kicks'', at the moderate $\Delta_e$~\cite{Mizrahi2014,  Ballance2016}. In particular, we demonstrate that a tailored adiabatic sequence with alternating up and down 2-photon chirps~\cite{Stack2011} (Fig.~\ref{fig:1b}(b)(c)) suppresses the coherent accumulation of spin leakages across the nearly degenerate hyperfine levels dressed by the cross-linearly polarized pulses (Fig.~\ref{fig:1}(b)). Furthermore, as to be clarified further, the time-reversal symmetry of the chirp-balanced scheme ensures robust cancellation of the dynamic, laser intensity-dependent diffraction phases, with $\Delta_e>\omega_{{\rm hfs},g}$ unbounded by traditional choices~\cite{Weiss1994, Wineland2003}, even in presence of non-perfect retro-reflection (Fig.~\ref{fig:1b}(b))~\cite{Perrin2019}. 

Our work suggests that fast, large-momentum atom optics can be synthesized for accurately manipulating multi-level atoms in the unconventional regime of Raman control~\cite{Weiss1994, Wineland2003}. 
Comparing with atom interferometric large-momentum beamsplitting techniques based on high-order Bragg diffraction~\cite{Plotkin-Swing2018, Asenbaum2020}, Bloch oscillations~\cite{Muller2009,Gebbe2021}, as well as traditional Raman schemes~\cite{McGuirk2000,   Kotru2015,Jaffe.2018}, the key advantage offered by our SDK technique is the ultra-high SU(2) operation speed. The high speed not only helps the composite technique to avoid the complexity of continuous motion itself, but also protects the precise intensity-error cancellation in presence of low-frequency noises, including those due to atomic motion in inhomogenuous laser beams. We notice that similar advantages are shared by the recently developed high speed clock atom interferometry~\cite{Rudolph2020, Wilkason2022}, where the atomic spins are defined directly on kHz-level optical lines, to achieve exceptionally high momentum transfer efficiencies. Similar to that technique,  to coherently drive the hyperfine spins within nanoseconds requires high laser intensities. The nanosecond SDK in this work is realized by weakly focusing the milli-Watt level pulses to a mesoscopic sample. To address larger samples for {\it e.g.} inertial sensing~\cite{Peters1997,Asenbaum2020} on a compact delay line, more advanced laser modulation techniques need to be developed for generating high power wideband optical waveforms.


In the following the main part of the paper is structured into three sections. In Sec.~\ref{sec:theory}, we set up a light-atom interaction framework where multi-level couplings are treated as perturbations to the effective 2-level Raman interaction within the ground state hyperfine manifold. Key notations and quantities for the experimental and numerical studies are defined. 
In Sec.~\ref{sec:expt}, we demonstrate the flexible matterwave control by programming counter-propagating pulses on an optical delay line with OAWG~\cite{He2020a}. Constrained by experimental resources, the characterization of the adiabatic SDK technique is limited to the inference of the SDK fidelity with atomic velocity and hyperfine population measurements.  
However, in Sec.~\ref{sec:controldynamics}, we  demonstrate efficient suppression of coherent spin-leakage accumulation and intensity-dependent dynamic phases by the chirp-alternating scheme, and provide numerical evidence that the chirp-balanced SDK scheme supports coherently control of multi-spinor matterwave with large momentum transfer, even with a moderate laser power in the 10~mW range as in this work. 




\section{Theoretical Model}\label{sec:theory}

\subsection{Spin-dependent kicks on a hyperfine manifold}

We consider the Fig.~\ref{fig:1} (a) alkaline-like atom subjected to Raman excitation on the D1 line. Similar conclusions holds for D2 excitations. The Zeeman-degenerate $a, b$ and $e$ hyperfine states with total angular momentum $F_{a,b}$  and $F'_{e}$ are labeled as $\ket{c_m}\equiv\ket{c,F_c, m}$ with magnetic quantum number $m\in\qty[-F_c, F_c]$. Here, with the nuclear spin $I>1/2$, there are $2F_c + 1$ Zeeman sublevels for each manifold $c\in\qty{a,b}$, and similarly for the $e$ manifold. The $|a_m\rangle\leftrightarrow|b_{n}\rangle$ hyperfine Raman transition is driven by pairs of counter-propagating laser pulses, $\vo{E}_{1,2}={\vo{e}}_{1,2}\mathcal{E}_{1,2} \me^{i\qty(\vb{k}_{1,2}\cdot\vb{r}-\omega_{1,2}t)}+c.c.$ with shaped slowly-varying amplitudes $\mathcal{E}_{1,2}(\vb{r},t)$ and Raman-resonant carrier frequency difference $\omega_2-\omega_1=\omega_{{\rm hfs},g}$. %

Although motion of the multi-level atom in the pulsed ${\bf E}_{1,2}$ fields is quite complicated (Appendix~\ref{App:FullH}), the internal state dynamics can be substantially simplified when the laser polarization ${\bf e}_{1,2}$ are chosen to be cross-linear (Fig.~\ref{fig:1b}(a))~\cite{Malossi2009, Zhou.2015} and the single-photon detuning $\Delta_e$ is much larger than the excited state hyperfine splitting $\omega_{{\rm hfs},e}$. In this case, the destructive interference of Raman couplings through the intermediate $|e_l\rangle$ levels prevent efficient $\Delta m=\pm 2$ transitions (The rule in Fig.~\ref{fig:1}(a) is that efficient Raman couplings are only composed by pairs of dipole couplings drawn with a same type of solid or dashed lines.)~\cite{Happer1972}. The residual ``leaking'' Raman Rabi frequency (Fig.~\ref{fig:1}(b)) scales as 
\begin{equation}\label{eq:leakrabi}
    \Omega^{\pm 2} = \mathcal{O}\qty(\frac{\omega_{\rm hfs,e}}{\Delta_e})\Omega_{\rm R},
\end{equation}
with $\Omega_{\rm R}=\Omega^{1*}_a\Omega_b^{2}/2\Delta_e$,  $\Omega_{a(b)}^{1(2)}= \frac{\mathcal{E}_{1(2)}}{\hbar}\frac{1}{\sqrt{3}}\abs{\bra{J_g}\!\abs{\vb{d}}\!\ket{J_e}}$~\cite{qao}.

Therefore, in the large $\Delta_e$ limit, the Raman dynamics in the ground-state manifold is decomposed into those within $2F_b+1$ copies of pseudo spin-$1/2$ sub-spaces $\qty{\ket{a_m}, \ket{b_m}}$, labeled by magnetic quantum number $m$ (See Fig.~\ref{fig:1}(b) for the $m=\pm 1$ examples.). For each $m$-spin, the Rabi frequency driven by the Raman pulse can be written as
\begin{equation}\label{eq:RamanRabi}
\Omega_{\rm R}^{(m)} =\chi^{(m)}\Omega_{\rm R}.
\end{equation}
The close-to-unity factor $\chi^{(m)}$ is determined by the associated Clebsh-Gordan coefficients, normalized at $\chi^{(0)} = 1$ and decreases slowly with $|m|$~\cite{Dunning.2014jm}. The Hamiltonian for the $m$-spin is given by
\begin{equation}
    H_{0}^{(m)}(\vb{r}, t)= \hbar \qty(\frac{\delta_0}{2} \sigma_z^{(m)} + \frac{\Omega_{\rm R}^{(m)}}{2}  e^{i\vb{k}_{\rm R}\cdot {\bf r}} \sigma_+^{(m)} + \mathrm{h.c.}).\label{eq:h0}
\end{equation}
Here the detuning $\delta_0$ is the ($m$-independent) differential Stark shift between the two hyperfine ground states that can be nullified at suitable $\Delta_e/\omega_{{\rm hfs},g}$ and $\Omega_{a,b}$ combinations~\cite{Weiss1994}. The Pauli matrices $\{\sigma_x^{(m)},\sigma_y^{(m)},\sigma_z^{(m)}\}$ are defined by $\sigma_+^{(m)}\equiv \sigma^{a_m b_m}$~\cite{Cidrim.2021}. For notation convenience in the following, we further define projection operators ${\bm 1}^{(m)}\equiv\sigma^{a_m a_m}+\sigma^{b_m b_m}$ and $\bar {\bm 1}^{(m)}=1-{\bm 1}^{(m)}$ respectively.

With the major part of the resonant Raman interaction identified, the D1 multi-level dynamics (Fig.~\ref{fig:1}(a)) governed by the effective light-atom interaction Hamiltonian, as detailed in Appendix~\ref{App:FullH}, can be written as
\begin{equation}
        H_{\rm eff}(\vb{r}, t) = \sum_{m=-F_b}^{F_b} H^{(m)}_{0}(\vb{r},t)+ V'(\vb{r}, t)\label{eq:heff}
    \end{equation}
to describe the Raman-dressed ground state dynamics. Apart from the $\Delta m=\pm 2$ couplings specified by Eq.~(\ref{eq:leakrabi}), the $V'$ term includes non-Hermitian and non-adiabatic $|e_l\rangle$ couplings, the $m-$sensitive light shifts~\cite{Happer1972}, and the ``counter-rotating'' couplings involving detuned Raman excitations (Eq.~(\ref{eq:h})). 

With the atomic position ${\bf r}$ parameter in Eq.~(\ref{eq:heff})  as a quantum mechanical operator acting on the external atomic wavefunction, the Eq.~(\ref{eq:heff}) Hamiltonian can propagate the spinor matterwave together with the kinetic $\hat P^2/2M$ operator, see Appendix~\ref{sec:numerical} for implementation ($\hat P=i\hbar \nabla$ and $M$ the atomic mass.). During nanosecond intervals, however, the matterwave dispersion for the laser-cooled atoms is negligibly small. It is therefore possible to design broadband matterwave controls for iterative applications. In particular, a spin-dependent kick is a transfer of photon momentum to atoms accompanied by a spin-flip~\cite{Mizrahi2013,Jaffe.2018}. We refer an ideal SDK as
    \begin{equation}\label{eq:sdkevo}
        U_{\rm K}(\vb{k}_{\rm R})  = \prod_{m=-F_b}^{F_b}\left(\me^{i\varphi^{+}}\me^{i\vb{k}_{\rm R} \cdot\vb{r}}\sigma_+^{(m)} - {\rm h.c.}+ \bar {\bm 1}^{(m)}\right).
    \end{equation}
 The $\varphi^+$ is a diffraction phase offset~\cite{Delhuille2003} which is generally Zeeman state $m$- and laser intensity $|\mathcal{E}|^2$-dependent. 

In the following our goal is to construct perfect $U_{\rm K}(\vb{k}_{\rm R})$ operation as in Eq.~(\ref{eq:sdkevo}) from the full Hamiltonian in Eq.~(\ref{eq:heff}), and furthermore to design multiple SDKs with suppressed $m$ and $|\mathcal{E}|^2$ dependence. For the purpose, we first define a fidelity $f_{\rm SDK}$ to qualify the implementation.

\subsection{Non-ideal SDK}
    

To qualify the Fig.~\ref{fig:1b} SDK implementation modelled by Eq.~(\ref{eq:heff}) Hamiltonian, we refer the imperfect realization of SDK as $\tilde U(\vb{k}_{\rm R};\eta)$ with $\eta$ to generally represent relevant Hamiltonian parameters. We define an average fidelity~\cite{Magesan2012} over the whole atomic ensemble for a single SDK acting on the $\{|b_m\rangle\}$,$\{|a_m\rangle\}$ states as
\begin{equation}
    f_{\rm SDK}=\left\langle \left |\left\langle  \langle c_m|U_{\rm K}^{\dagger}(\vb{k}_{\rm R}) \tilde U(\vb{k}_{\rm R};\eta) |c_m\rangle  \right\rangle_{z}\right |^2 \right\rangle_{\eta,c_m}. \label{equ:fSDK}
\end{equation}
Here the $\tilde U({\bm k}_{\rm R};\eta)|c_m\rangle$ is compared to $U_{\rm K}({\bm k}_{\rm R})|c_m\rangle$ across the atomic sample with an $\langle ... \rangle_z$ average.
An ensemble average of the mode-squared fidelity is then performed with $\langle ...\rangle_{\eta}$ over the Hamiltonian parameters of interest. Assuming light intensity hardly varies along $z$ over the wavelength-scale distance of interest (Fig.~\ref{fig:1b}(a)), the SDK fidelity defined in this way becomes insensitive to the diffraction phase $\varphi^+$ and therefore provides a convenient measure for the quality of controlling incoherent observables, such as for the recoil momentum and hyperfine population transfer to be experimentally measured next. The $\langle ...\rangle_{c_m}$ instead performs average over the $2(2 F_b+1)$ initial states of interest with $m=-F_b,...,F_b$ and $c=a,b$.

To quantify the leakage of an atomic state out of each spin sub-space $\{|a_m\rangle,|b_m\rangle\}$ by $V'({\bf r},t)$, we define an average spin leakage probability by the non-ideal SDK as
\begin{equation}
    \varepsilon_{\rm leak}=1-\left\langle  \langle c_m |\tilde U^{\dagger}(\vb{k}_{\rm R},\eta) {\bf 1}^{(m)}\tilde U(\vb{k}_{\rm R};\eta) |c_m\rangle \right\rangle_{\eta,c_m}.\label{equ:lSDK}
\end{equation}

By comparing Eq.~(\ref{equ:fSDK}) with Eq.~(\ref{equ:lSDK}) it is easy to verify $f_{\rm SDK}\leq 1-\varepsilon_{\rm leak}$, that is, any spin leakage leads to inefficient control. The leakage probability $\varepsilon_{\rm leak}$ by Eq.~(\ref{equ:lSDK}) includes contributions from $\Delta m=\pm 2$ transition as well as those due to spontaneous emission, as $\varepsilon_{\rm leak}=\varepsilon_{\Delta m}+\varepsilon_{\rm sp}$. Here the spontaneous emission probability $\varepsilon_{\rm sp}$ during the SDK control is obtained by evaluating $\tilde U(\vb{k}_{\rm R};\eta)$ with the stochastic wavefunction method~\cite{Dalibard1992,Carmichael1993} (Appendix~\ref{sec:numerical}). For the $\Delta m=\pm 2$ leakage, we find $\varepsilon_{\Delta m}\propto \omega_{\rm hfs,e}^2/\Delta_e^2$, as by Eq.~(\ref{eq:leakrabi}), and is thus suppressible at large $\Delta_e$ similar to the suppression of spontaneous emission. However, one should note that unlike spontaneous emission which simply leads to decoherence, the $\Delta m=\pm 2$ leakage is a coherent process. A sequence of such coherent leakages may quickly add up in amplitudes to become significant, even if $\varepsilon_{\Delta m}$ for a single SDK is negligibly small.

\subsection{Adiabatic SDK}\label{sec:asdk}

To achieve a uniformly high SDK fidelity across an intensity-varying sample volume, particularly for  $m-$spin on a hyperfine manifold with different $\Omega_{\rm R}^{(m)}$ (Fig.~\ref{fig:1}, Eq.~(\ref{eq:RamanRabi})), a standard technique is to exploit the geometric robustness of 2-level system by inducing an adiabatic rapid passage (Fig.~\ref{fig:1b}(a))~\cite{Bergmann1998, Vitanov.2001, Kral2007, Kotru2015,Jaffe.2018}. For a particular $\{|a_m\rangle,|b_m\rangle\}$ sub-spin, if we ignore $V'$ in Eq.~(\ref{eq:heff}), then the Raman adiabatic passage can easily be generated by the $H_0^{(m)}$ Hamiltonian. In particular, we parametrize the Rabi frequency of the two pulses as $\Omega_{a(b)}^{1(2)}=C_{a(b)}e^{i\phi_{1(2)}}$, and specifically consider the time-dependent phase difference and the amplitude profile as $\Delta\phi(t)=\phi_0 \sin(\pi t/\tau_{\rm c})$, $C_{a(b)}(t)=C_{a(b)}^{(0)} \sin (\pi t/\tau_{\rm c})$ respectively for $ t\in \qty[0,\tau_{\rm c} ]$ (Fig.~\ref{fig:measure}(b) inset). With a large enough 2-photon sweep frequency $\delta_{\rm swp}=\pi \phi_0/\tau_{\rm c}$, $|\delta_{\rm swp}|\gg \delta_0$ to let the 2-photon detuning $\delta_{\rm R}=\Delta\dot{\phi}$ cover the differential Stark shift $\delta_0$ (Fig.~\ref{fig:1b}(b)), a strong enough Raman Rabi amplitude $C_{\rm R}^{(0)}=C_a^{(0)} C_b^{(0)}/2\Delta_e =2 \chi^{(m)} \mathcal{A}_{\rm R}/\tau_{\rm c}$ with a peak Raman pulse area $\mathcal{A}_{\rm R} \gg 1$, and matched  magnitudes between $C_{\rm R}^{(0)}$ and $\delta_{\rm swp}$, SDK can be generated in a quasi-adiabatic manner~\cite{Miao2007,Guery-Odelin2019}, {\it i.e.}, with atomic state $|\psi_m(t)\rangle\approx c_a(0)e^{i\varphi_a(t)}|\tilde a_m(t)\rangle+c_b(0)e^{i\varphi_b(t)}|\tilde b_m(t)\rangle$ following the adiabatic basis $\{|\tilde a_m(t)\rangle,|\tilde b_m(t)\rangle \}$ which are simply the eigenstates of the instantaneous Hamiltonian. Population inversion are thus driven quasi-adiabatically during $0<t<\tau_{\rm c}$ with the efficiency insensitive to the laser intensity, detuning, and their slow deviations from the specific time-dependent forms. Putting back the ${\bf r}$-dependence, the diffraction phase accompanying the population inversion, $\tilde \varphi^+({\bf r})\equiv \tilde \varphi^++\vb{k}_{\rm R}\cdot {\bf r}$, is evaluated as $\tilde \varphi^+({\bf r})=\varphi_a(\tau_{\rm c})-\varphi_b(\tau_{\rm c})$ which includes not only a geometric phase $\varphi_{\rm G}=\pi/2+\vb{k}_{\rm R}\cdot {\bf r}$~\cite{Berry.1984,   Zhu2002}, but also a dynamic phase $\varphi_{{\rm D},m}\propto \int_0^{\tau_{\rm c}}  (\langle \tilde a_m| H_0^{(m)} |\tilde a_m \rangle-\langle \tilde b_m| H_0^{(m)}|\tilde b_m \rangle ) d\tau $ sensitive to the control parameters $\Omega_{a(b)}^{1(2)}$ characterized by the laser intensity profiles.






\subsection{Dynamic phase cancellation}\label{sec:dsdk}
The $U_{\rm K}({\bf k}_{\rm R})$ operation (Eq.~(\ref{eq:sdkevo})) prints matterwave with state $m$- and laser intensity $|\mathcal{E}|^2$-dependent diffraction phases $\varphi^+$ in general. For interferometrically useful coherent matterwave controls, these dynamic parts of $\varphi^+$ need to be suppressed. It is worth noting that the dynamic phase for the adiabatic SDK survives at vanishing $\delta_0$~\cite{Weiss1994,Zhu2002}. Therefore, coherent matterwave control with adiabatic SDK requires certain dynamic phase cancellation in general.

The dynamic phases for perfect SDKs can be cancelled in pairs. To avoid trivial operations during the pairing, the 2-photon wavevector $\vb{k}_{\rm R}$ can be inverted~\cite{McGuirk2000, Kotru2015,Jaffe.2018}, leading to
\begin{equation}\label{eq:dSDK}
    \begin{split}
    U^{(2 N)}_{\rm K}(\vb{k}_{\rm R})&=
    \underbrace{
    U_{\rm K}(-\vb{k}_{\rm R})U_{\rm K}(\vb{k}_{\rm R})...U_{\rm K}(-\vb{k}_{\rm R})U_{\rm K}(\vb{k}_{\rm R})}_{2 N}\\
    &= \prod_{m=-F_b}^{F_b}\left((-1)^N e^{i 2 N \vb{k}_{\rm R}\cdot {\bf r}\sigma_z^{(m)}}+\bar {\bf 1}^{(m)}\right).
    \end{split}
\end{equation}
Within each $m$-spin space span by $\{|a_m\rangle,|b_m\rangle\}$, the $U^{(2N )}_{\rm K}(\vb{k}_{\rm R})$ acts as a position-dependent phase gate to pattern the two components of arbitrary spinor matterwave with $\pm (2 N\vb{k}_{\rm R}\cdot {\bf r})$ phases, {\it i.e.}, to coherently transfer opposite photon recoil momenta to the spin components.  

Practically, however, swapping the k-vectors of ${\bf E}_{1,2}$ can affect other pulse parameters~\cite{GustavsonThesis,Canuel2006, Durfee2006, Perrin2019}. Here, the k-vector swapping is achieved on the delay line with retro-reflection (Fig.~\ref{fig:1b}) by programming the carrier frequencies $\omega_1\leftrightarrow \omega_2$ for the delayed pulses. The imperfect reflection with $\kappa<1$ generally leads to unbalanced diffraction phases~\cite{Peters1997} $\tilde \varphi^+$ associated with $\tilde U(\pm \vb{k}_{\rm R})$. For the spinor matterwave control, we quantify the impact of dynamic phase by comparing the phase of imperfectly controlled $\tilde U^{(2N)} |c_m\rangle$ with the ideal $U_{\rm K}^{(2N)}|c_m\rangle$ ($c=a,b$),
\begin{equation}
    \begin{array}{l}
    \varphi^{(2N)}_{{\rm D},m}= {\rm arg}[\langle a_m |U^{(2N)\dagger}_{\rm K} (\vb{k}_{\rm R}) \tilde U^{(2N)}(\vb{k}_{\rm R};\eta) |a_m\rangle]-\\
    ~~~~~~~~~~~~{\rm arg}[\langle b_m |U^{(2N)\dagger}_{\rm K} (\vb{k}_{\rm R}) \tilde U^{(2N)}(\vb{k}_{\rm R};\eta) |b_m\rangle]
    .  
    \end{array}\label{equ:dphase}
\end{equation}

\subsection{Chirp-alternating adiabatic SDK schemes}\label{sec:cpSDK}

A central goal of this work is to synthesize perfect $U^{(2 N)}_{\rm K}(\vb{k}_{\rm R})$ spinor matterwave phase gates with $2 N$ imperfect $\tilde U(\pm \vb{k}_{\rm R})$, even in presence of the unbalanced dynamic phase $\varphi_{{\rm D},m}^{(2 N)}$ and the $\Delta m=\pm 2$ leakage. For the purpose, we refer the experimentally realized non-ideal adiabatic SDKs with equal amount of positive ($\delta_{\rm swp}>0$) and negative ($\delta_{\rm swp}<0$) 2-photon sweeps as $\tilde U_{\rm u}(\pm\vb{k}_{\rm R})$ and $\tilde U_{\rm d}(\pm\vb{k}_{\rm R})$ respectively (Fig.~\ref{fig:1b}(b)). The associated control Hamiltonians are $H_{\rm u}({\bf r},t)$ and $H_{\rm d}({\bf r}, t)$. In Sec.~\ref{sec:uddu} we show that a chirp-alternating sequence
    \begin{equation}
        \tilde U_{\rm ud}^{(2 N)} (\vb{k}_{\rm R})= \underbrace{\tilde{U}_{\rm u}(-\vb{k}_{\rm R})\tilde{U}_{\rm d}(\vb{k}_{\rm R}) \cdots \tilde{U}_{\rm u}(-\vb{k}_{\rm R})\tilde{U}_{\rm d}(\vb{k}_{\rm R})}_{2 N}, \label{eq:nSDKud}
    \end{equation}
fairly efficiently suppress the $\Delta m=\pm 2$ leakage. Next, as detailed in Sec.~\ref{sec:uddu}, Sec.~\ref{sec:dphase}, by combining the $\tilde U_{\rm u d}^{(2 N)}$ and $\tilde U_{\rm d u}^{(2 N)}$ to form a balanced chirp-alternating sequence,
    \begin{equation}
        \tilde U_{\rm uddu}^{(4 N)} (\vb{k}_{\rm R})= \tilde U_{\rm u d}^{(2 N)} (\vb{k}_{\rm R}) \tilde U_{\rm d u}^{(2 N)} (\vb{k}_{\rm R}), \label{eq:nSDKuddu}
    \end{equation}
the dynamic phases by $\tilde U_{\rm d u}^{(2 N)}(\vb{k}_{\rm R})$ and $\tilde U_{\rm u d}^{(2 N)}(\vb{k}_{\rm R})$ robustly cancel each other. The Eq.~(\ref{eq:nSDKuddu}) scheme can therefore faithfully realize a $U^{(4N)}_{\rm K}({\bf k}_{\rm R})$ phase gate.

For comparison, here the ``traditional'' chirp-repeating sequences~\cite{lu2005,Kotru2015,Jaffe.2018} are defined as
\begin{equation}
    \tilde U_{\rm uu}^{(2 N)} (\vb{k}_{\rm R})= \underbrace{\tilde{U}_{\rm u}(-\vb{k}_{\rm R})\tilde{U}_{\rm u}(\vb{k}_{\rm R}) \cdots \tilde{U}_{\rm u}(-\vb{k}_{\rm R})\tilde{U}_{\rm u}(\vb{k}_{\rm R})}_{2 N}. \label{eq:nSDKuu}
\end{equation}
Similar $\tilde U_{\rm dd}^{(2 N)} (\vb{k}_{\rm R})$ controls are defined with $\mathrm{u}\rightarrow \mathrm{d}$ replacements.

To conclude this section, we note that while the D1 line is chosen in this work (Fig.~\ref{fig:1}), the conclusions on the ground state hyperfine control dynamics can be applied in a straightforward manner to the D2 line. Comparing with the D1 operation, Raman SDK at a same single-photon detuning $\Delta_e$ on the D2 line suffers more spontaneous emission loss due to the excitations of the cycling transitions~\cite{Sievers2015}. On the other hand, a reduced $m-$changing rate (Eq.~(\ref{eq:leakrabi})) is expected to help the D2 performance due to the typically smaller $\omega_{{\rm hfs},e}$ for the intermediate $P_{3/2}$ states.

\section{Experimental Implementation}\label{sec:expt}


\subsection{Nanosecond SDK on a delay line}\label{subsec:nsSDK}

The adiabatic SDK is implemented on the $^{85}$Rb $5S_{1/2}-5P_{1/2}$ D1 line as depicted in Fig.~\ref{fig:1},  with counter-propagating chirp pulses programmed by a wideband optical waveform generator~\cite{He2020a} on an optical delay line~\cite{He2020b}. The cross-linear polarization is realized by double-passing the light beam with a quarter waveplate before the end mirror (Fig.~\ref{fig:1b}(a)) which converts the incident ${\bf e}_x$ polarization to ${\bf e}_y$. With the OAWG output peak power limited to $P_{\rm max}\approx 20~$mW, the incident control pulse $\mathcal{E}_{1,2}$ is weakly focused to a waist radius of $w \approx \SI{13}{\micro m}$ to reach a peak Rabi frequency of $\Omega_{a(b)}\approx 2 \pi \times 2$~GHz. The imperfect retro-reflection with $\kappa\approx 0.7$ ($r=|\kappa|^2\approx 50\%$) reflectivity, primarily limited by increased focal beam size due to wavefront distortion, leads to decreased $\Omega_{b(a)}=\kappa \Omega_{b(a)}$ for the reflected pulses seen by the atomic sample. We set the single-photon detuning to be $\Delta_e=2\pi\times 10$~GHz to achieve a peak Raman Rabi frequency of $C_{\rm R}^{(0)}\approx 2\pi\times \kappa \SI{300}{MHz}$ estimated at the center of the Gaussian $\mathcal{E}_{1,2}$ beams. 
An $\tau_{\rm d}=140.37$~ns optical delay is introduced by the $L\approx 20$~m folded delay line, which is long enough to spatially resolve the counter-propagating nanosecond pulses. To form the counter-propagating $\mathcal{E}_{1,2}$ pulse pair, we pre-program $\mathcal{E}_{1,2}(t)$ and $\mathcal{E}_{2,1}(t-\tau_{\rm d})$, with a relative delay matching the optical delay line, to ensure the pulse pair with proper carrier frequency $\omega_{1,2}$ meeting head-on-head in the atomic cloud. To continue multiple SDKs, additional, individually shaped pulses with alternating $\omega_{1,2}$ can be applied, as in Fig.~\ref{fig:1b}(b), with a $T_{\rm rep}=\tau_{\rm d}$ periodicity.

Importantly, to periodically generate multiple Raman SDKs with the delay line, every shaped pulse contribute twice to the SDK sequence. To properly shape the Raman coupling $\Omega_{\rm R}(t)\propto \mathcal{E}^*_1(t)\mathcal{E}_2(t)$ (Eq.~(\ref{eq:RamanRabi})), therefore, one needs to program the incident $\mathcal{E}_{1(2)}'$ pulse according to the retro-reflected $\mathcal{E}_{2(1)}$ pulse. To clear out such pulse-history dependence in a long sequence, the retro-reflection cycles can be interrupted a few times by increasing the inter-pulse delays beyond $\tau_{\rm d}$. 
Here, if our goal is to alternate Raman chirps $\delta_{\rm R}$, as in Fig.~\ref{fig:1b}(b) and Eq.~(\ref{eq:nSDKuddu}),  then the frequency-sweeping amplitudes for every other incoming optical pulse needs to be increased by an additional $\delta_{\rm swp}$  (Sec.~\ref{sec:asdk}). As a result, in the balanced chirp-alternating scheme (Eq.~(\ref{eq:nSDKuddu})) the first $2N$ pulses becomes more and more chirped, before a reversal of the process to rewind back the rate. High optical chirping rates affect the $V'$-couplings (Eq.~(\ref{eq:heff})). In our experiments, the single-photon detuning $\Delta_e= 2\pi\times\SI{10}{GHz}$ is much larger than $\delta_{\rm swp}$, and we have numerically confirmed that the increasingly chirped waveforms do not significantly affect the Raman dynamics in the $\tilde U^{(4N)}_{\rm uddu}$ scheme up to $N=6$. Nevertheless, to avoid systematic errors associated with the digital OAWG pulse shaping~\cite{He2020a}, the chirp-rate accumulation is interrupted in this work by separating $\tilde U^{(4N)}_{\rm uddu}$ into $N$ sets of 4-pulse sequences, as described above.

We note that in the delay-line based SDK scheme, there are extra ``pre-pulses'' and ``post-pulses'' that interact with the atomic sample alone without any counter-propagating pulse to help driving the Raman transition ({\it e.g.} the first and last pulses in Fig.~\ref{fig:1b}b). Although these extra pulses impact negligibly the atomic hyperfine population and momentum transfer in this work, they cause extra dynamic phases which need to be precisely compensated for during future coherent matterwave controls. Similar to the frequency domain Stark shift compensation~\cite{Asenbaum2020}, for the nanosecond SDKs here one can fire additional pulses with suitable single-photon detunings to trim the overall dynamic phase (Eq.~(\ref{equ:dphase})). Within nanoseconds, cold atoms hardly move to change the local laser intensity. We therefore expect the dynamic phase compensation to function well in the time domain.




\subsection{Optimizing Adiabatic SDK}

\begin{figure}[htbp]
    \centering
    \includegraphics[width=\linewidth]{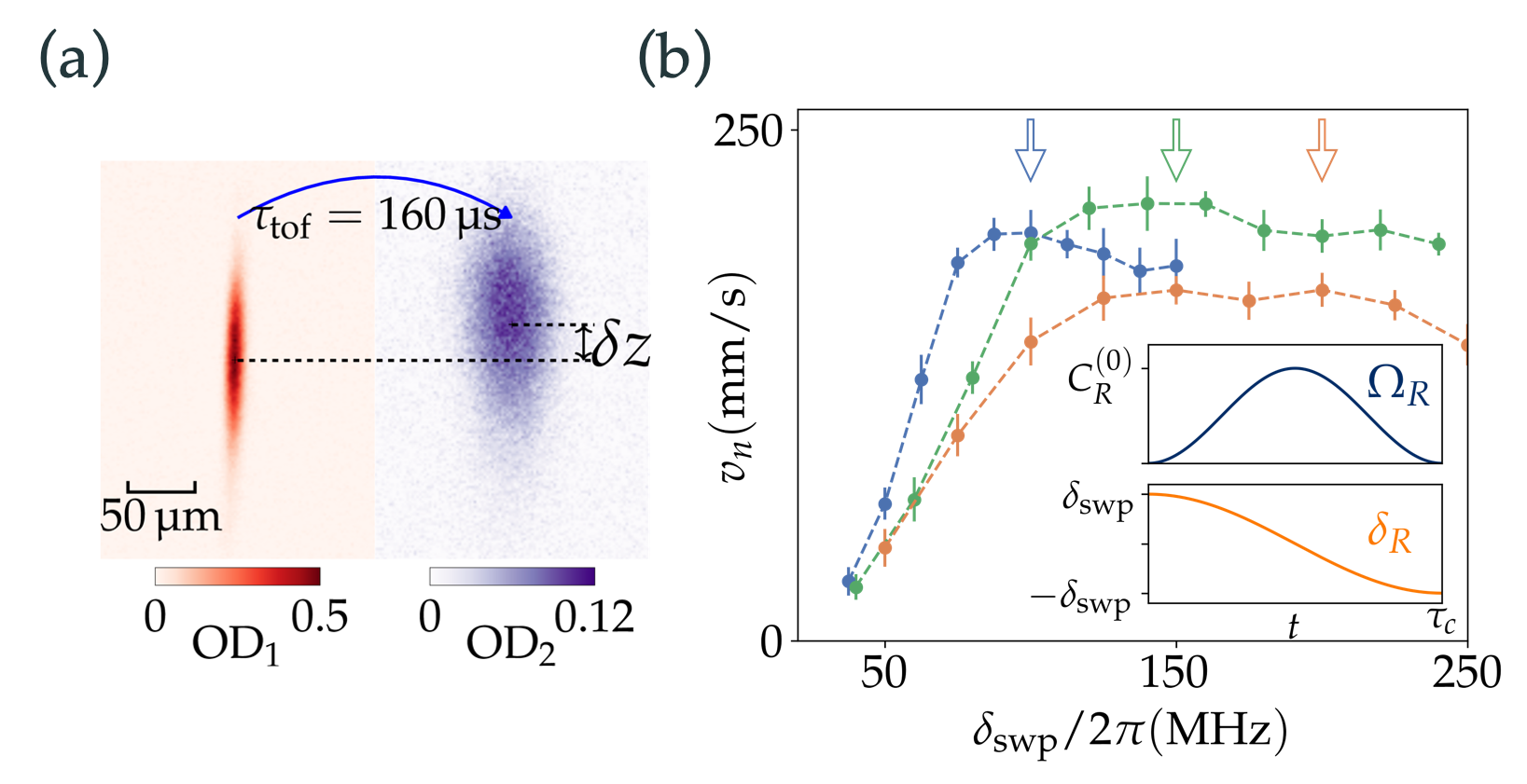}
    \caption{(a) SDK momentum transfer measurement with a double imaging method: For atomic sample subjecting $n$ SDKs ($n=25$ here) starting with $\tilde U_{\rm u}(\vb{k}_{\rm R})\tilde U_{\rm u}(-\vb{k}_{\rm R})$, a sequence of two absorption images are taken at $\bar t_1=10~\mu$s and $\bar t_2=\bar t_1+t_{\rm tof}$, separated by $\tau_{\rm tof}=\SI{160}{\mu s}$ time-of-flight and with a hyperfine repumping pulse along $x$ applied in between (see main text). The center-of-mass position shift $\delta z$ is retrieved  with a typical $\pm 2~\mu$m accuracy by fitting the optical depth (OD). The recoil velocity $v_n=\delta z/\tau_{\rm tof}$ is then estimated during typical adiabatic SDK parameter scan as in (b). Here, with $n=25$, $\tau_{\rm c}=60~$ns SDK pulses applied, $v_n$ is optimized as a function of sweep frequency $\delta_{\rm swp}$, and peak laser intensity parametrized by an estimated peak Raman pulse area $\mathcal{A}_{\rm R}= C_{\rm R}^{(0)}\tau_{\rm c}/2$. Blue, green and orange lines correspond to estimated peak $\mathcal{A}_{\rm R}$ of approximately  $6\pi, 9\pi, 12\pi$ respectively. The corresponding peak Rabi frequencies $C_{\rm R}^{(0)}$ are marked with arrows on the top along the $\delta_{\rm swp}$ axis. The error bars give statistical uncertainties of $v_n$ in repeated measurements.}
    \label{fig:measure}
\end{figure}

We prepare $N_A\sim 10^5$ $^{85}$Rb atoms in a compressed optical dipole trap at a temperature of $T\sim\SI{200}{\micro K}$~\cite{He2020a}. The atomic sample is optically pumped into the $F=2$, $|b_m\rangle$ hyperfine states, elongated along $z$, with a characteristic radius of $\sigma\approx\SI{7}{\micro m}$ in the x-y plane (Fig.~\ref{fig:measure}(a)). Immediately after the atoms released from the trap, multiple SDKs programmed on the optical delay line with alternating $\pm \vb{k}_{\rm R}$, $\vb{k}_{\rm R}=2 k_0 {\bf e}_z$ are applied to transfer photon momentum by repetitively inverting the atomic population between $F=2$, $|b_m\rangle$ and the $F=3$, $|a_m\rangle$ hyperfine states. Here $k_0=2\pi/\lambda$ is the wavenumber of the D1 line SDK pulses at  $\lambda=795$~nm. 

We use a double-imaging technique to characterize the performance of the adiabatic SDKs, by simultaneously measuring the spin-dependent momentum transfer and population inversion (Appendix~\ref{sec:abs}). Specifically, immediately after the last of $n$ SDK pulses, a probe pulse resonant to the D2 line $F=3-F'=4$ hyperfine transition is applied for $\tau_{\rm p}=\SI{20}{\micro s}$ to record the spatial distribution of atoms in state $|a\rangle$, in the $x-z$ plane, with calibrated absorption imaging~\cite{He2020b}. For atoms in $|b\rangle$, this probe is far-detuned and weak enough not to perturb the motion. Next, after a $\tau_{\rm tof}=\SI{160}{\mu }$s free-flight time, the 2nd $\tau_{\rm p}=20~\mu$s probe pulse is applied to image all the atoms. For the purpose, during the time of flight an additional $\SI{50}{\micro s}$ pulse along ${\bf e}_x$, resonant to $F=2-F'=3$ transition, repumps the $|b\rangle$ atoms to $|a\rangle$ for the 2nd imaging. By comparing atom number $N_a$ in state $|a\rangle$ and the total atom number $N_A=N_a+N_b$, inferred from the first and second images respectively, the probability of atoms ending up in $|a\rangle$ can be measured as a function of the number of SDKs $n$ as $\rho_{aa, n}=N_a/(N_a+N_b)$. In addition, by fitting both images to locate the center-of-mass vertical positions $z_{1,2}$, the atomic velocity $v_n=\delta z/ t_{\rm tof}$ can be retrieved to estimate the photon momentum transfer $ p_n=M v_n$ in unit of $\hbar k_{\rm eff}$. 

Typical $v_n$ measurement results are given in Fig.~\ref{fig:measure}(b). Here, for atoms prepared in $|b\rangle$ states subjected to an $n=25$ SDK sequence starting with $\tilde U_{\rm u}(\vb{k}_{\rm R})\tilde U_{\rm u}(-\vb{k}_{\rm R})$ (a $2N=24$ double-SDK followed by an additional kick to drive the final Raman transition), the atomic population is largely in $|a\rangle$ while $v_n$ is unidirectional along ${\bf e}_z$. We optimize $v_n$  by varying the peak Raman coupling amplitude $C_{\rm R}^{(0)}$ and sweep frequency $\delta_{\rm swp}$ of the adiabatic SDK pulses at fixed $\tau_{\rm c}$. As in Fig.~\ref{fig:measure}(b), for a fixed peak Raman pulse area $\mathcal{A}_{\rm R}$, we generally find  $\delta_{\rm swp}$ to be optimized  for efficient photon momentum transfer when it matches $C_{\rm R}^{(0)}$ (See the arrow markers in Fig.~\ref{fig:measure}(b)). However, unlike 2-level transfer~\cite{Miao2007} where an increased pulse area always leads to improved adiabaticity and population inversion robustness, here we find the peak $\mathcal{A}_{\rm R}\approx 9\pi$ reaches optimal to ensure the resilience of adiabatic SDK against the up to $50\%$ intensity variation in the setup. Larger $\mathcal{A}_{\rm R}$ is accompanied by slow decrease of $v_n$, due to increased probability of spontaneous emission. Here, for $\tau_{\rm c}=60~$ns, we need to attenuate $C^{(0)}_{\rm R}\approx 2\pi\times 150$~MHz to keep the peak $\mathcal{A}_{\rm R}\approx 9\pi$. By using the full $C^{(0)}_{\rm R}\approx 2\pi\times 200$~MHz available in this work, we are able to reduce $\tau_{\rm c}$ to 40~nanoseconds while maintaining nearly identical momentum transfer efficiency at $\delta_{\rm swp}=2\pi\times\SI{150}{MHz}$.


\subsection{Inference of $f_{\rm SDK}$}\label{sec:expInf}

\begin{figure}[htbp]
    \centering
    \includegraphics[width=\linewidth]{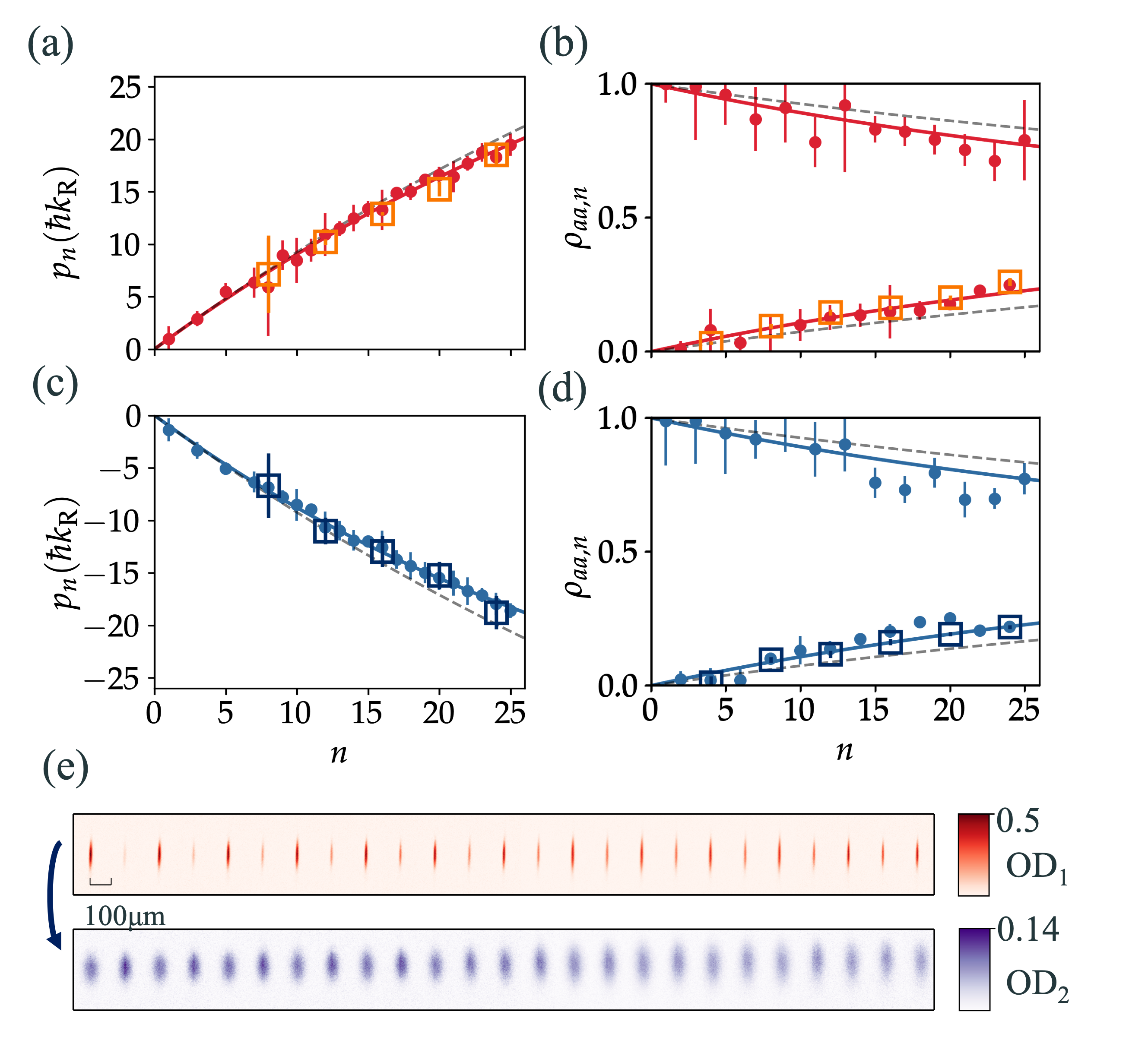}
    \caption{Measurements of recoil momentum and hyperfine population transfer by multiple SDKs optimized at $\tau_{\rm c}=40$~ns. Normalized momentum transfer $p_n$ and hyperfine population $\rho_{aa,n}$ vs kick number $n$ are plotted in (a,b)(c,d) respectively. The atoms are initialized in $F=2$, $|b_m\rangle$ states. Results for the $\tilde U_{\rm uu}^{(n)}(\pm {\bf k}_{\rm R})$ SDK sequence, starting with $\tilde U_{\rm u}(\vb{k}_{\rm R})\tilde U_{\rm u}(-\vb{k}_{\rm R})$ and  $\tilde U_{\rm u}(-\vb{k}_{\rm R})\tilde U_{\rm u}(\vb{k}_{\rm R})$, are plotted in top (a,b) and bottom (c,d) panels with disk symbols respectively. Results for the balanced chirp-alternating $\tilde U^{(4N)}_{\rm uddu}(\pm {\bf k}_{\rm R})$ (Eq.~(\ref{eq:nSDKuddu})) with $N=2-6$ are plotted via open squares. 
    For comparison, spontaneous-emission-limited $p_n$ and $\rho_{aa,n}$ according to numerical simulation are plotted with black dashed lines. 
    A phenomenological fit (solid red and blue curves) suggests hyperfine Raman transfer efficiency $f_{\rm R}= 98.8(3)\%$ for the experimental data. See discussions in the main text for inference of $f_{\rm SDK}=97.6(3)\%$ from the data. (e) Top: State-selective absorption images of atomic sample in $F=3$, $\{|a_m\rangle\}$ states, probed immediately after $n=1-25$ SDK $\tilde U_{\rm uu}^{(n)}({\bf k}_{\rm R})$ pulses. Bottom: Corresponding absorption images of optically repumped atomic samples with $\tau_{\rm tof}=\SI{160}{\micro s}$. Notice the $n$-odd measurements with higher ${\rm OD}_1$ suffer stronger probe-induced heating and thus wider ${\rm OD}_2$ distributions probed later, as expected.
    }
    \label{fig:2}
\end{figure}

With the optimal $\delta_{\rm swp}=2\pi\times\SI{150}{MHz}$ and peak $\mathcal{A}_{\rm R}\approx 9\pi$ at $\tau_{\rm c}=\SI{40}{ns}$, we now apply  $n=1-25$ SDKs to characterize the momentum transfer $p_n=M v_n$ and normalized population $\rho_{aa, n}$ as a function of kicking number $n$. Typical results are given in Fig.~\ref{fig:2}. Here the momentum change $p_n$ along ${\bf e}_z$ is again unidirectional along $z$ as in Figs.~\ref{fig:2}(a). The direction is conveniently reversed by programming $\tilde U_{\rm u}(-\vb{k}_{\rm R})$ first in the $\pm \vb{k}_{\rm R}$ alternating sequence, resulting in acceleration of atoms along $-{\bf e}_z$ instead as in Figs.~\ref{fig:2}(c). In contrast, the hyperfine population $\rho_{aa,n}$ is suppressed and revived after an even and odd number of SDKs respectively, as demonstrated in Fig.~\ref{fig:2}(c,d). In addition, we program a balanced chirp-alternating sequence $n=4 N$ that combines $\tilde U_{\rm u}(\pm \vb{k}_{\rm R})$ with $\tilde U_{\rm d}(\mp \vb{k}_{\rm R})$ as $\tilde U_{\rm u d d u}^{(4N)} (\pm \vb{k}_{\rm R})$ according to Eq.~(\ref{eq:nSDKuddu}), a sequence which will be detailed in Sec.~\ref{sec:controldynamics} for interferometric applications, with $p_n$ and $\rho_{aa,n}$ measurement results also given in Fig.~\ref{fig:2} to demonstrate similar momentum and population transfer efficiency.  

We estimate $f_{\rm SDK}$ by comparing the $p_n$ and $\rho_{aa,n}$ measurements as in Fig.~\ref{fig:2} with precise numerical modeling detailed in Appendix~\ref{sec:numerical}, taking into account the finite laser beam sizes and imperfect reflection with reflectivity $r=|\kappa|^2$. The comparison is facilitated by fitting both the measurement and simulating data according to a phenomenological model (Appendix~\ref{sec:mmodel}), which assumes that errors between successive SDKs are uncorrelated and are solely parametrized by $f_{\rm R}$, a hyperfine Raman population transfer efficiency. The model predicts exponentially reduced increments $|p_{n+1}-p_{n}|/\hbar k_{\rm eff}=|\rho_{aa,n+1}-\rho_{aa,n}|=f_{\rm R} (2 f_{\rm R}-1)^n$ for $f_{\rm R}\approx 1$ by each SDK. From measurement data in Fig.~\ref{fig:2}(a-d), $f_{\rm R}\approx 98.8\%$ can be estimated in both $\tilde U_{\rm u}(\vb{k}_{\rm R})$ and  $\tilde U_{\rm u}(-\vb{k}_{\rm R})$ kicks (Fig.~\ref{fig:2}(a,c)), slightly less than spontaneous-emission-limited $f_{\rm R}\approx 99.2\%$ predicted by numerical simulation of the experiments, assuming perfect reflection with $\kappa=1$. 

We simply attribute the slightly reduced $f_{\rm R}$ from the theoretical value to imperfect retro-reflection (Appendix~\ref{sec:inferFSDK}). In particular, we numerically find $f_{\rm R}$ reduces with $f_{\rm SDK}$ when $\mathcal{E}_{1,2}$ are unbalanced in amplitudes, so that both spin leakage and spontaneous emission are increasingly likely to occur. Taking into account the independently measured $r\approx 0.5$ in this work and with numerically matched $f_{\rm R}$, $\varepsilon_{\rm leak}\approx 2.5\%$ with $\varepsilon_{\Delta m}\approx 0.5\%$ and $\varepsilon_{\rm sp}\approx 2\%$ can be estimated respectively. We therefore infer from the combined analysis an $f_{\rm SDK}=97.6(3)\%$ for the adiabatic SDK in this experiment, slightly less than the spontaneous-emission-limited $f_{\rm SDK}\approx 98\%$ for the nearly perfect adiabatic SDK. The numbers can be improved further by increasing $\Delta_e$ to suppress spontaneous emission as well as the $\Delta m$-leakage. It is important to note that while the imperfect reflection only moderately reduce $f_{\rm SDK}$, the resulting imperfection of $\vb{k}_{\rm R}\leftrightarrow -\vb{k}_{\rm R}$ swapping in successive $\tilde U_{\rm u}(\pm \vb{k}_{\rm R})$ control can greatly compromise the cancellation of dynamic phase $\varphi_{\rm D}$ in double SDK (Eq.~(\ref{eq:dSDK}))~\cite{Kotru2015,Jaffe.2018}, a topic to be detailed in Sec.~\ref{sec:dphase}. 

The adiabatic SDK demonstrated in this experimental section is the fastest realization to neutral atoms to date~\cite{Kotru2015,Jaffe.2018}. By equipping a more powerful laser, the SDK time can be reduced to a few nanoseconds or less~\cite{Mizrahi2013}. A one-meter level compact delay line would then be able to resolve the counter-propagating pulses. Here, it is worth pointing out that the motion of the retro-reflecting mirror  (Fig.~\ref{fig:1b}(a)) shifts the SDK diffraction phases (Eq.~(\ref{eq:sdkevo}))~\cite{Peters1997}. For interferometric measurements with long interrogation time (Sec.~\ref{sec:AI}), the end mirror for the compact delay line can be vibration-isolated as those in traditional atom interferometry~\cite{Peters1997,Durfee2006,Canuel2006,Malossi2009,Zhou2015,Perrin2019,Hartmann2020}.

\section{Coherent control of spinor matterwave}\label{sec:controldynamics}

In the previous section, we experimentally characterize nanosecond adiabatic SDK by measuring the transfer of photon momentum and hyperfine population by a SDK sequence.  A natural question to ask is whether it is possible to exploit the technique for coherent spinor matterwave control.  As discussed in Sec.~\ref{sec:dsdk}, for ideal SDKs the double SDK sequence (Eq.~(\ref{eq:dSDK})) can be constructed to perform position-dependent phase gates. However, as to be illustrated in the following, the operation degrades rapidly with kick number $N$ in presence of coherent spin leakage among the $2F_b+1$ $\{|a_m\rangle,|b_m\rangle\}$ sub-spins (Sec.~\ref{sec:leak}), or when the non-perfect $\vb{k}_{\rm R}\leftrightarrow -\vb{k}_{\rm R}$ swapping introduces additional dynamic phases (Sec.~\ref{sec:dphase}).


Nevertheless, in this section we demonstrate that the $\Delta m$-leakage and dynamic phases can be efficiently suppressed by the balanced chirp-alternating SDK sequence $\tilde U^{(4N)}_{\rm uddu}(\vb{k}_{\rm R})$ introduced in Sec.~\ref{sec:cpSDK} (Eq.~(\ref{eq:nSDKuddu})) for faithful implementation of the $U^{(4 N)}_{\rm K}(\vb{k}_{\rm R})$ phase gate by Eq.~(\ref{eq:dSDK}) to finely enable high-efficiency large momentum transfer~\cite{McGuirk2000}.  We further numerically demonstrate the utility of the tailored adiabatic SDK sequence by simulating an area-enhanced atom interferometry sequence, using the experimental parameters both within and beyond this experimental work.

We notice control of spin leakage in quasi-two-level systems is an important topic in quantum control theory~\cite{Genov2013,Kang2018,Jo2019}. Previous studies on the topic typically involve a Morris–Shore transformation of the interaction matrix to decompose the multi-level dynamics~\cite{Shore2014}. However, as being schematically summarized in Fig.~\ref{fig:1}(b), the spin leakages are coherently driven through multiple paths with multiple Raman couplings to preclude a straightforward transformation, nor a direct application of the associated leakage-suppression techniques~\cite{Genov2013,Kang2018,Jo2019}.

\subsection{Average gate fidelity}

To evaluate an imperfect double-SDK sequence $\tilde U^{(2N)}(\vb{k}_{\rm R};\eta)$ as a quantum gate, we define an average fidelity for its performance on arbitrary spinor matterwave states of interest, {\it e.g.}, spatially within the focal laser beam (Fig.~\ref{fig:1b}(a)) and internally span the $\{|a_m\rangle,|b_m\rangle\}$ sub-spin space. The average fidelity is written as~\cite{Magesan2012}
\begin{equation}
    \mathcal{F}^{(2N)}_{m}=\left|\langle \psi_{m,j}|U^{(2N)\dagger}_{\rm K}(\vb{k}_{\rm R}) \tilde U^{(2N)}(\vb{k}_{\rm R};\eta) |\psi_{m,j}\rangle \right|^2_{\eta,j}.\label{equ:f2nSDK}
\end{equation}
Similar to Eq.~(\ref{equ:fSDK}), here $\eta$ represents the Hamiltonian parameter in Eq.~(\ref{eq:heff}) including the atomic position ${\bf r}$ of interest. For internal states the evaluation samples $|\psi_{m,j}\rangle$ as the six eigenstates of $\sigma_{x,y,z}^{(m)}$, listing as $\{|\psi_{m,j}\rangle\}=\{|b_m\rangle,|a_m\rangle,(|b_m\rangle \pm |a_m\rangle)/\sqrt{2},(|b_m\rangle\pm i |a_m\rangle)/\sqrt{2}\}$. With external motion ignored, $ \mathcal{F}^{(2N)}_{m}\approx 1$ ensures that the effectively instantaneous $\tilde U^{(2N)}({\bf k}_{\rm R};\eta)$ faithfully apply the phase gate to arbitrary spinor matterwave states  near the beam focus, with certain laser intensity of interest.

For the convenience of related discussions, we define an average $\Delta m$-leakage probability, similar to Eq.~(\ref{equ:lSDK}), as
\begin{equation}
    \varepsilon^{(n)}_{m,{\Delta m}}=1-\left\langle  \langle \tilde \psi_{m,j}^{(n)} |{\bf 1}^{(m)}|\tilde \psi_{m,j}^{(n)}\rangle \right\rangle_{\eta, j}\label{equ:l2nSDK}
\end{equation}
where $|\tilde \psi_{m,j}^{(n)}\rangle=\tilde U^{(n)}(\vb{k}_{\rm R};\eta) |\psi_{m,j}\rangle$ is defined as the final atomic state after the imperfect control. To exclude spontaneous emission, we simply renormalize to have $\langle \tilde \psi_{m,j}^{(n)}|\tilde \psi_{m,j}^{(n)}\rangle=1$ before the evaluation of the $\Delta m$-leakage. Similar to Eq.~({\ref{equ:l2nSDK}}), here the $\Delta m$-leakage probability is also evaluated for a specific $m$-sub-spin of interest. Similar to Eqs.~(\ref{equ:fSDK})(\ref{equ:lSDK}), here we  expect $\mathcal{F}^{(n)}_{m}\leq 1-\varepsilon^{(n)}_{m,{\Delta m}}$, {\it i.e.}, any spin leakage results in gate infidelity.

\subsection{Simple double-SDK: spin leakage}\label{sec:leak}
\begin{figure}[htbp]
    \centering
    \includegraphics[width=\linewidth]{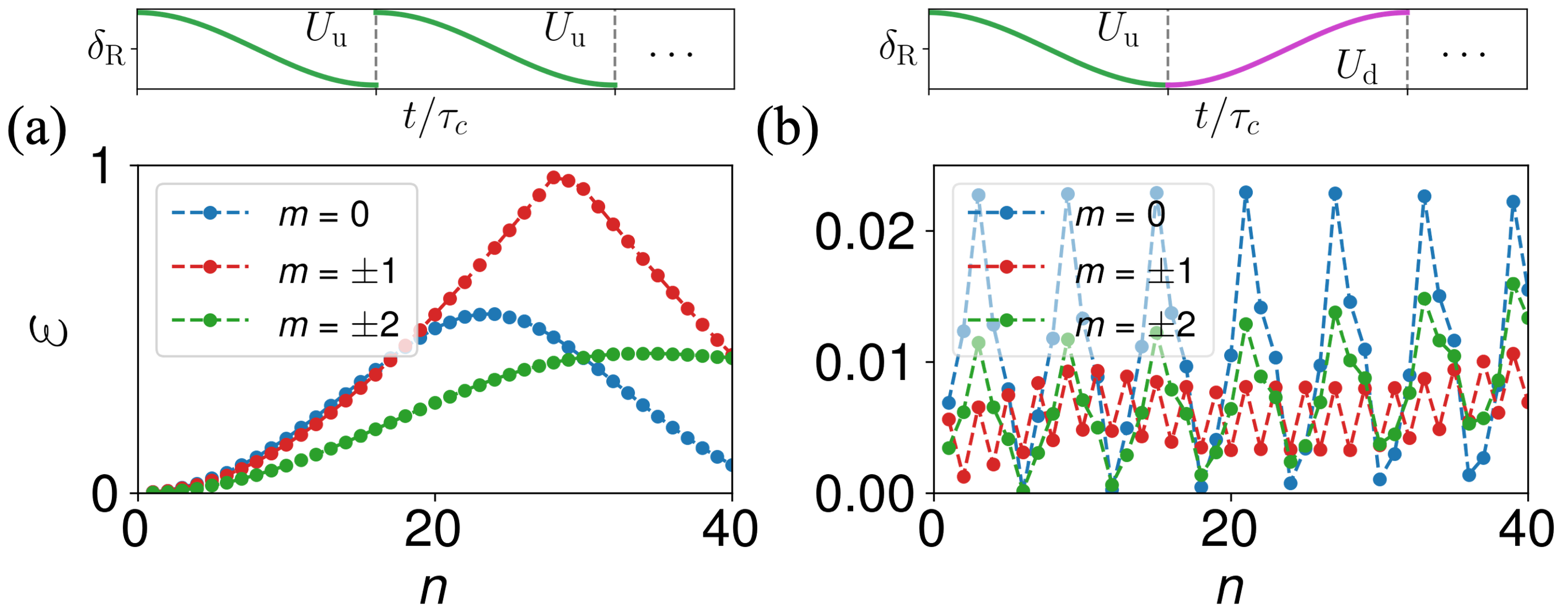}
    \caption{Numerical results of spin leakage $\varepsilon=\varepsilon^{(n)}_{m, \Delta m}$ for the traditional $\tilde U^{(n)}_{\rm uu}$ (left) and chirp-alternating $\tilde U^{(n)}_{\rm ud}$ (right) controls as a function of pulse number $n$, The peak Raman pulse area is $\mathcal{A}_{\rm R}=9\pi$. The corresponding 2-photon detuning profiles are illustrated on top of (a,b). Additional parameters for the simulations are according to the experimental Sec.~\ref{sec:expInf}.}
    \label{fig:leak-n}
\end{figure}
\begin{figure}[t]
    \centering
    \includegraphics[width=\linewidth]{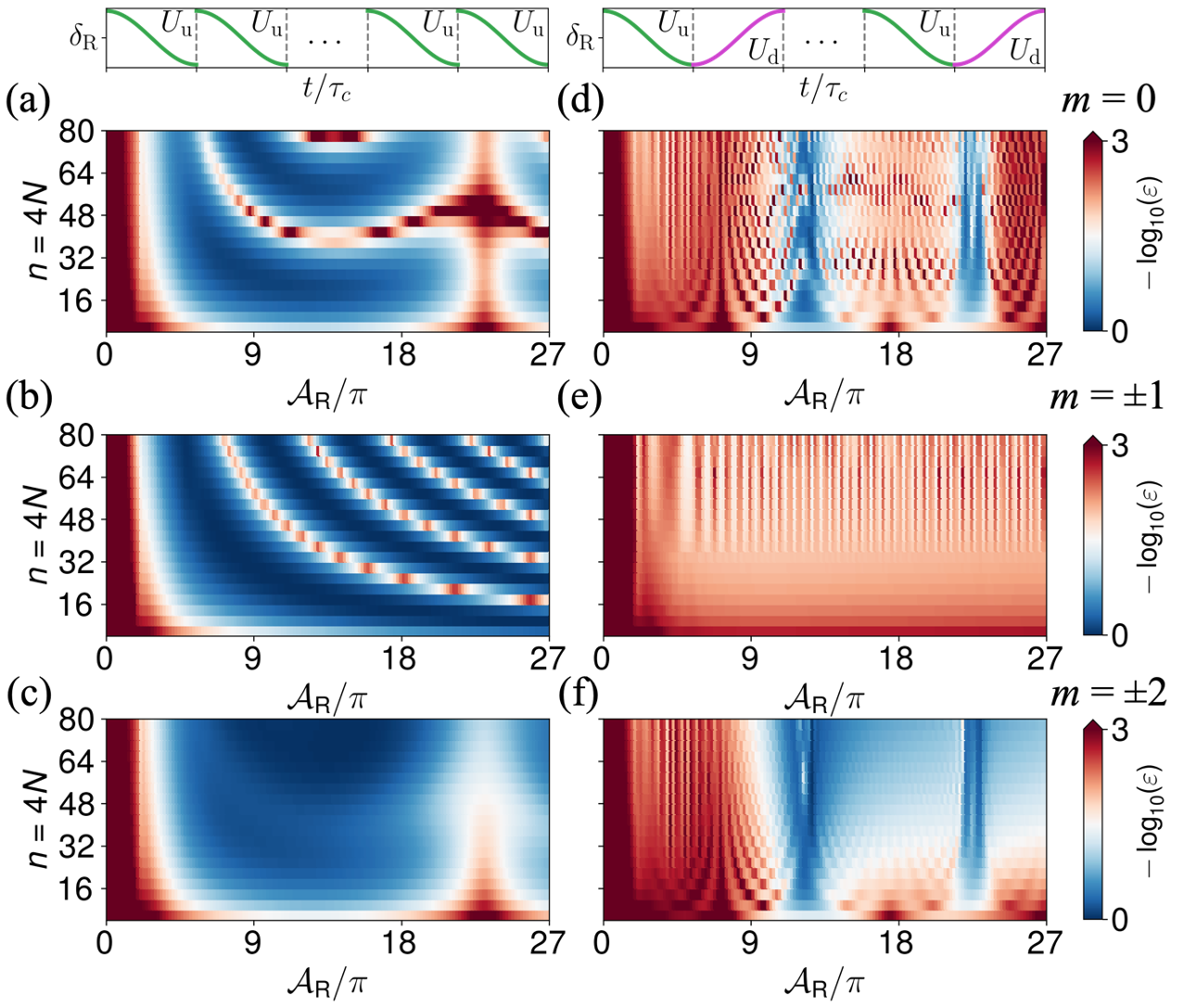}
    \caption{
    Numerical results of spin leakage $\varepsilon=\varepsilon^{(4N)}_{m, \Delta m}$ for  the traditional $\tilde U^{(4N)}_{\rm uu}$ (left) and  chirp-alternating $\tilde U^{(4N)}_{\rm ud}$ (right) as a function of kicking number $4 N$ and $\mathcal{A}_{\rm R}$, with $m=0$ (a,d), $m=\pm$ (b,e),  $m=\pm2$ (c,f). The 2-photon detuning profiles are illustrated on top of (a,d). Additional parameters for the simulations are according to the experimental Sec.~\ref{sec:expInf}.}    
     \label{fig:leak}
\end{figure}

We first consider the chirp-repeating double-SDK sequences $\tilde U^{(2N)}_{\rm uu}(\vb{k}_{\rm R})$ and $\tilde U^{(2N)}_{\rm dd}(\vb{k}_{\rm R})$ with Eq.~(\ref{eq:nSDKuu}). The simple sequences are widely applied in previous works to drive multiple recoil momentum transfer through both optical~\cite{Miao2007,Muniz2018} and Raman excitations~\cite{Kotru2015,Jaffe.2018}. In absence of the spin leakage, we expect both $\tilde U^{(2N)}_{\rm uu}(\vb{k}_{\rm R})$ and $\tilde U^{(2N)}_{\rm dd}(\vb{k}_{\rm R})$ to amplify the momentum transfer with cancelled dynamic phase~\cite{Kotru2015,Jaffe.2018}, as suggested by Eq.~(\ref{eq:dSDK}). 

\begin{figure}[t]
    \centering
    \includegraphics[width=\linewidth]{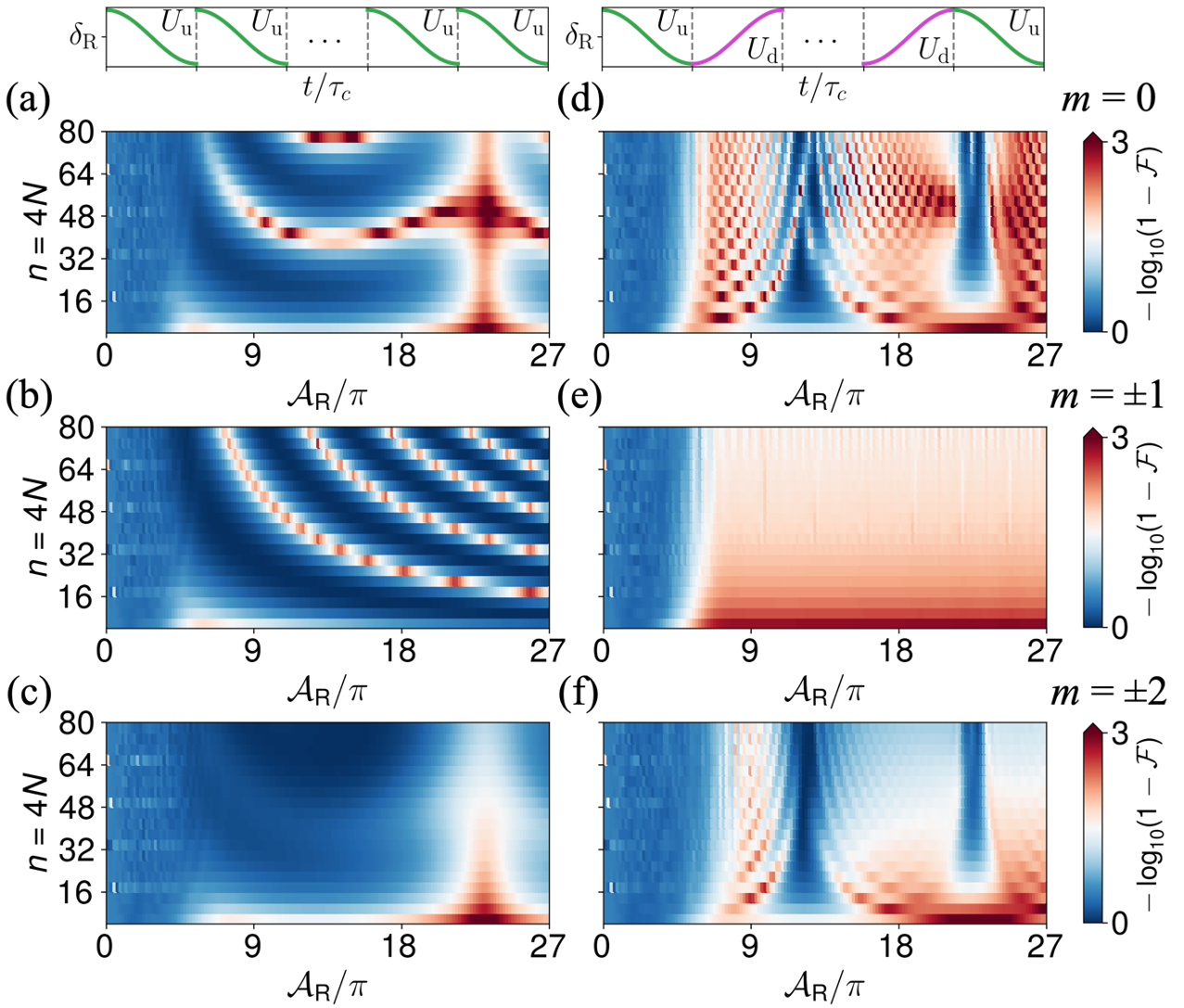}
    \caption{ Numerical results of gate infidelity $1-\mathcal{F}$ for  the traditional $\tilde U^{(4N)}_{\rm uu}$ (left panels) and  balanced chirp-alternating $\tilde U^{(4N)}_{\rm uddu}$ (right panels) controls. With $\mathcal{F}=\mathcal{F}_m^{(4N)}$ and $m=0$ (a,d), $m=\pm1$ (b,e),  $m=\pm2$ (c,f). The 2-photon detuning profiles are illustrated on top of (a,d). Additional parameters for the simulations are according to Sec.~\ref{sec:expInf}. } 
     \label{fig:gatefidelity}
\end{figure}

Here, we numerically investigate the $\Delta m$-leakage probability ${\varepsilon}^{(n)}_{m,{\Delta m}}$ and the gate fidelity $\mathcal{F}^{(n)}_m$, for an atom initialized in an $\{|a_{m}\rangle,|b_{m}\rangle\}$ sub-spin being subjected to $\tilde U^{(n)}_{\rm uu}(\vb{k}_{\rm R})$. With $\Delta_e\gg \delta_{\rm swp}$, the results for $\tilde U^{(n)}_{\rm dd}(\vb{k}_{\rm R})$ are effectively identical. The Hamiltonian parameters in the simulation follow the experimental setup described in Sec.~\ref{sec:expt} for the $^{85}$Rb D1 line scheme (in particular $\delta_{\rm swp}=2\pi\times 150$~MHz), except here the reflective coefficient is set as $\kappa=1$ for simplicity, and $\Gamma_e=0$ to focus on the coherent control dynamics.
To elucidate the role of laser intensity for the coherent control, we repeat the simulation while scanning the ${\bf E}_{1,2}$ laser intensities in proportion, parametrized by the Raman pulse area $\mathcal{A}_{\rm R}$ in the following. Typical results for $^{85}$Rb are shown in the left panels of Fig.~\ref{fig:leak-n}, Fig.~\ref{fig:leak} and Fig.~\ref{fig:gatefidelity}.

We first discuss the $\Delta m$-leakage probability $\varepsilon^{(n)}_{m,\Delta m}$ as a function of kick number $n$ for typical $\mathcal{A}_{\rm R}=9\pi$ presented in Fig.~\ref{fig:leak-n}(a). Here, for $n=1$, the tiny $\varepsilon^{(n)}_{m,\Delta m}$ for all the $m-$ sub-spins are close to the single kick leakage of $\sim 0.5\%$ as being inferred experimentally. However, with increased $n$, $\varepsilon^{(n)}_{m,\Delta m}$ increases rapidly to approach unity for merely $n\sim 20$, {\it i.e.}, atoms starting from any of the sub-spins $\{|a_m\rangle,|b_m\rangle\}$ have a substantial probability of ending up in a different $m-$subspace. 

To investigate the laser intensity dependence, the $\Delta m$-leakage ${\varepsilon}^{(2N)}_{m,\Delta m}$ is further plotted in Figs.~\ref{fig:leak}(a-c) (left panels) as a function of both the kicking number $n=4 N$ and the pulse area $\mathcal{A}_{\rm R}$, for $m=0,\pm1,\pm2$ sub-spins respectively. The choice of $n=4 N$ is for comparison with the balanced chirp-alternating sequence to be discussed shortly. We see that over a broad range of pulse area $\mathcal{A}_{\rm R}$, the spin-leakage probability $\varepsilon^{(4N)}_{m,{\Delta m}}$ increases rapidly with $4N$ in oscillatory fashions. There is hardly any continuous region of laser intensity with  ${\varepsilon}^{(4N)}_{m,\Delta m}<0.1$. 

It is important to note that although the $\Delta m$-leakages hardly affect the recoil momenta and hyperfine population transfers (Sec.~\ref{sec:expt}), they do limit the gate fidelity for faithful spinor matterwave control. The impact of spin leakage to average gate fidelity, $\mathcal{F}^{(4 N)}_m$, is demonstrated by comparing Figs.~\ref{fig:leak}(a-c) with Figs.~\ref{fig:gatefidelity}(a-c). Here we see the gate infidelity $1-\mathcal{F}^{(4 N)}_m$ closely follows $\varepsilon^{(4 N)}_{m,{\Delta m}}$ to hardly reach $0.1$ over most laser intensities, except when the laser intensity is too low to adiabatically drive the Raman transition at all  (with $\mathcal{A}_{\rm R}<3\pi $ here) where we instead  find $\mathcal{F}^{(4 N)}_m\approx 0.5$, as expected.



Clearly, the coherent accumulation of $\Delta m$-leakage error as in Figs.~\ref{fig:leak}(a-c) needs to be suppressed before the adiabatic SDK sequence can be exploited for coherent control of spinor matterwave. Traditional methods for such suppression include applying a moderate quantization field to lift the Zeeman degeneracy~\cite{Kasevich1992, Kotru2015,Jaffe.2018, foot:circular}. This Zeeman shift approach rules out the possibility of parallel multi-Zeeman spin wave coherent control prescribed by Eq.~(\ref{eq:dSDK}). Furthermore, since the Zeeman shifts need to be substantially larger than the SDK bandwidth, a uniform field at a kilo-Gauss level is likely required for the nanosecond operations, which is not favored in precision metrology with alkaline atoms. 

\subsection{Balanced chirp-alternating SDK scheme}\label{sec:uddu}

We now demonstrate that the direction of chirps in an adiabatic sequence can be programmed to suppress the coherent accumulation of $\Delta m$-leakage. Furthermore, by balancing $\tilde U_{\rm u d}^{(2 N)}$ with $\tilde U_{\rm d u}^{(2 N)}$ as those in Eq.~(\ref{eq:nSDKuddu}), the dynamic phase can be cancelled in $\tilde U_{\rm u d d u}^{(2 N)}$ to support faithful spinor matterwave phase gate. As to be clarified in the following, both the non-adiabatic spin-leakage suppression and dynamic phase cancellation are supported by time-reversals of the driven spin dynamics, much like those in the traditional spin-echo schemes, but is achieved here in the adiabatic SDK sequence to also ensure the laser intensity-error resilience. We note the benefit of alternating the chirp directions was also discovered in the context of 2-level atom slowing with adiabatic pulses~\cite{Stack2011}, but otherwise rarely explored before.

To understand why the accumulation of $\Delta m$-leakage error can be partly suppressed by the chirp-alternating $\tilde U_{\rm ud}^{(2 N)} (\vb{k}_{\rm R})$ or $\tilde U_{\rm du}^{(2 N)} (\vb{k}_{\rm R})$ sequence prescribed by Eq.~(\ref{eq:nSDKud}), we come back to Eq.~(\ref{eq:heff}) to better understand the $\Delta m=\pm 2$ leakage itself. In particular, for atom starting in $|a_m\rangle$ or $|b_m\rangle$ and subjected to a close-to-ideal adiabatic SDK control, with an $H_{\rm u}({\bf r},t)$ Hamiltonian during $0<t<\tau_{\rm c}$ as prescribed in Sec.~\ref{sec:cpSDK}, the $\Delta m=\pm 2$ transitions driven by $V'$ are often a result of non-adiabatic couplings among sub-spins with nearly equal Stark shifts (for example, the $m=\pm1$ sub-spins in Fig.~\ref{fig:1}b). By reversing the time-dependence of the Hamiltonian, here with $H_{\rm d}({\bf r},t)=H_{\rm u}({\bf r},\tau_{\rm d}-t)$ within $\tau_{\rm d}<t<\tau_{\rm d}+\tau_{\rm c}$, the sign of the  non-adiabatic couplings are reversed. Such a sign reversal would lead to complete cancellation of the non-adiabatic transitions if the adiabatic states involved in the couplings are truly degenerate. Here, for atoms being addressed by the cross-linear polarized light, the fact that the $m-$sub-spins are nearly degenerate makes the sign reversal efficient for the suppression of the unwanted leakages. 

The simple picture of coherent leakage suppression is verified with numerical simulation in Fig.~\ref{fig:leak-n}(b) for the chirp-alternating $ \tilde U_{\rm ud}^{(2 N)} (\vb{k}_{\rm R})$ sequence. Here, in contrast to the $ \tilde U_{\rm uu}^{(2 N)} (\vb{k}_{\rm R})$ case in Fig.~\ref{fig:leak-n}(a), the spin leakage $\varepsilon_{m, \Delta m}^{(n)}$ for all the $m=0,1,2$ oscillates and is overall efficiently suppressed. Comparing with $m=\pm 1$ where the leakage is suppressed for even $n$, the leakage suppression from the $m=0,\pm2$ sub-spin follows a more complicated pattern with an approximate periodicity of 4 to 5, suggesting more complex multi-level dynamics. We further investigate the leakage dynamics by plotting 
$\varepsilon_{m, \Delta m}^{(n)}$ in 2D vs $n$ and pulse area $\mathcal{A}_{\rm R}$. The results for $n=4 N$ are presented in Fig.~\ref{fig:leak}(d-f) to be compared with those in Fig.~\ref{fig:leak}(a-c). For the $m=\pm 1$ sub-spins, the $\Delta m$-leakage is suppressed to $\sim 1\%$, limited by their residual couplings to the $|a_m\rangle$ states with $m=\pm 3$ (Fig.~\ref{fig:1}(a)).  In Appendix~\ref{sec:diffatom} we demonstrate that the spin-leakage suppression becomes perfect in species with nuclear spin $I=1.5$ ($^{87}$Rb-like)
for the pair of $m=\pm 1$ sub-spins. On the other hand, for $m=0,\pm 2$ sub-spins here, there are stripes of $\mathcal{A}_{\rm R}$-region (around $\mathcal{A}_{\rm R}=12\pi$ and $\mathcal{A}_{\rm R}=22\pi$ here for example) where the leakage still accumulate with increased $n=4N$, even though the suppression still works fairly well ($\varepsilon_{m, \Delta m}^{(4N)}<10\%$) for most of other $\mathcal{A}_{\rm R}$. The intricate $\mathcal{A}_{\rm R}$-dependent $\varepsilon^{(2 N)}_{m, \Delta m}$ as in Figs.~\ref{fig:leak}(d)(f) suggests that the difference of dynamic phases among $m=0, \pm2$ sub-spins by each SDK is large enough to affect the coherent leakage cancellation~\cite{foot:circular}. 

We leave a detailed investigation of the intricate coupling dynamics for future work. Here, to construct faithful spinor matterwave control, we simply combine the $\tilde U_{\rm u d}^{(2 N)}$ and $\tilde U_{\rm d u}^{(2 N)}$ sequence in Eq.~(\ref{eq:nSDKuddu}) to form the balanced chirp-alternating SDK sequence. The idea is to exploit the time-reversal dynamics again to let the dynamic phases by the leakage-suppressing $\tilde U_{\rm d u}^{(2 N)}(\vb{k}_{\rm R})$ and $\tilde U_{\rm u d}^{(2 N)}(\vb{k}_{\rm R})$ cancel each other.

We evaluate the average gate fidelity for realizing the quantum gate for $U_{\rm K}^{(4 N)}(\vb{k}_{\rm R})$ (Eq.~(\ref{eq:dSDK})) with $\tilde U_{\rm uddu}^{(4 N)} (\vb{k}_{\rm R})$. The results of $\mathcal{F}_m^{(4N)}$, with otherwise identical Hamiltonian parameters as those in Fig.~\ref{fig:gatefidelity}(a-c), are shown in Fig.~\ref{fig:gatefidelity}(d-f). Similar to Figs.~\ref{fig:gatefidelity}(a-c), here the infidelity $1-\mathcal{F}_m^{(4N)}$ for the $\tilde U_{\rm uddu}^{(4 N)} (\vb{k}_{\rm R})$ sequence largely follows $\varepsilon_{m,\Delta m}^{(4N)}$ for $\tilde U_{\rm ud}^{(4 N)}$. Therefore, the improvement is most significant for the $m=\pm 1$ sub-spins to approach $\mathcal{F}_m^{(4N)}\approx 98\%$ at $4N=80$. For $m=0,\pm 2$ sub-spins, we also see improved gate fidelity $\mathcal{F}^{(4N)}_m> 90\%$ span a substantial range of intensity for $4 N$ up to 80, with lower $\mathcal{F}$ values for $\mathcal{A}_{\rm R}>6\pi $ confined around specific $\mathcal{A}_{\rm R}$ only (deep-blue area). By moderately increasing $\Delta_e$, in Appendix~\ref{sec:diffatom} we demonstrate $\mathcal{F}>99\%$ with more confined low-$\mathcal{F}$ region after many kicks. Experimentally, the narrow low-$\mathcal{F}$ intensity region should be avoided for faithful multi-Zeeman spinor matterwave control of $^{85}$Rb with the balanced chirp-alternating scheme.

Finally, it is interesting to note that for the nearly perfect SDK, the gate infidelity $1-\mathcal{F}^{(4N)}_m$ is dominantly due to the spin leakage, as demonstrated by the remarkably similar $1-\mathcal{F}^{(4N)}_m$ and $\varepsilon^{(4N)}_{m,\Delta m}$ data in Fig.~\ref{fig:leak} and Fig.~\ref{fig:gatefidelity}. This is a result of dynamic phase cancellation in the adiabatic limit for both the $\tilde U_{\rm uu}^{(4N)}$ and $\tilde U_{\rm uddu}^{(4N)}$ sequences, when the $\pm \vb{k}_{\rm R}$ swapping as detailed next is perfect.

\subsection{Robust cancellation of dynamic phase}\label{sec:dphase}

\begin{figure}[t]
    \centering
    \includegraphics[width=\linewidth]{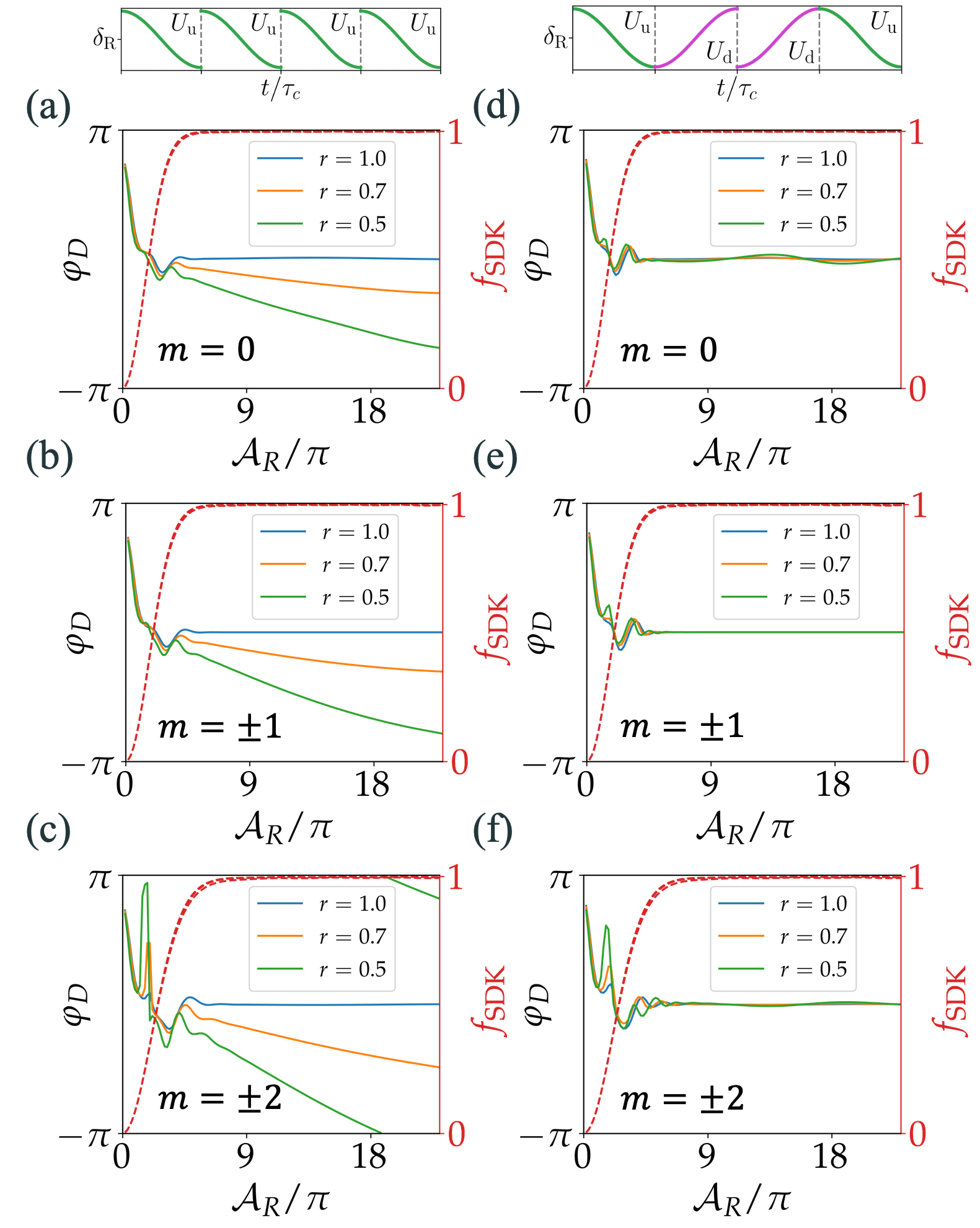}
    \caption{
        Unbalanced dynamic phase $\varphi_{\rm D}=\varphi^{(4)}_{{\rm D},m}$ according to Eq.~(\ref{equ:dphase}) vs pulse area $\mathcal{A}_{\rm R}$ for $m=0$ (a,d), $m=\pm 1$ (b,e) and $m=\pm 2$ (c,f) subjected to $\tilde U_{\rm uu}^{(4)}(\vb{k}_{\rm R})$ (left)  and  $\tilde U_{\rm uddu}^{(4)}(\vb{k}_{\rm R})$ (right) controls. The corresponding 2-photon detuning profiles are illustrated on top of (a,d). Parameters for the simulations are according to the experimental Sec.~\ref{sec:expInf}, except here the mirror reflectivity $r$ varies. The red dashed lines give $f_{\rm SDK}$ for single kicks (different $r$-curves closely overlap).  }
    \label{fig:dphase}
\end{figure}

The numerical results in Fig.~\ref{fig:gatefidelity} demonstrate that precise dynamic phase cancellation can be achieved by pairing $\tilde U^{(2N)}_{\rm ud}(\vb{k}_{\rm R})$ with $\tilde U^{(2N)}_{\rm du}(\vb{k}_{\rm R})$ into the balanced $\tilde U^{(4N)}_{\rm uddu}(\vb{k}_{\rm R})$. In fact, we find that the dynamic phase cancellation in the balanced chirp-alternating scheme is substantially more robust than the traditional double-SDK  by Eq.~(\ref{eq:dSDK}), as following.

As in Sec.~\ref{sec:dsdk}, the traditional method of dynamic phase cancellation~\cite{Jaffe.2018} requires perfect $\vb{k}_{\rm R}\leftrightarrow -\vb{k}_{\rm R}$ swapping for the successive $U_{\rm K}(\vb{k}_{\rm R})$ $U_{\rm K}(-\vb{k}_{\rm R})$ controls. Practically the k-vector swapping is typically accompanied by a modification of ${\bf E}_{1,2}$ intensity ratio. For example, in the retro-reflection setup (Fig.~\ref{fig:1b}(a)), the amplitude of the reflected beam is reduced by a $\kappa<1$ factor due to the imperfect reflection, leading to unbalanced dynamic phases $\varphi_{\rm D}$ associated with $\tilde U(\pm \vb{k}_{\rm R})$ to compromise their cancellation in the traditional double-SDK (Eq.~(\ref{eq:dSDK}))~\cite{Kotru2015,Jaffe.2018}. In fact, this systematic exists quite generally in retro-reflection setups since the 2-photon shift $\delta_0$ is sensitive to the laser intensities ratios~\cite{Weiss1994,GustavsonThesis}.


In contrast, here we notice that in the $\tilde U^{(4 N)}_{\rm uddu}(\vb{k}_{\rm R})$ sequence (Eq.~(\ref{eq:nSDKuddu})) the dynamic phase by any $\tilde U^{(2)}_{\rm du}(\vb{k}_{\rm R})$ pair is expected to be cancelled by a $\tilde U^{(2)}_{\rm ud}(\vb{k}_{\rm R})$ pair later. In the adiabatic limit the cancellation is guaranteed, since for free atom starting from any specific 2-level spin state, the sign of the chirp frequency $\delta_{\rm swp}$ dictates the adiabatic quantum number~\cite{Zhu2002} and thus the  sign of the dynamic phase in the adiabatic limit. 

To demonstrate the robust dynamic phase cancellation, in Fig.~\ref{fig:dphase}, we compare  $\varphi_{\rm D}$ according to Eq.~(\ref{equ:dphase}) for the $\tilde U^{(4)}_{\rm uu}(\vb{k}_{\rm R})$ and $\tilde U^{(4)}_{\rm uddu}(\vb{k}_{\rm R})$ controls. The $f_{\rm SDK}$ values are given in the same plots, with which we see that the high $f_{\rm SDK}$ in the quasi-adiabatic regime to be hardly affected even by a poor reflectivity $r=|\kappa|^2=0.5$. On the other hand, in contrast to Fig.~\ref{fig:dphase}(a-c) (left) where $r\approx 1$ is required for precise suppression of $\varphi_{\rm D}$ (blue line), in Fig.~\ref{fig:dphase}(d-f) (right) the $\varphi_{\rm D}$ is largely suppressed so long as $f_{\rm SDK}\approx 1$. It is worth noting that the residual variation of $\varphi_{\rm D,m}^{(4)}$ for $\tilde U^{(4)}_{\rm uddu}(\vb{k}_{\rm R})$ 
in Fig.~\ref{fig:dphase}(d-f) in the quasi-adiabatic regime are coupled to the coherent spin leakage dynamics (Fig.~\ref{fig:leak}(d-f)). Nevertheless, even for $r=0.5$ similar to the experimental situation in this work, the residual phase variations are still limited to $|\varphi_{\rm D,m}^{(4)}|<$50~mrad for the $m=0,\pm 2$ sub-spins, and $|\varphi_{{\rm D},m}^{(4)}|<$3~mrad for the $m=\pm 1$ sub-spins, respectively.

\subsection{Area-enhancing atom interferometry}\label{sec:AI}

So far in this section, we have shown that matterwave phase gate prescribed by the ideal double-SDK (Eq.~(\ref{eq:dSDK})) 
can be faithfully implemented by the balanced chirp-alternating sequence $\tilde U^{(4N)}_{\rm uddu}(\vb{k}_{\rm R})$, to coherently shift any $|a_m\rangle$, $|b_m\rangle$ components of hyperfine spinor matterwave with opposite $\pm 4N\hbar \vb{k}_{\rm R}$ momentum within nanoseconds. For control parameters in this experimental demonstration, our numerical results already suggest fairly high gate fidelity with efficient suppression of the coherent spin leakage and inhomogenuous dynamic phase. In future work, by increasing $\Delta_e/\Gamma$ and $\Delta_e/\Delta_{\rm hfs,e}$ ratios and the laser intensities $I_{1,2}$ in proportion
(Appendix~\ref{sec:diffatom}), the residual imperfections can be further suppressed to meet the exquisite requirements in the applications of quantum information processing~\cite{Garcia-Ripoll2003, Duan2004,
Schafer2018,Fluhmann2019} and quantum enhanced atom interferometry~\cite{Szigeti2020, Wu2020,Anders2021,Greve2021}

\begin{figure}[t]
    \centering
    \includegraphics[width=\linewidth]{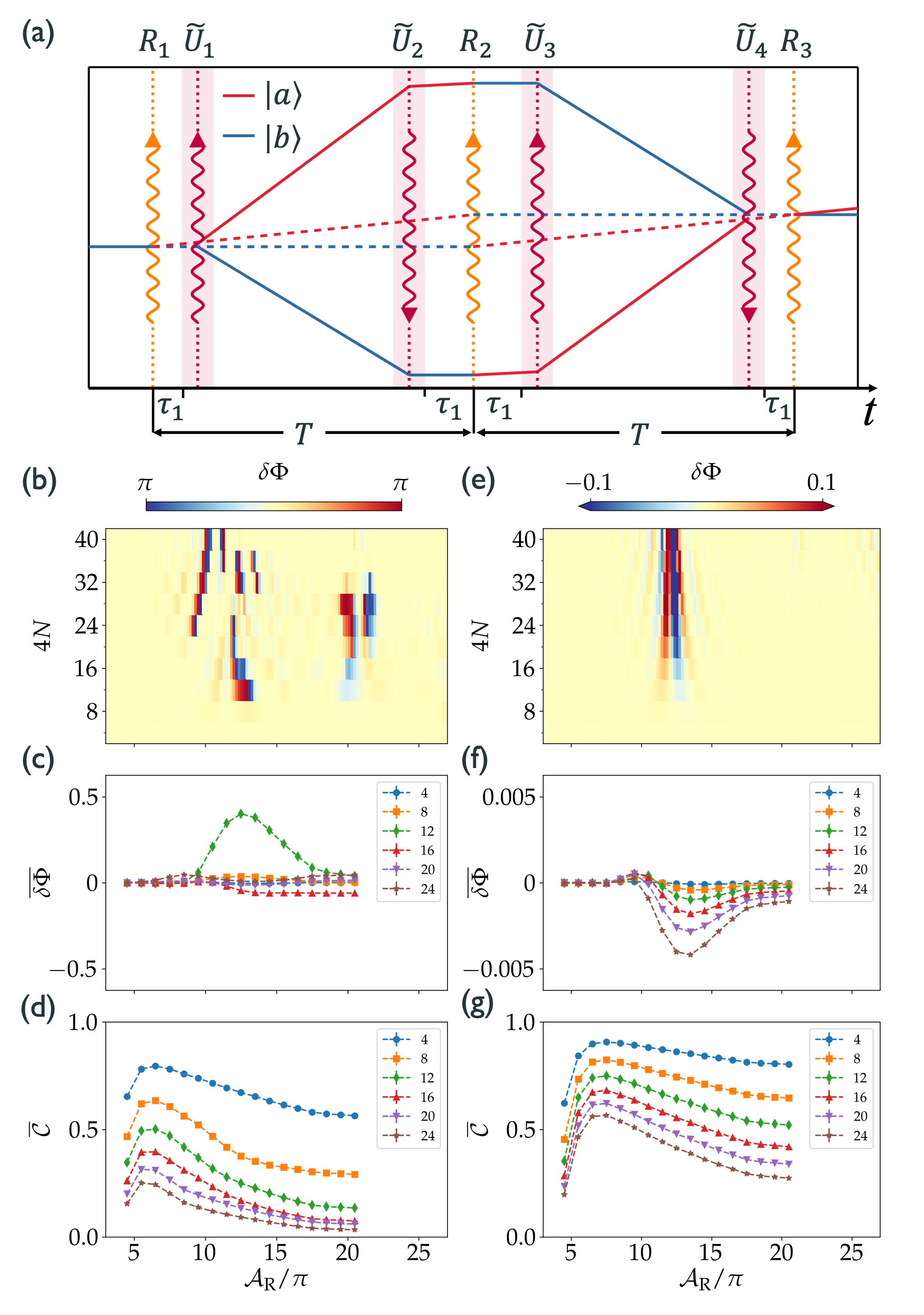}
    \caption{Enhancing the enclosed area of an atom interferometer with four $\tilde U_{\rm uddu}(\pm \vb{k}_{\rm R})$ SDK sequences. (a): Schematic of the interferometry scheme. The dashed lines at $t=0,T,2T$ represent regular Raman interferometry controls for $R_1=R_{\varphi}(\pi/2)$ splitter, $R_2=R_{\varphi}(\pi)$ mirror and $R_3=R_{\varphi}(\pi/2)$ combiner of the spinor matterwave respectively with $\varphi=\vb{k}_{\rm R}\cdot {\bf r}$. The atomic wavefunctions in $|b\rangle$ and $|a\rangle$ states are represented by the blue and red lines~\cite{McGuirk2000}. The four thick vertical lines at $t=\tau_1,T-\tau_1, T+\tau_1, 2T-\tau_1$ with red curved arrows represent $\tilde{U}_1=\tilde U_{\rm uddu}^{(4N)} (\vb{k}_{\rm R})$, $\tilde{U}_2=\tilde U_{\rm uddu}^{(4N)} (-\vb{k}_{\rm R})$, $\tilde{U}_3=\tilde U_{\rm uddu}^{(4N)} (-\vb{k}_{\rm R})$ and $\tilde{U}_4=\tilde U_{\rm uddu}^{(4N)} (\vb{k}_{\rm R})$ sequences respectively. We consider $\tau_1\ll T$.  By properly choosing $\tau_1/T$ ratio, spurious interference by multiple imperfect controls can be suppressed, and are not included in the simulations. (b-d): The interferometry phase offset $\delta \Phi$ and contrast $\mathcal{C}$ as a function of SDK number $n=4N$ and pulse area $\mathcal{A}_{\rm R}$. Here the single-photon detuning is chosen as $\Delta_e=-3.3~\omega_{{\rm hfs},g}$. The simulations average over $m=-2,-1,0,1,2$ states, and include $\Gamma_e=0.017~\omega_{{\rm hfs},e}$ as for the case of $^{85}$Rb. The phase offsets and interferometry contrasts for the simulated signals locally averaged over a $50\%$ intensity distribution are given in (c,d). The data in Fig. (e-g) are similar to Fig. (b-d), but with an increased single-photon detuning of $\Delta_e=-6.6~\omega_{{\rm hfs},g}$.}
    \label{fig:4}
\end{figure}

Here, to demonstrate the utility of the adiabatic SDK sequence for precision measurements, we numerically investigate a simple atom interferometry scheme~\cite{McGuirk2000,Kotru2015,Jaffe.2018}  where an  ``enclosed area'' $A$ is enhanced by the $\tilde U^{(4N)}_{\rm uddu}(\pm \vb{k}_{\rm R})$ sequences. As in the Mach-Zehnder configuration in Fig.~\ref{fig:4}(a), we consider the two spinor matterwave components forming a loop to interfere at $t=2T$ (the dashed lines). When sensing a constant force field, the difference of matterwave phase shifts along the two paths,  $\Delta \Phi\propto A$ associated with the path-dependent diffraction phase~\cite{Asenbaum2020}, is read out interferometrically~\cite{Muller2009,Bertoldi2019}. Here $A$ is the spatial-temporal ``area'' enclosed by the loop. When the duration of the pulsed rotations are negligibly short relative to the ``interrogation time'' $T$, then $A=v_{\rm R} T^2$ is easily evaluated. Here $v_{\rm R}=\hbar k_{\rm eff}/M$ is the photon recoil velocity. Clearly, when the atomic interferometer acts as a force sensor, its sensitivity increases with $A$. In fact, to achieve as large ``area'' $A$ as possible within a measurement time $T$ is of general importance to precision measurements with light pulse atom interferometry~\cite{Peters1997,McGuirk2000, Muller2009}.

More specifically, the enclosed area is defined as $A=\int_0^{2T} \Delta z(t) dt$ during a 3-pulse Raman interferometry sequence by integrating the relative displacement $\Delta z(t)$ between the two matterwave diffraction paths under the three operations as splitter ($t=0$), mirror ($t=T$) and combiner ($t=2T$).
We generally refer the idealized local spin rotations as $R_{\varphi}(\theta)=\cos(\theta/2){\bf 1}+i\sin(\theta/2)(e^{i\varphi}\sigma_+ + e^{-i\varphi}\sigma_-)$ for the Raman interferometer, for any spin state within $\{|a_m\rangle,|b_m\rangle\}$. Here $\varphi=\vb{k}_{\rm R}\cdot{\bf r}$ is the local Raman optical phase. Notice the spatial-dependent $R_{\varphi}(\theta)$ rotation can in principle be generated by the Eq.~(\ref{eq:heff}) Hamiltonian~\cite{foot:biasRotation} as phase-coherent ``half'' and ``full'' kicks.  The splitter and mirror operations in the standard light Raman interferometer can then be expressed as $R_1=R_{\varphi}(\pi/2)$, $R_2=R_{\varphi}(\pi)$ and $R_3=R_{\varphi}(\pi/2)$ respectively to manipulate the spin states while imparting the $\pm \hbar \vb{k}_{\rm R}$ photon recoil momentum.

We now consider enhancing the enclosed area $A$ of the standard 3-pulse Raman interferometer with the chirp-alternating SDK sequence. In particular, we consider the Fig.~\ref{fig:4}(a) scheme with the spinor matterwave diffraction paths marked with solid lines: a $\tilde U_1=\tilde U^{(4 N)}_{\rm uddu}(\vb{k}_{\rm R})$ is first applied at $t=\tau_1$ to increase the momentum displacement between the two interfering paths from $\Delta {\bf p}=\hbar \vb{k}_{\rm R}$ by $R_{\varphi}(\pi/2)$ to $\Delta {\bf p}=(2\times 4 N+1) \hbar \vb{k}_{\rm R}$ with the spin-dependent kicks. This $\Delta {\bf p}$-enhancement is followed by an opposite $\tilde U_2=\tilde U^{(4N)}_{\rm uddu}(-\vb{k}_{\rm R})$ at $t=T-\tau_1$ before the $R_2$-operation to recover the initial $\Delta {\bf p}$. To ensure that the interfering paths spatially overlap at $t=2 T$, an additional pair of opposite momentum boosts, $\tilde U_{3,4}$ are applied at $t=T+\tau_1$ after the $R_2$ and $t=2T-\tau_1$ before the $R_3$ operation respectively.  By properly choosing the $\tau_1/T$ ratio, spurious interference by imperfect $R_{1,2,3}$ and $\tilde U_{1,2,3,4}$ controls can be suppressed~\cite{Sidorenkov2020,Dubetsky2006}. 
With $T\gg\tau_1$, the enclosed area of the resulting interfering loop is enhanced to $A'=(2\times 4 N+1)A$.

For the numerical simulation, we consider at $t=0$ the atomic state to be initialized at certain $|b_m\rangle$ and subjected to the $\tilde U^{(4N)}_{\rm uddu}(\vb{k}_{\rm R})$-enhanced 3-pulse interferometry sequence. The output atomic state, right {\it before} the final matterwave combiner $R_3$, can then be written as $|\psi_m({\bf r},2T^-)\rangle=U_{\rm AI}|c_m\rangle$, with $U_{\rm AI}=U_f(\tau_1) \tilde U_4 U_f(T-2\tau_1)\tilde U_3 R'_2 \tilde U_2 U_f(T-2\tau_1)\tilde U_1 R'_1$ to be ${\bf r}$-dependent. Here $U_f(t)$ designates free propagation of matterwave for time $t$. The $R'_1=U_f(\tau_1) R_1$, $R'_2=U_f(\tau_1) R_2 U_f(\tau_1)$ takes into account the free propagation of matterwave between the standard $R_{j}$ and area-enhancing $\tilde U_j$ sequences. We numerically evaluate $|\psi_m(z,2T^-)\rangle$ for 1D spinor matterwave between $0<z<\lambda/2$ ($\lambda$ is the optical wavelength), as described in Appendix~\ref{sec:numerical}. To focus on the performance of SDK, we set $R_{1,3}$ as perfect $\pi/2$ pulses and $R_2$ as perfect $\pi$ mirror pulse respectively. A further simplification sets $\vb{k}_{\rm R}=0$ for the idealized $R_{1,2,3}$ controls, with which we numerically evaluate $\expval{\Sigma_j} = \langle \psi_m(z, 2T^-)|\Sigma_j|\psi_m(z, 2T^-)\rangle_{z,m}$ for an initially unpolarized atomic sample. Here $\Sigma_j=\sum_{m'=-F_b}^{F_b} \sigma_j^{(m')}$ are summed over all $m$ sub-spins for Pauli matrices with $j = x, y, z$. $\Sigma_j$ corresponds to observables of experimental measurements in which Zeeman sublevels are not resolved, as in most atom interferometry experiments with hyperfine state-dependent fluorescence readouts~\cite{Peters1997}.

With $U_f$ chosen as free 1D propagation, the values of $\tau_1$ and $T$ only affect contributions of spurious interfering paths into the final readouts~\cite{Sidorenkov2020,Dubetsky2006} in the simulation. With $R_{1,2,3}$ set as ideal, the spurious interfering paths are from imperfect $\tilde U_{1,2,3,4}$ diffractions only. Notably, since successive adiabatic SDKs here within each $\tilde U_j$ last merely tens of nanoseconds, the spatial displacements among the spurious interfering paths are negligibly small comparing with the typical coherence length of cold atom samples, and therefore do not alter the matterwave dynamics~\cite{Petitjean2007,Su2010}. We have randomly sampled the atomic initial position and velocity to numerically verify that the residual spurious interference are indeed suppressible. Practically, to generate the results in Fig.~\ref{fig:4}(b-d) with 
all spurious interference removed in an efficient manner, we simply apply a digital filter to remove unwanted diffraction orders after each $\tilde U_j$ sequence. 

We are particularly interested in the interferometry contrast $\mathcal{C}$ and diffraction phase offset $\delta \Phi$. The contrast $\mathcal{C}$ sets the quality of the final matterwave interference fringes. The phase offset $\delta \Phi$, stemming from the unbalanced dynamic phase by the four $\tilde U_{\rm uddu}$ sequences, enters the interferometry readout as  systematic bias against any precision measurements or controls. We  evaluate the interferometry contrast as $\mathcal{C} =\sqrt{\expval{\Sigma_x}^2+\expval{\Sigma_y}^2}$, keeping in mind the ideal $R_3$ rotates ${\bf \Sigma}=\{\Sigma_j\}$ by $90^{\circ}$ for the final population readouts.  The diffraction phase offset is instead evaluated as $\delta\Phi = \arg\qty[\expval{\Sigma_x}+i\expval{\Sigma_y}] - \Phi_0$ with $\Phi_0$ to be the relative phase between $\ket{a_m}$ and $\ket{b_m}$ right after the ideal $R_1$ splitter. Typical numerical results are presented in Figs.~\ref{fig:4}(b-d). The simulation is again performed on the $^{85}$Rb D1 line, here with spontaneous emission included. For Figs.~\ref{fig:4}(b-d) on the left panel, the control laser parameters are chosen close to this experimental work, with $\Delta_e=-3.3~\omega_{{\rm hfs},g}$ so both $\varepsilon_{\rm sp}$ and $\varepsilon_{\Delta m}$ are quite substantial. Nevertheless, we find four $n=4N=12$ chirp-alternating SDKs can be applied for a 25-fold enhancement of interferometry enclosed area, at a moderate expense of reduced contrast to $\mathcal{C}\approx 0.5$ (which is still highly useful~\cite{Peters1997}), with $\delta \Phi<0.01$ to be experimentally calibrated when necessary. 
Here, to avoid excessive spontaneous emission and coherent leakage (Fig.~\ref{fig:leak}(d-f), the peak $\mathcal{A}_{\rm R}\approx 6-8\pi$ needs to be chosen (Fig.~\ref{fig:4}(b-d)). On the other hand, by doubling the single photon detuning to $\Delta_e=-6.6~\omega_{{\rm hfs},g}$ (with laser intensity increased in proportion to maintain the Raman Rabi frequency),  $\varepsilon_{\rm sp}$ are halved, while the impact of $\varepsilon_{\Delta m}$ leakage are dramatically suppressed in the $\tilde U_{\rm uddu}$ sequence (Figs.~\ref{fig:4}(e-g), also see Appendix~\ref{sec:diffatom})~\cite{foot:AIc}. The further detuned $\tilde U_{\rm uddu}$ sequence should thus support up to 50-fold enhancement of interferometry enclosed area, with spontaneous-emission-limited $\mathcal{C}>0.5$ contrast.

\section{Discussions}\label{sec:conclusion}

Significant aspects of advanced quantum technology today are based on controlling alkaline atoms through their center-of-mass motion and ground-state hyperfine interaction. The two long-lived degrees of freedom can be entangled optically by transferring photon recoil momentum with Raman excitations. The tiny atomic recoil effect can be amplified by the repetitive application of such excitations. The spin-dependent large momentum transfer is expected to improve the scalability for precision measurements in atom interferometry~\cite{McGuirk2000,Kotru2015,Jaffe.2018} and for quantum information processing with trapped ions~\cite{Garcia-Ripoll2003, Duan2004,Mizrahi2013,Lo2015,Fluhmann2019}. 
Practically, unlike interrogating microscopically confined single ions where Raman excitation with multiple-THz single-photon detuning is feasible~\cite{Mizrahi2014,  Ballance2016}, addressing larger samples prefers efficient excitations at moderate single-photon detunings. The seemingly unavoidable imperfections associated with spin-leakages and dynamic phases need to be managed in non-traditional ways, for achieving the high speed, high fidelity control with intensity-error resilience required by the next generation quantum technology~\cite{Schafer2018,Fluhmann2019,Szigeti2020, Wu2020,Anders2021,Greve2021}.

In this work, we have demonstrated a novel configuration of adiabatic SDK implemented on an optical delay line, which is able to reach the speed limit of Raman SDK control~\cite{Mizrahi2013,Mizrahi2014}, featuring robust intensity-error resilience, while maintaining various advantages of optical retro-reflection established for precision atom interferometry. The experimental characterization of the technique is limited to the hyperfine population and spin-dependent momentum transfers, but we clarify in Sec.~\ref{sec:controldynamics} that high precision phase gates enabling spin-dependent large momentum transfer can be efficiently realized at the moderate single-photon detuning, by properly programming the chirp direction to suppress the accumulation of coherent errors.
We have provided numerical evidence that the chirp-balanced SDK scheme support faithful, parallel $\Delta m=0$ control of multi-Zeeman spinor matterwave, with giant spin-dependent forces applied within nanoseconds to rapidly shift the phase-space spin separation, even with the $\sim$10~mW laser power as in this work. Since within nanoseconds various low-frequency noises including those due to matterwave dispersion are negligible for cold atoms, we expect accurate implementation of the full scheme to be benchmarked in future interferometric measurements.

Our work suggests that Raman SDK can be dynamically perfected against multi-level couplings and dynamic phase broadening, for error-resilient parallel control of multiple hyperfine spinors within nanoseconds, with exquisite precision. Operating in the unconventional regime of Raman control~\cite{Weiss1994, Wineland2003}, the technique requires relatively moderate laser intensity when comparing with similar techniques for controlling microscopically confined ions~\cite{Mizrahi2013,   Schafer2018, Fluhmann2019}. Nevertheless, we note the high speed SDK demonstrated in this work is realized by weakly focusing the milli-Watt level output from OAWG~\cite{He2020a} to a mesoscopic sample to reach the required intensity. Our technique is immediately useful for coherently controlling mesoscopic ultra-cold samples for quantum simulation~\cite{Gross.2017,Blatt2012} and atom interferometry~\cite{wang2005, Hughes2009, Debs2011, Abend2016}. By improving the peak power of the nanosecond pulses~\cite{Macrae.2021,Wang.2022,Jayich2014, Ma2020,Liu2022a}, our method may drastically enhance the practical benefits of large-momentum beamsplitting in Raman atom interferometry~\cite{McGuirk2000,Kotru2015,Jaffe.2018}, and to support ultra-precise matterwave control for quantum enhanced technologies~\cite{Szigeti2020, Wu2020, Greve2021, Anders2021}.

\section*{Acknowledgements}
We thank Yidi Ma and Xing Huang for experimental assistance. We are grateful to Prof.~Yiqiu Ma for insightful comments to the manuscript, 
and to Prof.~Xiaopeng Li and Prof.~Haidong Yuan for helpful discussions. We acknowledge support from National Key Research Program of China under Grant No.~2017YFA0304204 and No.~2016YFA0302000, from NSFC under Grant No.~12074083.

\appendix
\renewcommand\thefigure{A\arabic{figure}}
\setcounter{figure}{0}  

\section{Full Hamiltonian}\label{App:FullH}

The theoretical analysis as well as numerical simulation in this work is based on the full light-atom interaction Hamiltonian on the D1 line. Following the notation in the main text, the effective, non-Hermitian Hamiltonian is written as 
    \begin{equation}
        \begin{split}
            H_{\rm eff}(\vb{r},t) = &\hbar\sum_{e}\qty(\omega_{e}-\omega_{e0}-i\Gamma_e/2)\sigma^{e_le_l} + \\
            &\hbar\sum_{c=a,b}\qty(\omega_{c}-\omega_{g0})\sigma^{c_mc_m}+\\
            & \frac{\hbar}{2}\sum_{c={a,b}}\sum_e\Omega^{j}_{c_me_l}(\vb{r}, t)\sigma^{c_me_l} + {\rm h.c.}
        \end{split}\label{eq:HD1}
    \end{equation}
    Here $\omega_{e0}, \omega_{g0}$ are decided by the energy of reference level in the excited and ground state manifolds respectively, chosen as the top hyperfine levels in this work. The laser Rabi frequency,
    \begin{equation}
        \begin{split}
            \Omega^{j}_{e_la_m\qty(b_n ) } (\vb{r},t)&\equiv -\frac{\mel{e_l}{\vb{d}\cdot\vo{e}_j\mathcal{E}_j(\vb{r},t)}{a_m\qty(b_n )}}{\hbar},
        \end{split}
    \end{equation}
  is accordingly written in the $\omega_{e0,g0}$ frame under the rotating wave approximation.     $\vb{d}$ to be the atomic electric dipole operator. The  $\sigma^{a_m e_l} = \op{a_m}{e_l}$, $\sigma^{e_l a_m} = \op{e_l}{a_m}$ are the raising and lowering operators between states  $|a_m\rangle$ and $|e_l\rangle$. Similar $\sigma$ operators are defined for all the other $|a_m\rangle$, $|b_n\rangle$ and $|e_l\rangle$ state combinations.

The full $H_{\rm eff}$ in Eq.~(\ref{eq:HD1}) is rewritten as Eq.~(\ref{eq:heff}) in the main text. For weak off-resonant pulses and to facilitate understanding of ground-state Raman interaction, we can also adiabatically eliminate the excited states to approximately have
    \begin{equation}\label{eq:h}
        \begin{split}
            H_{\rm eff}(\vb{r},t) &\approx \\
            &\hbar\sum_e\sum_{j=1,2}\frac{\Omega^{j}_{a_m e_l}\Omega^{j*}_{a_n e_l}}{4\qty(\nu_j-\omega_{e a})}\sigma^{a_m a_n } + \frac{\Omega^{j}_{b_m e_l}\Omega^{j*}_{b_n e_l}}{{4\qty(\nu_j-\omega_{e b})}}\sigma^{b_m b_n }\\
            & +\hbar\sum_e\frac{\Omega^1_{a_m e_l}\Omega^{2*}_{b_n e_l}}{4\Delta_e}\me^{i \vb{k}_{\rm R} \cdot\vb{r}}\sigma^{a_m b_n} + {\rm h.c.}\\
            & + \hbar\sum_e\frac{\Omega^2_{a_m e_l}\Omega^{1*}_{b_n e_l}}{4\Delta_e}\me^{i \vb{k}_{\rm R} \cdot\vb{r}}\me^{i2\omega_{a b}t}\sigma^{b_na_m} + {\rm h.c.},
        \end{split}
    \end{equation}
with the convention of summing over repeated $m, n, l$ indices~\cite{Cidrim.2021}. The single photon detuning is defined as  $\Delta_e = \omega_1 - \omega_{e a} = \omega_2 - \omega_{e b}$.  The vector $\vb{k}_{\rm R}=\vb{k}_2-\vb{k}_1$ is the k-vector associated with the Raman transition driven by the counter-propagating pulses.  

The Raman coupling associated with $\sigma^{b_n a_m}$  in line~2 of Eq.~(\ref{eq:h}) is accompanied by a $\pm\hbar\vb{k}_{\rm R}$ momentum transfer to the spinor matterwave.  Similarly, the ``counter-rotating'' term in line~3 of Eq.~(\ref{eq:h})  leads to an opposite, $\mp\hbar\vb{k}_{\rm R}$ momentum transfer. 
For smooth laser pulses with bandwidth $\delta\omega \ll \omega_{ab}$ to be discussed, this ``counter-rotating'' Raman process is energetically suppressed. The regime of resonant Raman interaction with directional momentum transfer is the focus of this work.

\section{Numerical model}\label{sec:numerical}


We numerically simulate the evolution of 1D spinor matterwaves with full D1 light-atom interactions~\cite{Sievers2015,Bruce2017} driven by the counter-propagating Raman pulses as in Fig.~\ref{fig:1} and \ref{fig:1b}. To account for radiation damping, 
we follow a stochastic wavefunction method~\cite{Dum1992} to evaluate the wavefunction $|\psi({\bf r}, t)\rangle$ for atom at location ${\bf r}$ under the non-Hermitian Hamiltonian $H_{\rm eff}$ (see Eq.~\eqref{eq:HD1} with non-Hermitian part $\hbar\sum_e i\Gamma_e\sigma^{e_le_l}/2$ in the first line). Here $\Gamma_e$ is the natural linewidth of the D1 line.  The simulations treat both the internal and external motion of the spinor matterwave quantum mechanically. For the purpose, we sample $|\psi({\bf r}, t)\rangle$ densely over a uniform grid within $0<z<\lambda/2$ and sparsely in the $x-y$ plane, and follow a split-operator method to evaluate internal/external atomic motion numerically with interleaved steps. 
Taking advantage of the short $\tau_{\rm c}$ for single SDK, the internal state dynamics is evaluated within a single step with frozen external motion under a local $|\psi({\bf r})\rangle$ basis, with atomic position ${\bf r}$ treated as a parameter of $H_{\rm eff}$. The evaluation of observables later is normalized by $\mathcal{N}=\sum_{\bf r} \langle \psi({\bf r}, t=0) |\psi({\bf r}, t=0)\rangle$, with corrections from stochastic contributions to be discussed shortly. Between SDKs, a Fourier transform along ${\bf e}_z$ can be performed to evolve the free-flying spinor matterwave along $z$ if necessary. To save computation resources, the relatively simple atomic dynamics in the x-y plane is ignored.
To evaluate momentum distribution of spinor matterwave, we simply perform a Fourier transform to the space dependent $\langle c_m|\psi({\bf r},t)\rangle$ for any specific spin state $|c_m\rangle$. 

Beyond the coherent evolution, the simulation complexity is substantially reduced by skipping the evaluation of stochastic trajectories heralded by a single ``quantum jump''~\cite{Dalibard1992,Carmichael1993}. Specifically, after each pulsed interaction, the trajectories suffering a quantum jump are simply assumed to repopulate $\{|a\rangle,|b\rangle\}$ in a uniform manner with properly shifted photon recoil momentum. Without further evolution, these trajectories contribute to the evaluation of incoherent, single-time observables such as hyperfine population and photon momentum  transfer. The overall probability of spontaneous emission is determined by the norm of the final wavefunction,  $\varepsilon_{\rm sp}=1-\frac{1}{\mathcal{N}}\sum_{\bf r}\langle \psi({\bf r},\tau_{\rm tot})|\psi({\bf r},\tau_{\rm tot})\rangle$, after a total evolution time $\tau_{\rm tot}$. The simplification is generally justified for evaluating coherent observables of interest, since the expectation values shifted stochastically lead to zero coherent contributions. For the incoherent observables such as average photon momentum and hyperfine population, the simplification is supported by the simple D1 structure under consideration here~\cite{Sievers2015}, where a single spontaneous emission effectively randomizes the following Raman interaction dynamics.  

\section{A Markovian model for \texorpdfstring{$f_{\rm SDK}$}{fSDK} estimation}\label{sec:mmodel}
    For atoms subjected to multiple SDKs, the dynamics of spinor matterwave that deviates from the ideal control can be depicted as diffusing in a ``momentum-lattice''~\cite{He2020b}. Our numerical simulation suggests that with fair efficiency of single adiabatic pulse  to achieve hyperfine transfer efficiency of $f_{\rm R}>95\%$, the resulting average momentum $p_n$ and population $\rho_{aa/bb,n}$ roughly follow a simple Markovian model. The model assumes that both the momentum and population transfer by the next kick are decided by the present population difference $\rho_{aa}-\rho_{bb}$ only. The details of the Markovian model are described as following.
    
    Suppose that after $n$ kicks, the atomic ensemble is with momentum $p_n$ (in unit of $\hbar \vb{k}_{\rm R}$) and population contrast $\mathcal{C}_n\equiv |\rho_{aa,n}-\rho_{bb,n}|$, then the next kick will impart momentum as
    \begin{equation}\label{eq:recform-momentum} 
        \Delta p_{n+1} = p_{n+1}-p_n = f_0(1-\varepsilon_{\rm sp}/2)\mathcal{C}_n,
    \end{equation}
    where $f_0$ is the hyperfine population transfer efficiency in absence of the spontaneous emission. Here we have assumed that during the single pulse process, the spontaneous emission occurs with a uniform distribution of probability, thus the associated population  recycled to the ground states acquires half of $\hbar \vb{k}_{\rm R}$ momentum on average. 
    
    Similarly, the population distribution can be written as
    \begin{equation}
        \begin{split}
            \rho_{aa,n+1} =& (1-\varepsilon_{\rm sp})\qty[(1-f_0)\frac{1+\mathcal{C}_n}{2} + f_0\frac{1-\mathcal{C}_n}{2}]\\
            &+ \varepsilon_{\rm sp}/2\\
            \rho_{bb,n+1} =& (1-\varepsilon_{\rm sp})\qty[f_0\frac{1+\mathcal{C}_n}{2} + (1-f_0)\frac{1-\mathcal{C}_n}{2}]\\
            &+ \varepsilon_{\rm sp}/2,
        \end{split}
    \end{equation}
    so that
    \begin{equation}\label{eq:recform-contrast}
        \mathcal{C}_{n+1} = (1-\varepsilon_{\rm sp})(2f_0-1)\mathcal{C}_n.
    \end{equation}

    We define Raman transfer efficiency as $f_{\rm R} = f_0(1-\varepsilon_{\rm sp}/2)$. When both $f_0$ and  $1-\varepsilon_{\rm sp}$ are close to unity, Eqs.~(\ref{eq:recform-momentum})(\ref{eq:recform-contrast}) can be approximated as
    \begin{equation}\label{eq:markovIter}
        \begin{split}
            &p_{n+1}-p_n = f_{\rm R}\mathcal{C}_n, \\
            &\mathcal{C}_{n+1} = (2f_{\rm R}-1)\mathcal{C}_n.
        \end{split}
    \end{equation}
 
    With the recursion relations by Eq.~(\ref{eq:markovIter}), we arrive at
    \begin{equation}
        \begin{split}
            &p_n = f_{\rm R}\frac{1-(2f_{\rm R}-1)^n}{1-(2f_{\rm R}-1)}, \\
            &\mathcal{C}_{n} = (2f_{\rm R}-1)^n.
        \end{split}
    \end{equation}

Finally, we remark that for the Raman SDK, there is a slight difference of spontaneous emission loss for single kicks between the $a\rightarrow b$ and $b\rightarrow a$ process. As we consider repetitive SDK with $n$ up to a quite large number (e.g. $n_{\rm max}=25$ in our experiment), we effectively set the same $\varepsilon_{\rm sp}$ parameter for the opposite population transfer processes.

\section{Absorption imaging analysis}\label{sec:abs}

    In Sec.~\ref{sec:expt} we have introduced the double imaging technique. This section provides details on deriving recoil momentum $p_n$ and population transfer $\rho_{aa}$ from the imaging data.    
    
        
Our atomic sample is prepared in a cross-dipole trap with slight asymmetry. When deriving the momentum transfer from the double images as in Fig.~\ref{fig:measure}, 
we found that neither before nor after the time-of-flight, the absorption profile can fit perfectly to a 2D Gaussian. In addition, due to the relatively short exposure time of 20$~\mu$s to the weak probe ($s=1$), there is substantial photon shot noise in single-shot images. To faithfully retrieve central position and atom number from each pair of double-image, we take the following procedure. First, we repeat the same type of measurements for $N=80$ shots, and do a principal components analysis to all pairs of double-images after background subtractions. The first three components are kept for the following analysis. We then do a 2D Gaussian fit, expanding the fitted Gaussian profiles to $1.5$ times the waist into a wide enough step-wise mask. The population ratios $\rho_{aa/bb}$ are evaluated by the ratio of total counts between the two images within the mask, where the center-of-mass (COM) positions $z_1$ and $z_2$ are also directly evaluated. 

We note that the first image senses atoms at $F=3$ ``$|a_m\rangle$'' states only. In other words, the atoms kicked to the ``visible'' (or ``invisible'', depending on whether the kicking number $n$ is an even or odd number) hyperfine levels are post-selected. Since there is a small interval $\tau_{\rm{p},1}$ between SDK and the first image, there is a bias to the COM position $z_1$ of the whole cloud due to this post-selection. To correct for the bias, we rewrite the position difference as $z_2-z_1 = \tilde{v}_n\tau_{\rm tof}+\rho_{bb,n}\tilde{v}_{b,n}\tau_{{\rm p},1}$. Here $\tilde{v}_n$ and $\tilde{v}_{b,n}$ are the mean velocity for the whole atomic ensemble and for the atoms in $F=2$ after $n$ SDKs, respectively. The correction to the post-selection induced velocity bias is then given by 
        \begin{equation}
            \tilde{v}_n = v_n\qty(1-\rho_{bb,n}v_{b,n}\frac{\tau_1}{\tau_{\rm tof}}) \equiv  v_n\qty(1-\xi\qty(n, v_n)\frac{\tau_{\rm{p},1}}{\tau_{\rm tof}}).
        \end{equation}
Here, $\xi\qty(n, v_n)$ depends on the relation between $v_n$ and $\rho_{bb,n}, v_{b,n}$. The relation can be approximated with the model in Appendix~\ref{sec:mmodel}. With $\tau_{\rm{p},1}\sim \SI{15}{\micro s}$ in our experiment, this correction is typically $1-\tilde{v}_n/v_n\sim\pm 5\%$ (the $\pm$ signs depend on $n$), which impacts $f_{\rm R}$ at $3\%$ level. The correction is model-dependent. We correct for the bias in our final estimation of $f_{\rm SDK}$ and leave $3\%$ as the dominant uncertainty in our $f_{\rm SDK}$ estimation.

\section{Impact of mirror optical loss to $f_{\rm SDK}$}\label{sec:inferFSDK}
\begin{figure}[htbp]
    \centering
    \includegraphics[width=\linewidth]{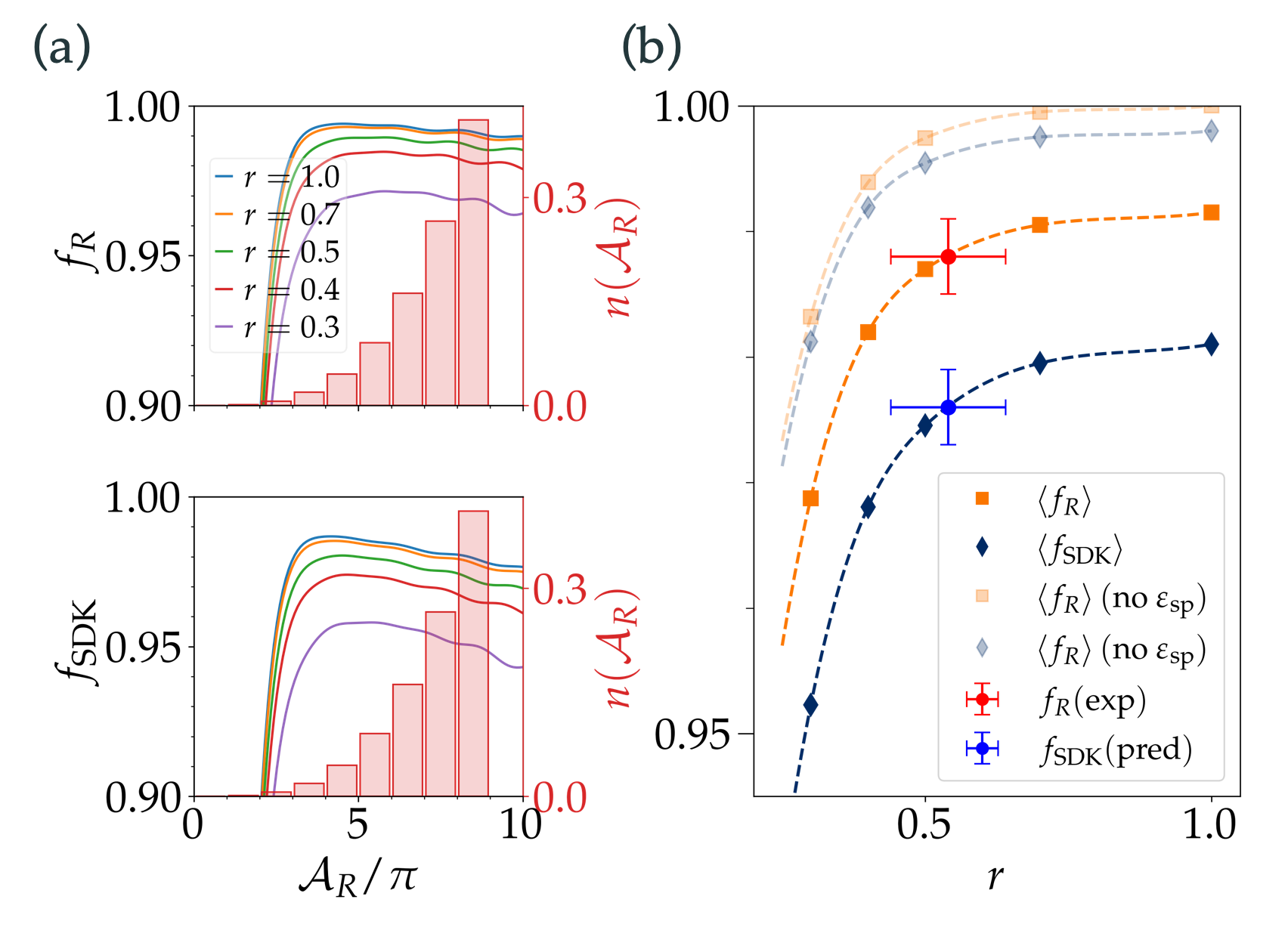}
    \caption{(a) Numerical simulations for $f_{\rm R}$ and $f_{\rm SDK}$ vs Raman pulse area $\mathcal{A}_{\rm R}$ for different reflectivity $r=\abs{\kappa}^2$ of the retro-reflection mirrors. The red bars show estimated distribution of atoms subject to different pulse areas under the Gaussian beam illumination in this experiment. (b) $n(\mathcal{A}_{\rm R})$-weighted average $f_{\rm R}$ and $f_{\rm SDK}$ vs reflectivity $r$. The experimentally measured $f_{\rm R}$ and inferred $f_{\rm SDK}$ are marked with error bars.} 
    \label{fig:A1}
\end{figure}  
    Numerical simulations  suggest that the reflectivity $r=\abs{\kappa}^2$ of the retro-reflection mirror (Fig.~\ref{fig:1b}) affects $f_{\rm SDK}$ at the $1\%$-level. To investigate the effect, we sample $0<r<1$  during the simulation, and calculate Raman transfer efficiency $f_{\rm R}$ and SDK fidelity $f_{\rm SDK}$ with peak Raman pulse area $\mathcal{A}_{\rm R}$ as in Fig.~\ref{fig:A1}(a). In light of the mixed state nature of the experimental measurements, here the results are again averaged over the initial states $|b_m\rangle$. For all the simulation, we evaluate peak Rabi frequency $\Omega_{a,b}$ as in previous work of electric dipole transition control~\cite{He2020b} where the laser beam waist and the atomic ensemble size are also carefully characterized. Based on the geometry parameters, we plot the fractions of atoms as a function of peak Raman pulse area in Fig.~\ref{fig:A1}(a) in histograms. The fractions are applied to weight the average over all pulse areas for the evaluation of $\expval{f_{\rm R}}$ and $\expval{f_{\rm SDK}}$ at various reflectivity $r$ in Fig.~\ref{fig:A1}(b). On the plot, the measured Raman transfer efficiency $f_{\rm R}$ suggest $r\sim 55(5)\%$. This reflectivity is consistent with experimental measurements on the ratio of Stark shifts by the incident and reflected beams. 
    

\section{Different nuclear spins}\label{sec:diffatom}

\begin{figure}[htbp]
    \centering
    \includegraphics[width=1 \linewidth]{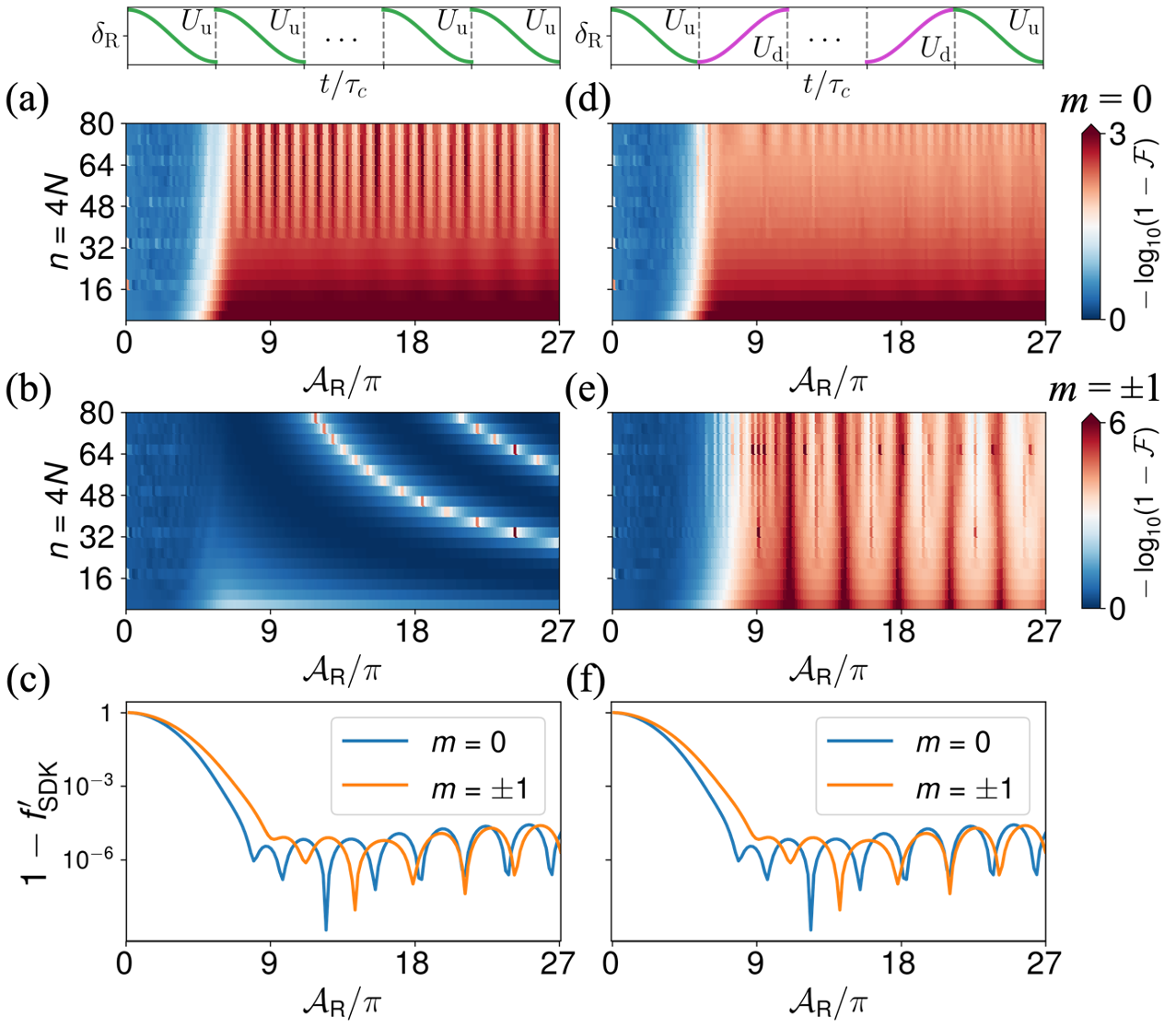}
    \caption{Numerical results of gate infidelity $1-\mathcal{F}$ for  the traditional $\tilde U^{(4N)}_{\rm uu}$ (left panels) and  balanced chirp-alternating $\tilde U^{(4N)}_{\rm uddu}$ (right panels) controls as those in Fig.~\ref{fig:gatefidelity}, $\mathcal{F}=\mathcal{F}^{(4N)}_m$, except for  $\Delta=-6.6\omega_{{\rm hfs,g}}$ here and for a fictitious $^{85}$Rb atom with nuclear spin $I=1.5$. The 2-photon detuning profiles are illustrated on top of (a,d). The exceptionally high fidelity demonstrated in Fig.~(e) is a result of $m=\pm 1$ sub-spin degeneracy in the Raman control dynamics. For comparison, $1-f'_{\rm SDK}$ for a single kick is plotted in both (c)(f) by setting $\Delta_{{\rm hfs},e}=0$ to suppress the spin-leakage according to Eq.~(\ref{eq:leakrabi}).} 
    \label{fig:Rb}
\end{figure}

\begin{figure}[htbp]
    \centering
    \includegraphics[width=1 \linewidth]{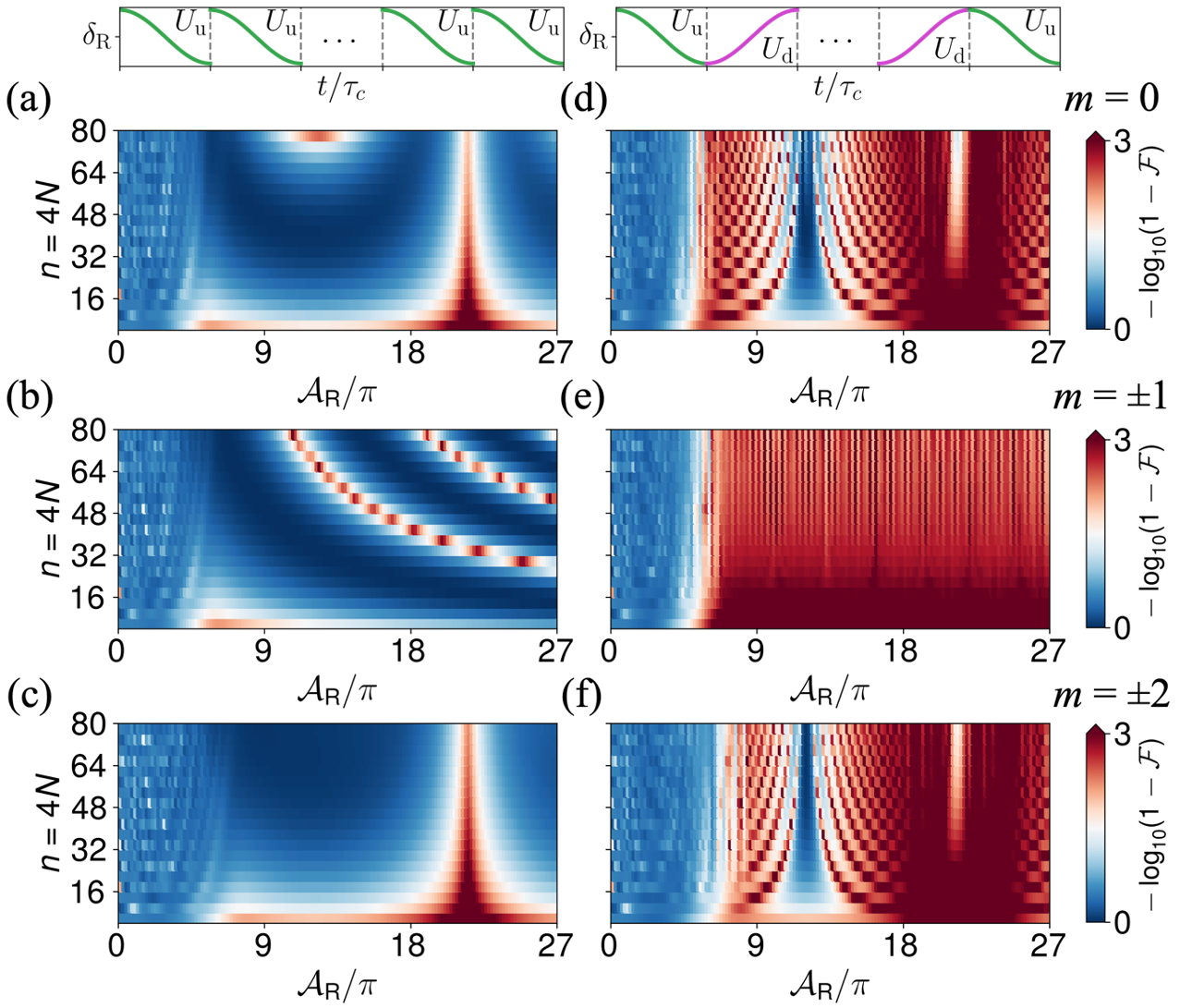}
    \caption{Numerical results of gate infidelity $1-\mathcal{F}$ for  the traditional $\tilde U^{(4N)}_{\rm uu}$ (left panels) and  balanced chirp-alternating $\tilde U^{(4N)}_{\rm uddu}$ (right panels) controls as those in Fig.~\ref{fig:gatefidelity}, $\mathcal{F}=\mathcal{F}^{(4N)}_m$, except for  $\Delta=-6.6\omega_{{\rm hfs,g}}$ here.  The corresponding 2-photon detuning profiles are illustrated on top of (a,d). } 
    \label{fig:Rb5b}
\end{figure}

\begin{figure}[htbp]
    \centering
    \includegraphics[width=1 \linewidth]{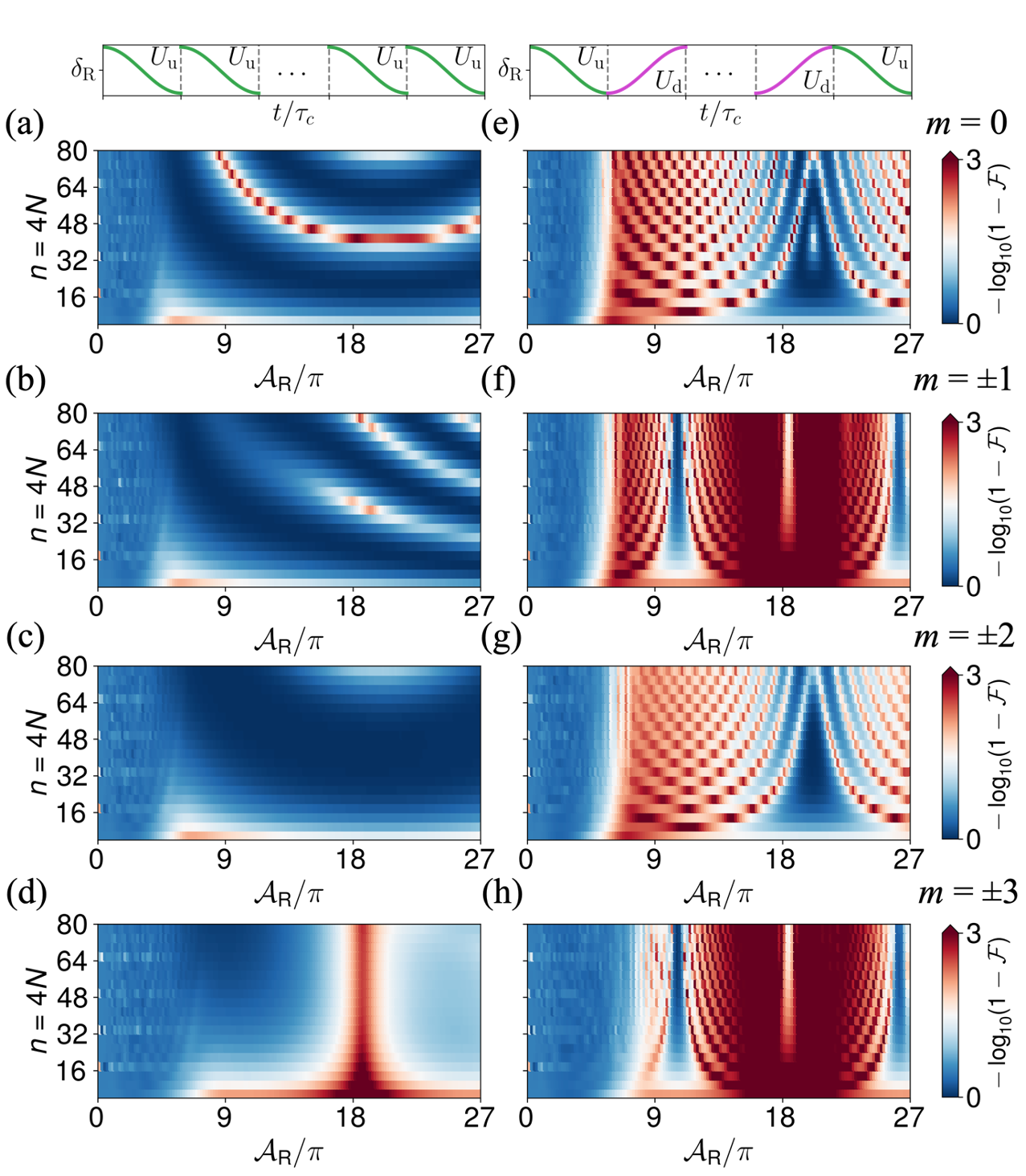}
    \caption{Numerical results of gate infidelity $1-\mathcal{F}$ for  the traditional $\tilde U^{(4N)}_{\rm uu}$ (left panels) and  balanced chirp-alternating $\tilde U^{(4N)}_{\rm uddu}$ (right panels) controls as those in Fig.~\ref{fig:gatefidelity}, $\mathcal{F}=\mathcal{F}^{(4N)}_m$, except for  $\Delta=-6.6\omega_{{\rm hfs,g}}$ here and for a fictitious $^{85}$Rb atom with nuclear spin $I=3.5$. The corresponding 2-photon detuning profiles are illustrated on top of (a,e).} 
    \label{fig:Cs}
\end{figure}

In the main text we have discussed the nanosecond SDK for $^{85}$Rb atoms featuring $m=0,\pm1,\pm2$ weakly-coupled sub-spins (Fig.~\ref{fig:1}). In particular, in Sec.~\ref{sec:controldynamics} we have shown that within the experimentally explored parameter regime, a balanced chirp-alternating sequence $\tilde U_{\rm uddu}^{(4N)}({\bf k}_{\rm R})$ (Eq.~(\ref{eq:nSDKuddu})) quite efficiently suppress the leakage among the five sub-spins to achieve phase gate fidelity $\mathcal{F}_m^{(4N)}\approx 90\sim98\%$ at large $N$, for most of light intensities (Fig.~\ref{fig:gatefidelity}). As clarified in Sec.~\ref{sec:uddu}, the leakage suppression exploits the approximate time-reversal symmetry to cancel the non-adiabatic tunnelings among the sub-spins. Practically, the leakage dynamics supported by the non-degenerate sub-spins is complex enough to merit future study by itself. Instead of attempting a full understanding of the dynamics, in this section we provide numerical examples of coherent matterwave control for other alkaline species. The simulations are according to those outlined in Appendix~\ref{sec:numerical} and in parallel to Fig.~\ref{fig:gatefidelity}, in particular with spontaneous emission ignored by setting $\Gamma_e=0$ on the D1 line. The examples with nuclear spin $I=1.5, 2.5, 3.5$ are given in Fig.~\ref{fig:Rb}, Fig.~\ref{fig:Rb5b}, Fig.~\ref{fig:Cs} respectively. To compare with the Fig.~\ref{fig:gatefidelity} results, we set the hyperfine splittings $\omega_{{\rm hfs},g}$, $\omega_{{\rm hfs},e}$ to be identical to $^{85}$Rb for all the simulations. In addition, we choose $\Delta_e=-6.6\omega_{{\rm hfs},g}$, twice as those for Fig.~\ref{fig:gatefidelity}, to demonstrate the substantially enhanced spin-leakage suppression.

Figure~\ref{fig:Rb} compares the SDK phase gate fidelity by the traditional $\tilde U^{(4N)}_{\rm uu}$ (left) and the chirp-alternating $\tilde U^{(4N)}_{\rm uddu}$ (right) sequences, for the $^{87}$Rb-like atom. With $F_b=1$ for $I=1.5$ (Fig.~\ref{fig:1}(a)), there are three sub-spins with $m=0, \pm1$. The $m=0$ sub-spin is only coupled to single $|a_m\rangle$ states with $m=\pm 2$ (similar to the $m=\pm 3$ end-couplings in Fig.~\ref{fig:1}(a)). The gate infidelity for $\tilde U^{(4N)}_{\rm uu}$ is slightly compromised by the resulting leakage ($\mathcal{F}_m^{(4N)}$ between $98.5\%$ and $99.9\%$ at $4N=80$ ), as in Fig.~\ref{fig:Rb}(a). For comparison, in absence sub-spin leakage to fix, the gate fidelity for the chirp-alternating $\tilde U^{(4N)}_{\rm uddu}$ in Fig.~\ref{fig:Rb}(d) is even slightly worse ($\mathcal{F}_m^{(4N)}$between $98.5\%$ and $99.5\%$ at $4N=80$). Nevertheless, we note the $\tilde U^{(4N)}_{\rm uddu}$ is still preferred practically, not only because it supports the $m=\pm 1$ sub-spin performance to be discussed next, but also due to the  robust dynamic phase cancellation (Sec.~\ref{sec:dphase}).
 
In contrast, the resonant leakages between the degenerate $m=\pm 1$ sub-spins is substantial which strongly affect the gate fidelity for $\tilde{U}_{{\rm uu}}^{(4N)}$. This is shown in Fig.~\ref{fig:Rb}(b) similar to the Fig.~\ref{fig:gatefidelity}(b) results. Interestingly, in absence of the $m=\pm3$ end-couplings for $^{85}$Rb (Fig.~\ref{fig:1}(a)), here the leakage-suppression by chirp-alternating the $\tilde U_{\rm uddu}^{(4N)}$ sequence is essentially perfect for the $^{87}$Rb-like atom. As in Fig.~\ref{fig:Rb}(e), the $\mathcal{F}_m^{(4N)}$ reaches $99.99\%$ level at large $N$, which appears to be only limited by the SU(2) non-adiabaticity of single adiabatic SDK~\cite{Guery-Odelin2019}, as suggested by $f'_{\rm SDK}$ in Fig.~\ref{fig:Rb}(c,f) evaluated according to Eq.~(\ref{equ:fSDK}), after setting $\omega_{{\rm hfs},e}=0$ to remove the $m$-changing couplings (Eq.~(\ref{eq:leakrabi})).

Next, Figure~\ref{fig:Rb5b} investigates SDK phase gate fidelity for $^{85}$Rb featuring five nearly degenerate sub-spins with $m=0,\pm1, \pm2$, as in the main text. Here the single-photon detuning $\Delta_e=-6.6\omega_{{\rm hfs},g}$ is doubled to be compared with the Fig.~\ref{fig:gatefidelity} results. We see that for both $\tilde U_{\rm uu}^{(4N)}$ and $\tilde U_{\rm uddu}^{(4N)}$ sequences,  at the larger $\Delta_e$ the intensity-dependent $\mathcal{F}^{(4N)}_m$ oscillatory features due to the spin-leakages are slowed (with respect to the increasing $n=4N$) and sharpened. With the $\Omega^{\pm 2}$ couplings halved (Eq.~(\ref{eq:leakrabi})) relative to the Fig.~\ref{fig:gatefidelity} case, the infidelity for the $\tilde U_{\rm uddu}^{(4N)}$ sequence is now below $1\%$ at the large $N$, for most of large laser intensities. The low-$\mathcal{F}$ region are now narrowed within $11<\mathcal{A}_R/\pi <13$.

Finally, Figure~\ref{fig:Cs} investigates SDK phase gate fidelity for $^{133}$Cs-like atom featuring seven nearly degenerate sub-spins with $m=0,\pm1, \pm2,\pm3$. With the large number of sub-spins involved, the spin-leakage dynamics are more complex. Nevertheless, comparing with the gate fidelity for $\tilde U_{\rm uu}^{(4N)}$ in Figs.~\ref{fig:Cs}(a-d), in Figs.~\ref{fig:Cs}(e-h) the drastic fidelity enhancement due to the spin-leakage suppression by $\tilde U_{\rm uddu}^{(4N)}$ is evident. For the seven sub-spin system, the strips of low$-\mathcal{F}$ intensity region become quite widely distributed. Consequently, comparing to the $^{87}$Rb-like atom with Fig.~\ref{fig:Rb}(d)(e) and the $^{85}$Rb-like atom with Fig.~\ref{fig:Rb5b}(d-f), it becomes more difficult to find a suitable range of laser intensities for the high-fidelity multi-$m$ SDK phase gates to operate. 
More generally, as the nuclear spin $I$ increases, it becomes more difficult to achieve $m-$independent phase gate prescribed by Eq.~(\ref{eq:dSDK}) in the main text with composite SDK. To improve the aspect, variations of additional pulse parameters beyond the simple chirp-alternation as in this work may be explored. In addition, in measurements where coherent control of a single $m-$spinor matterwave, such as the $m=0$ clock states, is all that is needed, then circularly polarized nanosecond pulses can be applied to naturally suppress the $m$-changing transitions during the spinor matterwave control~\cite{foot:circular}.  To this end, we expect the delay-line based nanosecond SDK to fascilitate the accurate implementation of $m-$preserving Raman control with circularly polarized pulses~\cite{GustavsonThesis}.

\bibliography{nfc,sdkexpt}

\begin{thebibliography}{89}%
\makeatletter
\providecommand \@ifxundefined [1]{%
 \@ifx{#1\undefined}
}%
\providecommand \@ifnum [1]{%
 \ifnum #1\expandafter \@firstoftwo
 \else \expandafter \@secondoftwo
 \fi
}%
\providecommand \@ifx [1]{%
 \ifx #1\expandafter \@firstoftwo
 \else \expandafter \@secondoftwo
 \fi
}%
\providecommand \natexlab [1]{#1}%
\providecommand \enquote  [1]{``#1''}%
\providecommand \bibnamefont  [1]{#1}%
\providecommand \bibfnamefont [1]{#1}%
\providecommand \citenamefont [1]{#1}%
\providecommand \href@noop [0]{\@secondoftwo}%
\providecommand \href [0]{\begingroup \@sanitize@url \@href}%
\providecommand \@href[1]{\@@startlink{#1}\@@href}%
\providecommand \@@href[1]{\endgroup#1\@@endlink}%
\providecommand \@sanitize@url [0]{\catcode `\\12\catcode `\$12\catcode
  `\&12\catcode `\#12\catcode `\^12\catcode `\_12\catcode `\%12\relax}%
\providecommand \@@startlink[1]{}%
\providecommand \@@endlink[0]{}%
\providecommand \url  [0]{\begingroup\@sanitize@url \@url }%
\providecommand \@url [1]{\endgroup\@href {#1}{\urlprefix }}%
\providecommand \urlprefix  [0]{URL }%
\providecommand \Eprint [0]{\href }%
\providecommand \doibase [0]{https://doi.org/}%
\providecommand \selectlanguage [0]{\@gobble}%
\providecommand \bibinfo  [0]{\@secondoftwo}%
\providecommand \bibfield  [0]{\@secondoftwo}%
\providecommand \translation [1]{[#1]}%
\providecommand \BibitemOpen [0]{}%
\providecommand \bibitemStop [0]{}%
\providecommand \bibitemNoStop [0]{.\EOS\space}%
\providecommand \EOS [0]{\spacefactor3000\relax}%
\providecommand \BibitemShut  [1]{\csname bibitem#1\endcsname}%
\let\auto@bib@innerbib\@empty
\bibitem [{\citenamefont {Kasevich}\ and\ \citenamefont
  {Chu}(1991)}]{Kasevich.1991}%
  \BibitemOpen
  \bibfield  {author} {\bibinfo {author} {\bibfnamefont {M.}~\bibnamefont
  {Kasevich}}\ and\ \bibinfo {author} {\bibfnamefont {S.}~\bibnamefont {Chu}},\
  }\href {https://doi.org/10.1103/physrevlett.67.181} {\bibfield  {journal}
  {\bibinfo  {journal} {Physical Review Letters}\ }\textbf {\bibinfo {volume}
  {67}},\ \bibinfo {pages} {181} (\bibinfo {year} {1991})}\BibitemShut
  {NoStop}%
\bibitem [{\citenamefont {Featonby}\ \emph {et~al.}(1996)\citenamefont
  {Featonby}, \citenamefont {Summy}, \citenamefont {Martin}, \citenamefont
  {Wu}, \citenamefont {Zetie}, \citenamefont {Foot},\ and\ \citenamefont
  {Burnett}}]{Featonby.1996}%
  \BibitemOpen
  \bibfield  {author} {\bibinfo {author} {\bibfnamefont {P.~D.}\ \bibnamefont
  {Featonby}}, \bibinfo {author} {\bibfnamefont {G.~S.}\ \bibnamefont {Summy}},
  \bibinfo {author} {\bibfnamefont {J.~L.}\ \bibnamefont {Martin}}, \bibinfo
  {author} {\bibfnamefont {H.}~\bibnamefont {Wu}}, \bibinfo {author}
  {\bibfnamefont {K.~P.}\ \bibnamefont {Zetie}}, \bibinfo {author}
  {\bibfnamefont {C.~J.}\ \bibnamefont {Foot}},\ and\ \bibinfo {author}
  {\bibfnamefont {K.}~\bibnamefont {Burnett}},\ }\href
  {https://doi.org/10.1103/physreva.53.373} {\bibfield  {journal} {\bibinfo
  {journal} {Physical Review A}\ }\textbf {\bibinfo {volume} {53}},\ \bibinfo
  {pages} {373} (\bibinfo {year} {1996})}\BibitemShut {NoStop}%
\bibitem [{\citenamefont {McGuirk}\ \emph {et~al.}(2000)\citenamefont
  {McGuirk}, \citenamefont {Snadden},\ and\ \citenamefont
  {Kasevich}}]{McGuirk2000}%
  \BibitemOpen
  \bibfield  {author} {\bibinfo {author} {\bibfnamefont {J.~M.}\ \bibnamefont
  {McGuirk}}, \bibinfo {author} {\bibfnamefont {M.~J.}\ \bibnamefont
  {Snadden}},\ and\ \bibinfo {author} {\bibfnamefont {M.~A.}\ \bibnamefont
  {Kasevich}},\ }\href {https://doi.org/10.1103/physrevlett.85.4498} {\bibfield
   {journal} {\bibinfo  {journal} {Physical Review Letters}\ }\textbf {\bibinfo
  {volume} {85}},\ \bibinfo {pages} {4498} (\bibinfo {year}
  {2000})}\BibitemShut {NoStop}%
\bibitem [{\citenamefont {Metcalf}\ and\ \citenamefont {van~der
  Straten}(1999)}]{MetcalfBook}%
  \BibitemOpen
  \bibfield  {author} {\bibinfo {author} {\bibfnamefont {H.~J.}\ \bibnamefont
  {Metcalf}}\ and\ \bibinfo {author} {\bibfnamefont {P.}~\bibnamefont {van~der
  Straten}},\ }\href@noop {} {\bibfield  {journal} {\bibinfo  {journal} {Laser
  Cooling and Trapping( Springer-Verlag)}\ } (\bibinfo {year}
  {1999})}\BibitemShut {NoStop}%
\bibitem [{\citenamefont {Mizrahi}\ \emph {et~al.}(2013)\citenamefont
  {Mizrahi}, \citenamefont {Senko}, \citenamefont {Neyenhuis}, \citenamefont
  {Johnson}, \citenamefont {Campbell}, \citenamefont {Conover},\ and\
  \citenamefont {Monroe}}]{Mizrahi2013}%
  \BibitemOpen
  \bibfield  {author} {\bibinfo {author} {\bibfnamefont {J.}~\bibnamefont
  {Mizrahi}}, \bibinfo {author} {\bibfnamefont {C.}~\bibnamefont {Senko}},
  \bibinfo {author} {\bibfnamefont {B.}~\bibnamefont {Neyenhuis}}, \bibinfo
  {author} {\bibfnamefont {K.~G.}\ \bibnamefont {Johnson}}, \bibinfo {author}
  {\bibfnamefont {W.~C.}\ \bibnamefont {Campbell}}, \bibinfo {author}
  {\bibfnamefont {C.~W.~S.}\ \bibnamefont {Conover}},\ and\ \bibinfo {author}
  {\bibfnamefont {C.}~\bibnamefont {Monroe}},\ }\href
  {https://doi.org/10.1103/PhysRevLett.110.203001} {\bibfield  {journal}
  {\bibinfo  {journal} {Physical Review Letters}\ }\textbf {\bibinfo {volume}
  {110}},\ \bibinfo {pages} {203001} (\bibinfo {year} {2013})}\BibitemShut
  {NoStop}%
\bibitem [{\citenamefont {Jaffe}\ \emph {et~al.}(2018)\citenamefont {Jaffe},
  \citenamefont {Xu}, \citenamefont {Haslinger}, \citenamefont {Müller},\ and\
  \citenamefont {Hamilton}}]{Jaffe.2018}%
  \BibitemOpen
  \bibfield  {author} {\bibinfo {author} {\bibfnamefont {M.}~\bibnamefont
  {Jaffe}}, \bibinfo {author} {\bibfnamefont {V.}~\bibnamefont {Xu}}, \bibinfo
  {author} {\bibfnamefont {P.}~\bibnamefont {Haslinger}}, \bibinfo {author}
  {\bibfnamefont {H.}~\bibnamefont {Müller}},\ and\ \bibinfo {author}
  {\bibfnamefont {P.}~\bibnamefont {Hamilton}},\ }\href
  {https://doi.org/10.1103/physrevlett.121.040402} {\bibfield  {journal}
  {\bibinfo  {journal} {Physical Review Letters}\ }\textbf {\bibinfo {volume}
  {121}},\ \bibinfo {pages} {040402} (\bibinfo {year} {2018})},\ \Eprint
  {https://arxiv.org/abs/1803.09024} {1803.09024} \BibitemShut {NoStop}%
\bibitem [{\citenamefont {Harty}\ \emph {et~al.}(2014)\citenamefont {Harty},
  \citenamefont {Allcock}, \citenamefont {Ballance}, \citenamefont {Guidoni},
  \citenamefont {Janacek}, \citenamefont {Linke}, \citenamefont {Stacey},\ and\
  \citenamefont {Lucas}}]{Harty2014}%
  \BibitemOpen
  \bibfield  {author} {\bibinfo {author} {\bibfnamefont {T.~P.}\ \bibnamefont
  {Harty}}, \bibinfo {author} {\bibfnamefont {D.~T.~C.}\ \bibnamefont
  {Allcock}}, \bibinfo {author} {\bibfnamefont {C.~J.}\ \bibnamefont
  {Ballance}}, \bibinfo {author} {\bibfnamefont {L.}~\bibnamefont {Guidoni}},
  \bibinfo {author} {\bibfnamefont {H.~A.}\ \bibnamefont {Janacek}}, \bibinfo
  {author} {\bibfnamefont {N.~M.}\ \bibnamefont {Linke}}, \bibinfo {author}
  {\bibfnamefont {D.~N.}\ \bibnamefont {Stacey}},\ and\ \bibinfo {author}
  {\bibfnamefont {D.~M.}\ \bibnamefont {Lucas}},\ }\href
  {https://doi.org/10.1103/PhysRevLett.113.220501} {\bibfield  {journal}
  {\bibinfo  {journal} {Phys. Rev. Lett.}\ }\textbf {\bibinfo {volume} {113}},\
  \bibinfo {pages} {220501} (\bibinfo {year} {2014})}\BibitemShut {NoStop}%
\bibitem [{\citenamefont {Ballance}\ \emph {et~al.}(2016)\citenamefont
  {Ballance}, \citenamefont {Harty}, \citenamefont {Linke}, \citenamefont
  {Sepiol},\ and\ \citenamefont {Lucas}}]{Ballance2016}%
  \BibitemOpen
  \bibfield  {author} {\bibinfo {author} {\bibfnamefont {C.~J.}\ \bibnamefont
  {Ballance}}, \bibinfo {author} {\bibfnamefont {T.~P.}\ \bibnamefont {Harty}},
  \bibinfo {author} {\bibfnamefont {N.~M.}\ \bibnamefont {Linke}}, \bibinfo
  {author} {\bibfnamefont {M.~A.}\ \bibnamefont {Sepiol}},\ and\ \bibinfo
  {author} {\bibfnamefont {D.~M.}\ \bibnamefont {Lucas}},\ }\href
  {https://doi.org/10.1103/PhysRevLett.117.060504} {\bibfield  {journal}
  {\bibinfo  {journal} {Phys. Rev. Lett.}\ }\textbf {\bibinfo {volume} {117}},\
  \bibinfo {pages} {060504} (\bibinfo {year} {2016})}\BibitemShut {NoStop}%
\bibitem [{\citenamefont {Wang}\ \emph {et~al.}(2016)\citenamefont {Wang},
  \citenamefont {Kumar}, \citenamefont {Wu},\ and\ \citenamefont
  {Weiss}}]{Wang2016}%
  \BibitemOpen
  \bibfield  {author} {\bibinfo {author} {\bibfnamefont {Y.}~\bibnamefont
  {Wang}}, \bibinfo {author} {\bibfnamefont {A.}~\bibnamefont {Kumar}},
  \bibinfo {author} {\bibfnamefont {T.-Y.}\ \bibnamefont {Wu}},\ and\ \bibinfo
  {author} {\bibfnamefont {D.~S.}\ \bibnamefont {Weiss}},\ }\href
  {https://doi.org/10.1126/science.aaf2581} {\bibfield  {journal} {\bibinfo
  {journal} {Science}\ }\textbf {\bibinfo {volume} {352}},\ \bibinfo {pages}
  {1562} (\bibinfo {year} {2016})},\ \Eprint {https://arxiv.org/abs/1601.03639}
  {1601.03639} \BibitemShut {NoStop}%
\bibitem [{\citenamefont {Kotru}\ \emph {et~al.}(2015)\citenamefont {Kotru},
  \citenamefont {Butts}, \citenamefont {Kinast},\ and\ \citenamefont
  {Stoner}}]{Kotru2015}%
  \BibitemOpen
  \bibfield  {author} {\bibinfo {author} {\bibfnamefont {K.}~\bibnamefont
  {Kotru}}, \bibinfo {author} {\bibfnamefont {D.~L.}\ \bibnamefont {Butts}},
  \bibinfo {author} {\bibfnamefont {J.~M.}\ \bibnamefont {Kinast}},\ and\
  \bibinfo {author} {\bibfnamefont {R.~E.}\ \bibnamefont {Stoner}},\ }\href
  {https://doi.org/10.1103/PhysRevLett.115.103001} {\bibfield  {journal}
  {\bibinfo  {journal} {Phys. Rev. Lett.}\ }\textbf {\bibinfo {volume} {115}},\
  \bibinfo {pages} {103001} (\bibinfo {year} {2015})}\BibitemShut {NoStop}%
\bibitem [{\citenamefont {Garc{\'{i}}a-Ripoll}\ \emph
  {et~al.}(2003)\citenamefont {Garc{\'{i}}a-Ripoll}, \citenamefont {Zoller},\
  and\ \citenamefont {Cirac}}]{Garcia-Ripoll2003}%
  \BibitemOpen
  \bibfield  {author} {\bibinfo {author} {\bibfnamefont {J.~J.}\ \bibnamefont
  {Garc{\'{i}}a-Ripoll}}, \bibinfo {author} {\bibfnamefont {P.}~\bibnamefont
  {Zoller}},\ and\ \bibinfo {author} {\bibfnamefont {J.~I.}\ \bibnamefont
  {Cirac}},\ }\href {https://doi.org/10.1103/PhysRevLett.91.157901} {\bibfield
  {journal} {\bibinfo  {journal} {Phys. Rev. Lett.}\ }\textbf {\bibinfo
  {volume} {91}},\ \bibinfo {pages} {157901} (\bibinfo {year}
  {2003})}\BibitemShut {NoStop}%
\bibitem [{\citenamefont {Duan}(2004)}]{Duan2004}%
  \BibitemOpen
  \bibfield  {author} {\bibinfo {author} {\bibfnamefont {L.~M.}\ \bibnamefont
  {Duan}},\ }\href {https://doi.org/10.1103/PhysRevLett.93.100502} {\bibfield
  {journal} {\bibinfo  {journal} {Phys. Rev. Lett.}\ }\textbf {\bibinfo
  {volume} {93}},\ \bibinfo {pages} {100502} (\bibinfo {year} {2004})},\
  \Eprint {https://arxiv.org/abs/0401185} {arXiv:0401185 [quant-ph]}
  \BibitemShut {NoStop}%
\bibitem [{\citenamefont {Lo}\ \emph {et~al.}(2015)\citenamefont {Lo},
  \citenamefont {Kienzler}, \citenamefont {{De Clercq}}, \citenamefont
  {Marinelli}, \citenamefont {Negnevitsky}, \citenamefont {Keitch},\ and\
  \citenamefont {Home}}]{Lo2015}%
  \BibitemOpen
  \bibfield  {author} {\bibinfo {author} {\bibfnamefont {H.~Y.}\ \bibnamefont
  {Lo}}, \bibinfo {author} {\bibfnamefont {D.}~\bibnamefont {Kienzler}},
  \bibinfo {author} {\bibfnamefont {L.}~\bibnamefont {{De Clercq}}}, \bibinfo
  {author} {\bibfnamefont {M.}~\bibnamefont {Marinelli}}, \bibinfo {author}
  {\bibfnamefont {V.}~\bibnamefont {Negnevitsky}}, \bibinfo {author}
  {\bibfnamefont {B.~C.}\ \bibnamefont {Keitch}},\ and\ \bibinfo {author}
  {\bibfnamefont {J.~P.}\ \bibnamefont {Home}},\ }\href
  {https://doi.org/10.1038/nature14458} {\bibfield  {journal} {\bibinfo
  {journal} {Nature}\ }\textbf {\bibinfo {volume} {521}},\ \bibinfo {pages}
  {336} (\bibinfo {year} {2015})},\ \Eprint {https://arxiv.org/abs/1412.7100}
  {arXiv:1412.7100} \BibitemShut {NoStop}%
\bibitem [{\citenamefont {Ma}\ \emph {et~al.}(2020)\citenamefont {Ma},
  \citenamefont {Huang}, \citenamefont {Wang}, \citenamefont {Ji},
  \citenamefont {He}, \citenamefont {Qiu}, \citenamefont {Zhao}, \citenamefont
  {Wang},\ and\ \citenamefont {Wu}}]{Ma2020}%
  \BibitemOpen
  \bibfield  {author} {\bibinfo {author} {\bibfnamefont {Y.}~\bibnamefont
  {Ma}}, \bibinfo {author} {\bibfnamefont {X.}~\bibnamefont {Huang}}, \bibinfo
  {author} {\bibfnamefont {X.}~\bibnamefont {Wang}}, \bibinfo {author}
  {\bibfnamefont {L.}~\bibnamefont {Ji}}, \bibinfo {author} {\bibfnamefont
  {Y.}~\bibnamefont {He}}, \bibinfo {author} {\bibfnamefont {L.}~\bibnamefont
  {Qiu}}, \bibinfo {author} {\bibfnamefont {J.}~\bibnamefont {Zhao}}, \bibinfo
  {author} {\bibfnamefont {Y.}~\bibnamefont {Wang}},\ and\ \bibinfo {author}
  {\bibfnamefont {S.}~\bibnamefont {Wu}},\ }\href
  {https://doi.org/10.1364/oe.389700} {\bibfield  {journal} {\bibinfo
  {journal} {Opt. Express}\ }\textbf {\bibinfo {volume} {28}},\ \bibinfo
  {pages} {17171} (\bibinfo {year} {2020})},\ \Eprint
  {https://arxiv.org/abs/2004.05320} {arXiv:2004.05320} \BibitemShut {NoStop}%
\bibitem [{\citenamefont {He}\ \emph {et~al.}(2020{\natexlab{a}})\citenamefont
  {He}, \citenamefont {Ji}, \citenamefont {Wang}, \citenamefont {Qiu},
  \citenamefont {Zhao}, \citenamefont {Ma}, \citenamefont {Huang},
  \citenamefont {Wu},\ and\ \citenamefont {Chang}}]{He2020b}%
  \BibitemOpen
  \bibfield  {author} {\bibinfo {author} {\bibfnamefont {Y.}~\bibnamefont
  {He}}, \bibinfo {author} {\bibfnamefont {L.}~\bibnamefont {Ji}}, \bibinfo
  {author} {\bibfnamefont {Y.}~\bibnamefont {Wang}}, \bibinfo {author}
  {\bibfnamefont {L.}~\bibnamefont {Qiu}}, \bibinfo {author} {\bibfnamefont
  {J.}~\bibnamefont {Zhao}}, \bibinfo {author} {\bibfnamefont {Y.}~\bibnamefont
  {Ma}}, \bibinfo {author} {\bibfnamefont {X.}~\bibnamefont {Huang}}, \bibinfo
  {author} {\bibfnamefont {S.}~\bibnamefont {Wu}},\ and\ \bibinfo {author}
  {\bibfnamefont {D.~E.}\ \bibnamefont {Chang}},\ }\href
  {http://arxiv.org/abs/2010.16066} {\bibfield  {journal} {\bibinfo  {journal}
  {Phys. Rev. Res.}\ }\textbf {\bibinfo {volume} {2}},\ \bibinfo {pages}
  {043418} (\bibinfo {year} {2020}{\natexlab{a}})},\ \Eprint
  {https://arxiv.org/abs/2010.16066} {arXiv:2010.16066} \BibitemShut {NoStop}%
\bibitem [{\citenamefont {Sch{\"{a}}fer}\ \emph {et~al.}(2018)\citenamefont
  {Sch{\"{a}}fer}, \citenamefont {Ballance}, \citenamefont {Thirumalai},
  \citenamefont {Stephenson}, \citenamefont {Ballance}, \citenamefont
  {Steane},\ and\ \citenamefont {Lucas}}]{Schafer2018}%
  \BibitemOpen
  \bibfield  {author} {\bibinfo {author} {\bibfnamefont {V.~M.}\ \bibnamefont
  {Sch{\"{a}}fer}}, \bibinfo {author} {\bibfnamefont {C.~J.}\ \bibnamefont
  {Ballance}}, \bibinfo {author} {\bibfnamefont {K.}~\bibnamefont
  {Thirumalai}}, \bibinfo {author} {\bibfnamefont {L.~J.}\ \bibnamefont
  {Stephenson}}, \bibinfo {author} {\bibfnamefont {T.~G.}\ \bibnamefont
  {Ballance}}, \bibinfo {author} {\bibfnamefont {A.~M.}\ \bibnamefont
  {Steane}},\ and\ \bibinfo {author} {\bibfnamefont {D.~M.}\ \bibnamefont
  {Lucas}},\ }\href {https://doi.org/10.1038/nature25737} {\bibfield  {journal}
  {\bibinfo  {journal} {Nature}\ }\textbf {\bibinfo {volume} {555}},\ \bibinfo
  {pages} {75} (\bibinfo {year} {2018})},\ \Eprint
  {https://arxiv.org/abs/1709.06952} {arXiv:1709.06952} \BibitemShut {NoStop}%
\bibitem [{\citenamefont {Fl{\"{u}}hmann}\ \emph {et~al.}(2019)\citenamefont
  {Fl{\"{u}}hmann}, \citenamefont {Nguyen}, \citenamefont {Marinelli},
  \citenamefont {Negnevitsky}, \citenamefont {Mehta},\ and\ \citenamefont
  {Home}}]{Fluhmann2019}%
  \BibitemOpen
  \bibfield  {author} {\bibinfo {author} {\bibfnamefont {C.}~\bibnamefont
  {Fl{\"{u}}hmann}}, \bibinfo {author} {\bibfnamefont {T.~L.}\ \bibnamefont
  {Nguyen}}, \bibinfo {author} {\bibfnamefont {M.}~\bibnamefont {Marinelli}},
  \bibinfo {author} {\bibfnamefont {V.}~\bibnamefont {Negnevitsky}}, \bibinfo
  {author} {\bibfnamefont {K.}~\bibnamefont {Mehta}},\ and\ \bibinfo {author}
  {\bibfnamefont {J.~P.}\ \bibnamefont {Home}},\ }\href
  {https://doi.org/10.1038/s41586-019-0960-6} {\bibfield  {journal} {\bibinfo
  {journal} {Nature}\ }\textbf {\bibinfo {volume} {566}},\ \bibinfo {pages}
  {513} (\bibinfo {year} {2019})},\ \Eprint {https://arxiv.org/abs/1807.01033}
  {arXiv:1807.01033} \BibitemShut {NoStop}%
\bibitem [{\citenamefont {Szigeti}\ \emph {et~al.}(2020)\citenamefont
  {Szigeti}, \citenamefont {Nolan}, \citenamefont {Close},\ and\ \citenamefont
  {Haine}}]{Szigeti2020}%
  \BibitemOpen
  \bibfield  {author} {\bibinfo {author} {\bibfnamefont {S.~S.}\ \bibnamefont
  {Szigeti}}, \bibinfo {author} {\bibfnamefont {S.~P.}\ \bibnamefont {Nolan}},
  \bibinfo {author} {\bibfnamefont {J.~D.}\ \bibnamefont {Close}},\ and\
  \bibinfo {author} {\bibfnamefont {S.~A.}\ \bibnamefont {Haine}},\ }\href
  {https://doi.org/10.1103/PhysRevLett.125.100402} {\bibfield  {journal}
  {\bibinfo  {journal} {Phys. Rev. Lett.}\ }\textbf {\bibinfo {volume} {125}},\
  \bibinfo {pages} {100402} (\bibinfo {year} {2020})},\ \Eprint
  {https://arxiv.org/abs/2005.00368} {arXiv:2005.00368} \BibitemShut {NoStop}%
\bibitem [{\citenamefont {Wu}\ \emph {et~al.}(2020)\citenamefont {Wu},
  \citenamefont {Krishnakumar}, \citenamefont {Martínez-Rincón},
  \citenamefont {Malia}, \citenamefont {Hosten},\ and\ \citenamefont
  {Kasevich}}]{Wu2020}%
  \BibitemOpen
  \bibfield  {author} {\bibinfo {author} {\bibfnamefont {Y.}~\bibnamefont
  {Wu}}, \bibinfo {author} {\bibfnamefont {R.}~\bibnamefont {Krishnakumar}},
  \bibinfo {author} {\bibfnamefont {J.}~\bibnamefont {Martínez-Rincón}},
  \bibinfo {author} {\bibfnamefont {B.~K.}\ \bibnamefont {Malia}}, \bibinfo
  {author} {\bibfnamefont {O.}~\bibnamefont {Hosten}},\ and\ \bibinfo {author}
  {\bibfnamefont {M.~A.}\ \bibnamefont {Kasevich}},\ }\href
  {https://doi.org/10.1103/physreva.102.012224} {\bibfield  {journal} {\bibinfo
   {journal} {Physical Review A}\ }\textbf {\bibinfo {volume} {102}},\ \bibinfo
  {pages} {012224} (\bibinfo {year} {2020})},\ \Eprint
  {https://arxiv.org/abs/1912.08334} {1912.08334} \BibitemShut {NoStop}%
\bibitem [{\citenamefont {Anders}\ \emph {et~al.}(2021)\citenamefont {Anders},
  \citenamefont {Idel}, \citenamefont {Feldmann}, \citenamefont {Bondarenko},
  \citenamefont {Loriani}, \citenamefont {Lange}, \citenamefont {Peise},
  \citenamefont {Gersemann}, \citenamefont {Meyer-Hoppe}, \citenamefont
  {Abend}, \citenamefont {Gaaloul}, \citenamefont {Schubert}, \citenamefont
  {Schlippert}, \citenamefont {Santos}, \citenamefont {Rasel},\ and\
  \citenamefont {Klempt}}]{Anders2021}%
  \BibitemOpen
  \bibfield  {author} {\bibinfo {author} {\bibfnamefont {F.}~\bibnamefont
  {Anders}}, \bibinfo {author} {\bibfnamefont {A.}~\bibnamefont {Idel}},
  \bibinfo {author} {\bibfnamefont {P.}~\bibnamefont {Feldmann}}, \bibinfo
  {author} {\bibfnamefont {D.}~\bibnamefont {Bondarenko}}, \bibinfo {author}
  {\bibfnamefont {S.}~\bibnamefont {Loriani}}, \bibinfo {author} {\bibfnamefont
  {K.}~\bibnamefont {Lange}}, \bibinfo {author} {\bibfnamefont
  {J.}~\bibnamefont {Peise}}, \bibinfo {author} {\bibfnamefont
  {M.}~\bibnamefont {Gersemann}}, \bibinfo {author} {\bibfnamefont
  {B.}~\bibnamefont {Meyer-Hoppe}}, \bibinfo {author} {\bibfnamefont
  {S.}~\bibnamefont {Abend}}, \bibinfo {author} {\bibfnamefont
  {N.}~\bibnamefont {Gaaloul}}, \bibinfo {author} {\bibfnamefont
  {C.}~\bibnamefont {Schubert}}, \bibinfo {author} {\bibfnamefont
  {D.}~\bibnamefont {Schlippert}}, \bibinfo {author} {\bibfnamefont
  {L.}~\bibnamefont {Santos}}, \bibinfo {author} {\bibfnamefont
  {E.}~\bibnamefont {Rasel}},\ and\ \bibinfo {author} {\bibfnamefont
  {C.}~\bibnamefont {Klempt}},\ }\href
  {https://doi.org/10.1103/PhysRevLett.127.140402} {\bibfield  {journal}
  {\bibinfo  {journal} {Phys. Rev. Lett.}\ }\textbf {\bibinfo {volume} {127}},\
  \bibinfo {pages} {140402} (\bibinfo {year} {2021})},\ \Eprint
  {https://arxiv.org/abs/2010.15796} {arXiv:2010.15796} \BibitemShut {NoStop}%
\bibitem [{\citenamefont {Greve}\ \emph {et~al.}(2021)\citenamefont {Greve},
  \citenamefont {Luo}, \citenamefont {Wu},\ and\ \citenamefont
  {Thompson}}]{Greve2021}%
  \BibitemOpen
  \bibfield  {author} {\bibinfo {author} {\bibfnamefont {G.~P.}\ \bibnamefont
  {Greve}}, \bibinfo {author} {\bibfnamefont {C.}~\bibnamefont {Luo}}, \bibinfo
  {author} {\bibfnamefont {B.}~\bibnamefont {Wu}},\ and\ \bibinfo {author}
  {\bibfnamefont {J.~K.}\ \bibnamefont {Thompson}},\ }\href@noop {} {\bibfield
  {journal} {\bibinfo  {journal} {arXiv}\ } (\bibinfo {year} {2021})},\ \Eprint
  {https://arxiv.org/abs/2110.14027} {2110.14027} \BibitemShut {NoStop}%
\bibitem [{\citenamefont {Genov}\ \emph {et~al.}(2014)\citenamefont {Genov},
  \citenamefont {Schraft}, \citenamefont {Halfmann},\ and\ \citenamefont
  {Vitanov}}]{Genov2014}%
  \BibitemOpen
  \bibfield  {author} {\bibinfo {author} {\bibfnamefont {G.~T.}\ \bibnamefont
  {Genov}}, \bibinfo {author} {\bibfnamefont {D.}~\bibnamefont {Schraft}},
  \bibinfo {author} {\bibfnamefont {T.}~\bibnamefont {Halfmann}},\ and\
  \bibinfo {author} {\bibfnamefont {N.~V.}\ \bibnamefont {Vitanov}},\ }\href
  {https://doi.org/10.1103/PhysRevLett.113.043001} {\bibfield  {journal}
  {\bibinfo  {journal} {physical review letter}\ }\textbf {\bibinfo {volume}
  {113}},\ \bibinfo {pages} {043001} (\bibinfo {year} {2014})}\BibitemShut
  {NoStop}%
\bibitem [{\citenamefont {Low}\ \emph {et~al.}(2016)\citenamefont {Low},
  \citenamefont {Yoder},\ and\ \citenamefont {Chuang}}]{Low2016}%
  \BibitemOpen
  \bibfield  {author} {\bibinfo {author} {\bibfnamefont {G.~H.}\ \bibnamefont
  {Low}}, \bibinfo {author} {\bibfnamefont {T.~J.}\ \bibnamefont {Yoder}},\
  and\ \bibinfo {author} {\bibfnamefont {I.~L.}\ \bibnamefont {Chuang}},\
  }\href {https://doi.org/10.1103/PhysRevX.6.041067} {\bibfield  {journal}
  {\bibinfo  {journal} {Physical Review X}\ }\textbf {\bibinfo {volume} {6}},\
  \bibinfo {pages} {041067} (\bibinfo {year} {2016})},\ \Eprint
  {https://arxiv.org/abs/1603.03996} {arXiv:1603.03996} \BibitemShut {NoStop}%
\bibitem [{\citenamefont {Saywell}\ \emph {et~al.}(2018)\citenamefont
  {Saywell}, \citenamefont {Kuprov}, \citenamefont {Goodwin}, \citenamefont
  {Carey},\ and\ \citenamefont {Freegarde}}]{Saywell2018}%
  \BibitemOpen
  \bibfield  {author} {\bibinfo {author} {\bibfnamefont {J.~C.}\ \bibnamefont
  {Saywell}}, \bibinfo {author} {\bibfnamefont {I.}~\bibnamefont {Kuprov}},
  \bibinfo {author} {\bibfnamefont {D.}~\bibnamefont {Goodwin}}, \bibinfo
  {author} {\bibfnamefont {M.}~\bibnamefont {Carey}},\ and\ \bibinfo {author}
  {\bibfnamefont {T.}~\bibnamefont {Freegarde}},\ }\href
  {https://doi.org/10.1103/PhysRevA.98.023625} {\bibfield  {journal} {\bibinfo
  {journal} {Phys. Rev. A}\ }\textbf {\bibinfo {volume} {98}},\ \bibinfo
  {pages} {023620} (\bibinfo {year} {2018})},\ \Eprint
  {https://arxiv.org/abs/1804.04625} {arXiv:1804.04625} \BibitemShut {NoStop}%
\bibitem [{\citenamefont {Weiss}\ \emph {et~al.}(1994)\citenamefont {Weiss},
  \citenamefont {Young},\ and\ \citenamefont {Chu}}]{Weiss1994}%
  \BibitemOpen
  \bibfield  {author} {\bibinfo {author} {\bibfnamefont {D.~S.}\ \bibnamefont
  {Weiss}}, \bibinfo {author} {\bibfnamefont {B.~C.}\ \bibnamefont {Young}},\
  and\ \bibinfo {author} {\bibfnamefont {S.}~\bibnamefont {Chu}},\ }\href
  {https://doi.org/10.1007/bf01081393} {\bibfield  {journal} {\bibinfo
  {journal} {Applied Physics B}\ }\textbf {\bibinfo {volume} {59}},\ \bibinfo
  {pages} {217} (\bibinfo {year} {1994})}\BibitemShut {NoStop}%
\bibitem [{\citenamefont {Happer}(1972)}]{Happer1972}%
  \BibitemOpen
  \bibfield  {author} {\bibinfo {author} {\bibfnamefont {W.}~\bibnamefont
  {Happer}},\ }\href@noop {} {\bibfield  {journal} {\bibinfo  {journal} {Rev.
  Mod. Phys.}\ }\textbf {\bibinfo {volume} {44}},\ \bibinfo {pages} {169}
  (\bibinfo {year} {1972})}\BibitemShut {NoStop}%
\bibitem [{\citenamefont {Kasevich}\ and\ \citenamefont
  {Chu}(1992)}]{Kasevich1992}%
  \BibitemOpen
  \bibfield  {author} {\bibinfo {author} {\bibfnamefont {M.}~\bibnamefont
  {Kasevich}}\ and\ \bibinfo {author} {\bibfnamefont {S.}~\bibnamefont {Chu}},\
  }\href {https://doi.org/10.1007/BF00325375} {\bibfield  {journal} {\bibinfo
  {journal} {Appl. Phys. B}\ }\textbf {\bibinfo {volume} {54}},\ \bibinfo
  {pages} {321} (\bibinfo {year} {1992})}\BibitemShut {NoStop}%
\bibitem [{\citenamefont {Durfee}\ \emph {et~al.}(2006)\citenamefont {Durfee},
  \citenamefont {Shaham},\ and\ \citenamefont {Kasevich}}]{Durfee2006}%
  \BibitemOpen
  \bibfield  {author} {\bibinfo {author} {\bibfnamefont {D.~S.}\ \bibnamefont
  {Durfee}}, \bibinfo {author} {\bibfnamefont {Y.~K.}\ \bibnamefont {Shaham}},\
  and\ \bibinfo {author} {\bibfnamefont {M.~A.}\ \bibnamefont {Kasevich}},\
  }\href {https://doi.org/10.1103/PhysRevLett.97.240801} {\bibfield  {journal}
  {\bibinfo  {journal} {Phys. Rev. Lett.}\ }\textbf {\bibinfo {volume} {97}},\
  \bibinfo {pages} {240801} (\bibinfo {year} {2006})},\ \Eprint
  {https://arxiv.org/abs/0510215} {arXiv:0510215 [quant-ph]} \BibitemShut
  {NoStop}%
\bibitem [{\citenamefont {Canuel}\ \emph {et~al.}(2006)\citenamefont {Canuel},
  \citenamefont {Leduc}, \citenamefont {Holleville}, \citenamefont {Gauguet},
  \citenamefont {Fils}, \citenamefont {Virdis}, \citenamefont {Clairon},
  \citenamefont {Dimarcq}, \citenamefont {Bord{\'{e}}}, \citenamefont
  {Landragin},\ and\ \citenamefont {Bouyer}}]{Canuel2006}%
  \BibitemOpen
  \bibfield  {author} {\bibinfo {author} {\bibfnamefont {B.}~\bibnamefont
  {Canuel}}, \bibinfo {author} {\bibfnamefont {F.}~\bibnamefont {Leduc}},
  \bibinfo {author} {\bibfnamefont {D.}~\bibnamefont {Holleville}}, \bibinfo
  {author} {\bibfnamefont {A.}~\bibnamefont {Gauguet}}, \bibinfo {author}
  {\bibfnamefont {J.}~\bibnamefont {Fils}}, \bibinfo {author} {\bibfnamefont
  {A.}~\bibnamefont {Virdis}}, \bibinfo {author} {\bibfnamefont
  {A.}~\bibnamefont {Clairon}}, \bibinfo {author} {\bibfnamefont
  {N.}~\bibnamefont {Dimarcq}}, \bibinfo {author} {\bibfnamefont {C.~J.}\
  \bibnamefont {Bord{\'{e}}}}, \bibinfo {author} {\bibfnamefont
  {A.}~\bibnamefont {Landragin}},\ and\ \bibinfo {author} {\bibfnamefont
  {P.}~\bibnamefont {Bouyer}},\ }\href
  {https://doi.org/10.1103/PhysRevLett.97.010402} {\bibfield  {journal}
  {\bibinfo  {journal} {Phys. Rev. Lett.}\ }\textbf {\bibinfo {volume} {97}},\
  \bibinfo {pages} {010402} (\bibinfo {year} {2006})},\ \Eprint
  {https://arxiv.org/abs/0604061} {arXiv:0604061 [physics]} \BibitemShut
  {NoStop}%
\bibitem [{\citenamefont {Malossi}\ \emph {et~al.}(2009)\citenamefont
  {Malossi}, \citenamefont {Bodart}, \citenamefont {Merlet}, \citenamefont
  {Landragin}, \citenamefont {Santos},\ and\ \citenamefont
  {Paris}}]{Malossi2009}%
  \BibitemOpen
  \bibfield  {author} {\bibinfo {author} {\bibfnamefont {N.}~\bibnamefont
  {Malossi}}, \bibinfo {author} {\bibfnamefont {Q.}~\bibnamefont {Bodart}},
  \bibinfo {author} {\bibfnamefont {S.}~\bibnamefont {Merlet}}, \bibinfo
  {author} {\bibfnamefont {A.}~\bibnamefont {Landragin}}, \bibinfo {author}
  {\bibfnamefont {F.~P.~D.}\ \bibnamefont {Santos}},\ and\ \bibinfo {author}
  {\bibfnamefont {O.~D.}\ \bibnamefont {Paris}},\ }\href@noop {} {\bibfield
  {journal} {\bibinfo  {journal} {Phys. Rev. A}\ }\textbf {\bibinfo {volume}
  {81}},\ \bibinfo {pages} {013617} (\bibinfo {year} {2009})},\ \Eprint
  {https://arxiv.org/abs/arXiv:0909.0102v2} {arXiv:arXiv:0909.0102v2}
  \BibitemShut {NoStop}%
\bibitem [{\citenamefont {Zhou}\ \emph
  {et~al.}(2015{\natexlab{a}})\citenamefont {Zhou}, \citenamefont {Long},
  \citenamefont {Tang}, \citenamefont {Chen}, \citenamefont {Gao},
  \citenamefont {Peng}, \citenamefont {Duan}, \citenamefont {Zhong},
  \citenamefont {Xiong}, \citenamefont {Wang}, \citenamefont {Zhang},\ and\
  \citenamefont {Zhan}}]{Zhou2015}%
  \BibitemOpen
  \bibfield  {author} {\bibinfo {author} {\bibfnamefont {L.}~\bibnamefont
  {Zhou}}, \bibinfo {author} {\bibfnamefont {S.}~\bibnamefont {Long}}, \bibinfo
  {author} {\bibfnamefont {B.}~\bibnamefont {Tang}}, \bibinfo {author}
  {\bibfnamefont {X.}~\bibnamefont {Chen}}, \bibinfo {author} {\bibfnamefont
  {F.}~\bibnamefont {Gao}}, \bibinfo {author} {\bibfnamefont {W.}~\bibnamefont
  {Peng}}, \bibinfo {author} {\bibfnamefont {W.}~\bibnamefont {Duan}}, \bibinfo
  {author} {\bibfnamefont {J.}~\bibnamefont {Zhong}}, \bibinfo {author}
  {\bibfnamefont {Z.}~\bibnamefont {Xiong}}, \bibinfo {author} {\bibfnamefont
  {J.}~\bibnamefont {Wang}}, \bibinfo {author} {\bibfnamefont {Y.}~\bibnamefont
  {Zhang}},\ and\ \bibinfo {author} {\bibfnamefont {M.}~\bibnamefont {Zhan}},\
  }\href {https://doi.org/10.1103/PhysRevLett.115.013004} {\bibfield  {journal}
  {\bibinfo  {journal} {Phys. Rev. Lett.}\ }\textbf {\bibinfo {volume} {115}},\
  \bibinfo {pages} {013004} (\bibinfo {year} {2015}{\natexlab{a}})},\ \Eprint
  {https://arxiv.org/abs/1503.00401} {arXiv:1503.00401} \BibitemShut {NoStop}%
\bibitem [{\citenamefont {Perrin}\ \emph {et~al.}(2019)\citenamefont {Perrin},
  \citenamefont {Bernard}, \citenamefont {Bidel}, \citenamefont {Bonnin},
  \citenamefont {Zahzam}, \citenamefont {Blanchard}, \citenamefont {Bresson},\
  and\ \citenamefont {Cadoret}}]{Perrin2019}%
  \BibitemOpen
  \bibfield  {author} {\bibinfo {author} {\bibfnamefont {I.}~\bibnamefont
  {Perrin}}, \bibinfo {author} {\bibfnamefont {J.}~\bibnamefont {Bernard}},
  \bibinfo {author} {\bibfnamefont {Y.}~\bibnamefont {Bidel}}, \bibinfo
  {author} {\bibfnamefont {A.}~\bibnamefont {Bonnin}}, \bibinfo {author}
  {\bibfnamefont {N.}~\bibnamefont {Zahzam}}, \bibinfo {author} {\bibfnamefont
  {C.}~\bibnamefont {Blanchard}}, \bibinfo {author} {\bibfnamefont
  {A.}~\bibnamefont {Bresson}},\ and\ \bibinfo {author} {\bibfnamefont
  {M.}~\bibnamefont {Cadoret}},\ }\href
  {https://doi.org/10.1103/PhysRevA.100.053618} {\bibfield  {journal} {\bibinfo
   {journal} {Phys. Rev. A}\ }\textbf {\bibinfo {volume} {100}},\ \bibinfo
  {pages} {053618} (\bibinfo {year} {2019})},\ \Eprint
  {https://arxiv.org/abs/1907.04403} {arXiv:1907.04403} \BibitemShut {NoStop}%
\bibitem [{\citenamefont {Hartmann}\ \emph {et~al.}(2020)\citenamefont
  {Hartmann}, \citenamefont {Jenewein}, \citenamefont {Giese}, \citenamefont
  {Abend}, \citenamefont {Roura}, \citenamefont {Rasel},\ and\ \citenamefont
  {Schleich}}]{Hartmann2020}%
  \BibitemOpen
  \bibfield  {author} {\bibinfo {author} {\bibfnamefont {S.}~\bibnamefont
  {Hartmann}}, \bibinfo {author} {\bibfnamefont {J.}~\bibnamefont {Jenewein}},
  \bibinfo {author} {\bibfnamefont {E.}~\bibnamefont {Giese}}, \bibinfo
  {author} {\bibfnamefont {S.}~\bibnamefont {Abend}}, \bibinfo {author}
  {\bibfnamefont {A.}~\bibnamefont {Roura}}, \bibinfo {author} {\bibfnamefont
  {E.~M.}\ \bibnamefont {Rasel}},\ and\ \bibinfo {author} {\bibfnamefont
  {W.~P.}\ \bibnamefont {Schleich}},\ }\href
  {https://doi.org/10.1103/PhysRevA.101.053610} {\bibfield  {journal} {\bibinfo
   {journal} {Phys. Rev. A}\ }\textbf {\bibinfo {volume} {101}},\ \bibinfo
  {pages} {53610} (\bibinfo {year} {2020})},\ \Eprint
  {https://arxiv.org/abs/1911.12169} {arXiv:1911.12169} \BibitemShut {NoStop}%
\bibitem [{\citenamefont {Peters}\ \emph {et~al.}(1997)\citenamefont {Peters},
  \citenamefont {Chung}, \citenamefont {Young}, \citenamefont {Hensley},\ and\
  \citenamefont {Chu}}]{Peters1997}%
  \BibitemOpen
  \bibfield  {author} {\bibinfo {author} {\bibfnamefont {A.}~\bibnamefont
  {Peters}}, \bibinfo {author} {\bibfnamefont {K.~Y.}\ \bibnamefont {Chung}},
  \bibinfo {author} {\bibfnamefont {B.}~\bibnamefont {Young}}, \bibinfo
  {author} {\bibfnamefont {J.}~\bibnamefont {Hensley}},\ and\ \bibinfo {author}
  {\bibfnamefont {S.}~\bibnamefont {Chu}},\ }\href
  {https://doi.org/10.1098/rsta.1997.0121} {\bibfield  {journal} {\bibinfo
  {journal} {Phil. Trans. R. Soc. Lond. A}\ }\textbf {\bibinfo {volume}
  {355}},\ \bibinfo {pages} {2223} (\bibinfo {year} {1997})}\BibitemShut
  {NoStop}%
\bibitem [{\citenamefont {Kral}\ \emph {et~al.}(2007)\citenamefont {Kral},
  \citenamefont {Thanopulos},\ and\ \citenamefont {Shapiro}}]{Kral2007}%
  \BibitemOpen
  \bibfield  {author} {\bibinfo {author} {\bibfnamefont {P.}~\bibnamefont
  {Kral}}, \bibinfo {author} {\bibfnamefont {I.}~\bibnamefont {Thanopulos}},\
  and\ \bibinfo {author} {\bibfnamefont {M.}~\bibnamefont {Shapiro}},\ }\href
  {https://doi.org/10.1103/RevModPhys.79.53} {\bibfield  {journal} {\bibinfo
  {journal} {Rev. Mod. Phys.}\ }\textbf {\bibinfo {volume} {79}},\ \bibinfo
  {pages} {53} (\bibinfo {year} {2007})}\BibitemShut {NoStop}%
\bibitem [{\citenamefont {He}\ \emph {et~al.}(2020{\natexlab{b}})\citenamefont
  {He}, \citenamefont {Ji}, \citenamefont {Wang}, \citenamefont {Qiu},
  \citenamefont {Zhao}, \citenamefont {Ma}, \citenamefont {Huang},
  \citenamefont {Wu},\ and\ \citenamefont {Chang}}]{He2020a}%
  \BibitemOpen
  \bibfield  {author} {\bibinfo {author} {\bibfnamefont {Y.}~\bibnamefont
  {He}}, \bibinfo {author} {\bibfnamefont {L.}~\bibnamefont {Ji}}, \bibinfo
  {author} {\bibfnamefont {Y.}~\bibnamefont {Wang}}, \bibinfo {author}
  {\bibfnamefont {L.}~\bibnamefont {Qiu}}, \bibinfo {author} {\bibfnamefont
  {J.}~\bibnamefont {Zhao}}, \bibinfo {author} {\bibfnamefont {Y.}~\bibnamefont
  {Ma}}, \bibinfo {author} {\bibfnamefont {X.}~\bibnamefont {Huang}}, \bibinfo
  {author} {\bibfnamefont {S.}~\bibnamefont {Wu}},\ and\ \bibinfo {author}
  {\bibfnamefont {D.~E.}\ \bibnamefont {Chang}},\ }\href
  {https://doi.org/10.1103/PhysRevLett.125.213602} {\bibfield  {journal}
  {\bibinfo  {journal} {Phys. Rev. Lett.}\ }\textbf {\bibinfo {volume} {125}},\
  \bibinfo {pages} {213602} (\bibinfo {year} {2020}{\natexlab{b}})}\BibitemShut
  {NoStop}%
\bibitem [{\citenamefont {Estey}\ \emph {et~al.}(2015)\citenamefont {Estey},
  \citenamefont {Yu},\ and\ \citenamefont {M{\"{u}}ller}}]{Estey2015}%
  \BibitemOpen
  \bibfield  {author} {\bibinfo {author} {\bibfnamefont {B.}~\bibnamefont
  {Estey}}, \bibinfo {author} {\bibfnamefont {C.}~\bibnamefont {Yu}},\ and\
  \bibinfo {author} {\bibfnamefont {H.}~\bibnamefont {M{\"{u}}ller}},\ }\href
  {https://doi.org/10.1103/PhysRevLett.115.083002} {\bibfield  {journal}
  {\bibinfo  {journal} {Phys. Rev. Lett.}\ }\textbf {\bibinfo {volume} {115}},\
  \bibinfo {pages} {083002} (\bibinfo {year} {2015})}\BibitemShut {NoStop}%
\bibitem [{\citenamefont {Kristensen}\ \emph {et~al.}(2021)\citenamefont
  {Kristensen}, \citenamefont {Jaffe}, \citenamefont {Xu}, \citenamefont
  {Panda},\ and\ \citenamefont {M{\"{u}}ller}}]{Kristensen2021}%
  \BibitemOpen
  \bibfield  {author} {\bibinfo {author} {\bibfnamefont {S.~L.}\ \bibnamefont
  {Kristensen}}, \bibinfo {author} {\bibfnamefont {M.}~\bibnamefont {Jaffe}},
  \bibinfo {author} {\bibfnamefont {V.}~\bibnamefont {Xu}}, \bibinfo {author}
  {\bibfnamefont {C.~D.}\ \bibnamefont {Panda}},\ and\ \bibinfo {author}
  {\bibfnamefont {H.}~\bibnamefont {M{\"{u}}ller}},\ }\href
  {https://doi.org/10.1103/PhysRevA.103.023715} {\bibfield  {journal} {\bibinfo
   {journal} {Phys. Rev. A}\ }\textbf {\bibinfo {volume} {103}},\ \bibinfo
  {pages} {023715} (\bibinfo {year} {2021})},\ \Eprint
  {https://arxiv.org/abs/2011.02946} {arXiv:2011.02946} \BibitemShut {NoStop}%
\bibitem [{\citenamefont {Mizrahi}\ \emph {et~al.}(2014)\citenamefont
  {Mizrahi}, \citenamefont {Neyenhuis}, \citenamefont {Johnson}, \citenamefont
  {Campbell}, \citenamefont {Senko}, \citenamefont {Hayes},\ and\ \citenamefont
  {Monroe}}]{Mizrahi2014}%
  \BibitemOpen
  \bibfield  {author} {\bibinfo {author} {\bibfnamefont {J.}~\bibnamefont
  {Mizrahi}}, \bibinfo {author} {\bibfnamefont {B.}~\bibnamefont {Neyenhuis}},
  \bibinfo {author} {\bibfnamefont {K.~G.}\ \bibnamefont {Johnson}}, \bibinfo
  {author} {\bibfnamefont {W.~C.}\ \bibnamefont {Campbell}}, \bibinfo {author}
  {\bibfnamefont {C.}~\bibnamefont {Senko}}, \bibinfo {author} {\bibfnamefont
  {D.}~\bibnamefont {Hayes}},\ and\ \bibinfo {author} {\bibfnamefont
  {C.}~\bibnamefont {Monroe}},\ }\href
  {https://doi.org/10.1007/s00340-013-5717-6} {\bibfield  {journal} {\bibinfo
  {journal} {Appl. Phys. B}\ }\textbf {\bibinfo {volume} {114}},\ \bibinfo
  {pages} {45} (\bibinfo {year} {2014})}\BibitemShut {NoStop}%
\bibitem [{\citenamefont {Stack}\ \emph {et~al.}(2011)\citenamefont {Stack},
  \citenamefont {Elgin}, \citenamefont {Anisimov},\ and\ \citenamefont
  {Metcalf}}]{Stack2011}%
  \BibitemOpen
  \bibfield  {author} {\bibinfo {author} {\bibfnamefont {D.}~\bibnamefont
  {Stack}}, \bibinfo {author} {\bibfnamefont {J.}~\bibnamefont {Elgin}},
  \bibinfo {author} {\bibfnamefont {P.~M.}\ \bibnamefont {Anisimov}},\ and\
  \bibinfo {author} {\bibfnamefont {H.}~\bibnamefont {Metcalf}},\ }\href
  {https://doi.org/10.1103/PhysRevA.84.013420} {\bibfield  {journal} {\bibinfo
  {journal} {Phys. Rev. A}\ }\textbf {\bibinfo {volume} {013420}},\ \bibinfo
  {pages} {013420} (\bibinfo {year} {2011})}\BibitemShut {NoStop}%
\bibitem [{\citenamefont {Wineland}\ \emph {et~al.}(2003)\citenamefont
  {Wineland}, \citenamefont {Barrett}, \citenamefont {Britton}, \citenamefont
  {Chiaverini}, \citenamefont {DeMarco}, \citenamefont {Itano}, \citenamefont
  {Jelenkovic}, \citenamefont {Langer}, \citenamefont {Leibfried},
  \citenamefont {Meyer}, \citenamefont {Rosenband},\ and\ \citenamefont
  {Schatz}}]{Wineland2003}%
  \BibitemOpen
  \bibfield  {author} {\bibinfo {author} {\bibfnamefont {D.~J.}\ \bibnamefont
  {Wineland}}, \bibinfo {author} {\bibfnamefont {M.}~\bibnamefont {Barrett}},
  \bibinfo {author} {\bibfnamefont {J.}~\bibnamefont {Britton}}, \bibinfo
  {author} {\bibfnamefont {J.}~\bibnamefont {Chiaverini}}, \bibinfo {author}
  {\bibfnamefont {B.}~\bibnamefont {DeMarco}}, \bibinfo {author} {\bibfnamefont
  {W.~M.}\ \bibnamefont {Itano}}, \bibinfo {author} {\bibfnamefont
  {B.}~\bibnamefont {Jelenkovic}}, \bibinfo {author} {\bibfnamefont
  {C.}~\bibnamefont {Langer}}, \bibinfo {author} {\bibfnamefont
  {D.}~\bibnamefont {Leibfried}}, \bibinfo {author} {\bibfnamefont
  {V.}~\bibnamefont {Meyer}}, \bibinfo {author} {\bibfnamefont
  {T.}~\bibnamefont {Rosenband}},\ and\ \bibinfo {author} {\bibfnamefont
  {T.}~\bibnamefont {Schatz}},\ }\href {https://doi.org/10.1364/qim.2013.th1.1}
  {\bibfield  {journal} {\bibinfo  {journal} {Phil. Trans. R. Soc. Lond. A}\
  }\textbf {\bibinfo {volume} {361}},\ \bibinfo {pages} {1349} (\bibinfo {year}
  {2003})}\BibitemShut {NoStop}%
\bibitem [{\citenamefont {Plotkin-Swing}\ \emph {et~al.}(2018)\citenamefont
  {Plotkin-Swing}, \citenamefont {Gochnauer}, \citenamefont {McAlpine},
  \citenamefont {Cooper}, \citenamefont {Jamison},\ and\ \citenamefont
  {Gupta}}]{Plotkin-Swing2018}%
  \BibitemOpen
  \bibfield  {author} {\bibinfo {author} {\bibfnamefont {B.}~\bibnamefont
  {Plotkin-Swing}}, \bibinfo {author} {\bibfnamefont {D.}~\bibnamefont
  {Gochnauer}}, \bibinfo {author} {\bibfnamefont {K.~E.}\ \bibnamefont
  {McAlpine}}, \bibinfo {author} {\bibfnamefont {E.~S.}\ \bibnamefont
  {Cooper}}, \bibinfo {author} {\bibfnamefont {A.~O.}\ \bibnamefont
  {Jamison}},\ and\ \bibinfo {author} {\bibfnamefont {S.}~\bibnamefont
  {Gupta}},\ }\href {https://doi.org/10.1103/PhysRevLett.121.133201} {\bibfield
   {journal} {\bibinfo  {journal} {Phys. Rev. Lett.}\ }\textbf {\bibinfo
  {volume} {121}},\ \bibinfo {pages} {133201} (\bibinfo {year} {2018})},\
  \Eprint {https://arxiv.org/abs/1712.06738} {arXiv:1712.06738} \BibitemShut
  {NoStop}%
\bibitem [{\citenamefont {Asenbaum}\ \emph {et~al.}(2020)\citenamefont
  {Asenbaum}, \citenamefont {Overstreet}, \citenamefont {Kim}, \citenamefont
  {Curti},\ and\ \citenamefont {Kasevich}}]{Asenbaum2020}%
  \BibitemOpen
  \bibfield  {author} {\bibinfo {author} {\bibfnamefont {P.}~\bibnamefont
  {Asenbaum}}, \bibinfo {author} {\bibfnamefont {C.}~\bibnamefont
  {Overstreet}}, \bibinfo {author} {\bibfnamefont {M.}~\bibnamefont {Kim}},
  \bibinfo {author} {\bibfnamefont {J.}~\bibnamefont {Curti}},\ and\ \bibinfo
  {author} {\bibfnamefont {M.~A.}\ \bibnamefont {Kasevich}},\ }\href
  {https://doi.org/10.1103/PhysRevLett.125.191101} {\bibfield  {journal}
  {\bibinfo  {journal} {Phys. Rev. Lett.}\ }\textbf {\bibinfo {volume} {125}},\
  \bibinfo {pages} {191101} (\bibinfo {year} {2020})},\ \Eprint
  {https://arxiv.org/abs/2005.11624} {arXiv:2005.11624} \BibitemShut {NoStop}%
\bibitem [{\citenamefont {M{\"{u}}ller}\ \emph {et~al.}(2009)\citenamefont
  {M{\"{u}}ller}, \citenamefont {Chiow}, \citenamefont {Herrmann},\ and\
  \citenamefont {Chu}}]{Muller2009}%
  \BibitemOpen
  \bibfield  {author} {\bibinfo {author} {\bibfnamefont {H.}~\bibnamefont
  {M{\"{u}}ller}}, \bibinfo {author} {\bibfnamefont {S.~W.}\ \bibnamefont
  {Chiow}}, \bibinfo {author} {\bibfnamefont {S.}~\bibnamefont {Herrmann}},\
  and\ \bibinfo {author} {\bibfnamefont {S.}~\bibnamefont {Chu}},\ }\href
  {https://doi.org/10.1103/PhysRevLett.102.240403} {\bibfield  {journal}
  {\bibinfo  {journal} {Phys. Rev. Lett.}\ }\textbf {\bibinfo {volume} {102}},\
  \bibinfo {pages} {240403} (\bibinfo {year} {2009})},\ \Eprint
  {https://arxiv.org/abs/0903.4192} {arXiv:0903.4192} \BibitemShut {NoStop}%
\bibitem [{\citenamefont {Gebbe}\ \emph {et~al.}(2021)\citenamefont {Gebbe},
  \citenamefont {Siem{\ss}}, \citenamefont {Gersemann}, \citenamefont
  {M{\"{u}}ntinga}, \citenamefont {Herrmann}, \citenamefont {L{\"{a}}mmerzahl},
  \citenamefont {Ahlers}, \citenamefont {Gaaloul}, \citenamefont {Schubert},
  \citenamefont {Hammerer}, \citenamefont {Abend},\ and\ \citenamefont
  {Rasel}}]{Gebbe2021}%
  \BibitemOpen
  \bibfield  {author} {\bibinfo {author} {\bibfnamefont {M.}~\bibnamefont
  {Gebbe}}, \bibinfo {author} {\bibfnamefont {J.~N.}\ \bibnamefont
  {Siem{\ss}}}, \bibinfo {author} {\bibfnamefont {M.}~\bibnamefont
  {Gersemann}}, \bibinfo {author} {\bibfnamefont {H.}~\bibnamefont
  {M{\"{u}}ntinga}}, \bibinfo {author} {\bibfnamefont {S.}~\bibnamefont
  {Herrmann}}, \bibinfo {author} {\bibfnamefont {C.}~\bibnamefont
  {L{\"{a}}mmerzahl}}, \bibinfo {author} {\bibfnamefont {H.}~\bibnamefont
  {Ahlers}}, \bibinfo {author} {\bibfnamefont {N.}~\bibnamefont {Gaaloul}},
  \bibinfo {author} {\bibfnamefont {C.}~\bibnamefont {Schubert}}, \bibinfo
  {author} {\bibfnamefont {K.}~\bibnamefont {Hammerer}}, \bibinfo {author}
  {\bibfnamefont {S.}~\bibnamefont {Abend}},\ and\ \bibinfo {author}
  {\bibfnamefont {E.~M.}\ \bibnamefont {Rasel}},\ }\href
  {https://doi.org/10.1038/s41467-021-22823-8} {\bibfield  {journal} {\bibinfo
  {journal} {Nat. Commun.}\ }\textbf {\bibinfo {volume} {12}},\ \bibinfo
  {pages} {1544} (\bibinfo {year} {2021})},\ \Eprint
  {https://arxiv.org/abs/1907.08416} {arXiv:1907.08416} \BibitemShut {NoStop}%
\bibitem [{\citenamefont {Rudolph}\ \emph {et~al.}(2020)\citenamefont
  {Rudolph}, \citenamefont {Wilkason}, \citenamefont {Nantel}, \citenamefont
  {Swan}, \citenamefont {Holland}, \citenamefont {Jiang}, \citenamefont
  {Garber}, \citenamefont {Carman},\ and\ \citenamefont {Hogan}}]{Rudolph2020}%
  \BibitemOpen
  \bibfield  {author} {\bibinfo {author} {\bibfnamefont {J.}~\bibnamefont
  {Rudolph}}, \bibinfo {author} {\bibfnamefont {T.}~\bibnamefont {Wilkason}},
  \bibinfo {author} {\bibfnamefont {M.}~\bibnamefont {Nantel}}, \bibinfo
  {author} {\bibfnamefont {H.}~\bibnamefont {Swan}}, \bibinfo {author}
  {\bibfnamefont {C.~M.}\ \bibnamefont {Holland}}, \bibinfo {author}
  {\bibfnamefont {Y.}~\bibnamefont {Jiang}}, \bibinfo {author} {\bibfnamefont
  {B.~E.}\ \bibnamefont {Garber}}, \bibinfo {author} {\bibfnamefont {S.~P.}\
  \bibnamefont {Carman}},\ and\ \bibinfo {author} {\bibfnamefont {J.~M.}\
  \bibnamefont {Hogan}},\ }\href
  {https://doi.org/10.1103/PhysRevLett.124.083604} {\bibfield  {journal}
  {\bibinfo  {journal} {Phys. Rev. Lett.}\ }\textbf {\bibinfo {volume} {124}},\
  \bibinfo {pages} {83604} (\bibinfo {year} {2020})},\ \Eprint
  {https://arxiv.org/abs/1910.05459} {arXiv:1910.05459} \BibitemShut {NoStop}%
\bibitem [{\citenamefont {Wilkason}\ \emph {et~al.}(2022)\citenamefont
  {Wilkason}, \citenamefont {Nantel}, \citenamefont {Rudolph}, \citenamefont
  {Jiang}, \citenamefont {Garber}, \citenamefont {Swan}, \citenamefont
  {Carman}, \citenamefont {Abe},\ and\ \citenamefont {Hogan}}]{Wilkason2022}%
  \BibitemOpen
  \bibfield  {author} {\bibinfo {author} {\bibfnamefont {T.}~\bibnamefont
  {Wilkason}}, \bibinfo {author} {\bibfnamefont {M.}~\bibnamefont {Nantel}},
  \bibinfo {author} {\bibfnamefont {J.}~\bibnamefont {Rudolph}}, \bibinfo
  {author} {\bibfnamefont {Y.}~\bibnamefont {Jiang}}, \bibinfo {author}
  {\bibfnamefont {B.~E.}\ \bibnamefont {Garber}}, \bibinfo {author}
  {\bibfnamefont {H.}~\bibnamefont {Swan}}, \bibinfo {author} {\bibfnamefont
  {S.~P.}\ \bibnamefont {Carman}}, \bibinfo {author} {\bibfnamefont
  {M.}~\bibnamefont {Abe}},\ and\ \bibinfo {author} {\bibfnamefont {J.~M.}\
  \bibnamefont {Hogan}},\ }\href@noop {} {\bibfield  {journal} {\bibinfo
  {journal} {arXiv:2205.06965}\ } (\bibinfo {year} {2022})}\BibitemShut
  {NoStop}%
\bibitem [{\citenamefont {Zhou}\ \emph
  {et~al.}(2015{\natexlab{b}})\citenamefont {Zhou}, \citenamefont {Long},
  \citenamefont {Tang}, \citenamefont {Chen}, \citenamefont {Gao},
  \citenamefont {Peng}, \citenamefont {Duan}, \citenamefont {Zhong},
  \citenamefont {Xiong}, \citenamefont {Wang}, \citenamefont {Zhang},\ and\
  \citenamefont {Zhan}}]{Zhou.2015}%
  \BibitemOpen
  \bibfield  {author} {\bibinfo {author} {\bibfnamefont {L.}~\bibnamefont
  {Zhou}}, \bibinfo {author} {\bibfnamefont {S.}~\bibnamefont {Long}}, \bibinfo
  {author} {\bibfnamefont {B.}~\bibnamefont {Tang}}, \bibinfo {author}
  {\bibfnamefont {X.}~\bibnamefont {Chen}}, \bibinfo {author} {\bibfnamefont
  {F.}~\bibnamefont {Gao}}, \bibinfo {author} {\bibfnamefont {W.}~\bibnamefont
  {Peng}}, \bibinfo {author} {\bibfnamefont {W.}~\bibnamefont {Duan}}, \bibinfo
  {author} {\bibfnamefont {J.}~\bibnamefont {Zhong}}, \bibinfo {author}
  {\bibfnamefont {Z.}~\bibnamefont {Xiong}}, \bibinfo {author} {\bibfnamefont
  {J.}~\bibnamefont {Wang}}, \bibinfo {author} {\bibfnamefont {Y.}~\bibnamefont
  {Zhang}},\ and\ \bibinfo {author} {\bibfnamefont {M.}~\bibnamefont {Zhan}},\
  }\href {https://doi.org/10.1103/physrevlett.115.013004} {\bibfield  {journal}
  {\bibinfo  {journal} {Physical Review Letters}\ }\textbf {\bibinfo {volume}
  {115}},\ \bibinfo {pages} {013004} (\bibinfo {year} {2015}{\natexlab{b}})},\
  \Eprint {https://arxiv.org/abs/1503.00401} {1503.00401} \BibitemShut
  {NoStop}%
\bibitem [{\citenamefont {Steck}(2016)}]{qao}%
  \BibitemOpen
  \bibfield  {author} {\bibinfo {author} {\bibfnamefont {D.~A.}\ \bibnamefont
  {Steck}}} (\bibinfo {year} {2016}),\ \bibinfo {note} {available online at
  http://steck.us/teaching (revision 0.11.5, 27 November 2016)}\BibitemShut
  {NoStop}%
\bibitem [{\citenamefont {Dunning}\ \emph {et~al.}(2014)\citenamefont
  {Dunning}, \citenamefont {Gregory}, \citenamefont {Bateman}, \citenamefont
  {Cooper}, \citenamefont {Himsworth}, \citenamefont {Jones},\ and\
  \citenamefont {Freegarde}}]{Dunning.2014jm}%
  \BibitemOpen
  \bibfield  {author} {\bibinfo {author} {\bibfnamefont {A.}~\bibnamefont
  {Dunning}}, \bibinfo {author} {\bibfnamefont {R.}~\bibnamefont {Gregory}},
  \bibinfo {author} {\bibfnamefont {J.}~\bibnamefont {Bateman}}, \bibinfo
  {author} {\bibfnamefont {N.}~\bibnamefont {Cooper}}, \bibinfo {author}
  {\bibfnamefont {M.}~\bibnamefont {Himsworth}}, \bibinfo {author}
  {\bibfnamefont {J.~A.}\ \bibnamefont {Jones}},\ and\ \bibinfo {author}
  {\bibfnamefont {T.}~\bibnamefont {Freegarde}},\ }\href
  {https://doi.org/10.1103/physreva.90.033608} {\bibfield  {journal} {\bibinfo
  {journal} {Physical Review A}\ }\textbf {\bibinfo {volume} {90}},\ \bibinfo
  {pages} {033608} (\bibinfo {year} {2014})},\ \Eprint
  {https://arxiv.org/abs/1406.2916} {1406.2916} \BibitemShut {NoStop}%
\bibitem [{\citenamefont {Cidrim}\ \emph {et~al.}(2021)\citenamefont {Cidrim},
  \citenamefont {Orioli}, \citenamefont {Sanner}, \citenamefont {Hutson},
  \citenamefont {Ye}, \citenamefont {Bachelard},\ and\ \citenamefont
  {Rey}}]{Cidrim.2021}%
  \BibitemOpen
  \bibfield  {author} {\bibinfo {author} {\bibfnamefont {A.}~\bibnamefont
  {Cidrim}}, \bibinfo {author} {\bibfnamefont {A.~P.}\ \bibnamefont {Orioli}},
  \bibinfo {author} {\bibfnamefont {C.}~\bibnamefont {Sanner}}, \bibinfo
  {author} {\bibfnamefont {R.~B.}\ \bibnamefont {Hutson}}, \bibinfo {author}
  {\bibfnamefont {J.}~\bibnamefont {Ye}}, \bibinfo {author} {\bibfnamefont
  {R.}~\bibnamefont {Bachelard}},\ and\ \bibinfo {author} {\bibfnamefont
  {A.~M.}\ \bibnamefont {Rey}},\ }\href
  {https://doi.org/10.1103/physrevlett.127.013401} {\bibfield  {journal}
  {\bibinfo  {journal} {Physical Review Letters}\ }\textbf {\bibinfo {volume}
  {127}},\ \bibinfo {pages} {013401} (\bibinfo {year} {2021})}\BibitemShut
  {NoStop}%
\bibitem [{\citenamefont {Delhuille}\ \emph {et~al.}(2003)\citenamefont
  {Delhuille}, \citenamefont {Miffre}, \citenamefont {Robilliard},
  \citenamefont {Vigue}, \citenamefont {Bu},\ and\ \citenamefont
  {Champenois}}]{Delhuille2003}%
  \BibitemOpen
  \bibfield  {author} {\bibinfo {author} {\bibfnamefont {R.}~\bibnamefont
  {Delhuille}}, \bibinfo {author} {\bibfnamefont {A.}~\bibnamefont {Miffre}},
  \bibinfo {author} {\bibfnamefont {C.}~\bibnamefont {Robilliard}}, \bibinfo
  {author} {\bibfnamefont {J.}~\bibnamefont {Vigue}}, \bibinfo {author}
  {\bibfnamefont {M.}~\bibnamefont {Bu}},\ and\ \bibinfo {author}
  {\bibfnamefont {C.}~\bibnamefont {Champenois}},\ }\href
  {https://doi.org/10.1103/PhysRevA.68.013607} {\bibfield  {journal} {\bibinfo
  {journal} {Phys. Rev. A}\ }\textbf {\bibinfo {volume} {68}},\ \bibinfo
  {pages} {013607} (\bibinfo {year} {2003})}\BibitemShut {NoStop}%
\bibitem [{\citenamefont {Magesan}\ \emph {et~al.}(2012)\citenamefont
  {Magesan}, \citenamefont {Gambetta},\ and\ \citenamefont
  {Emerson}}]{Magesan2012}%
  \BibitemOpen
  \bibfield  {author} {\bibinfo {author} {\bibfnamefont {E.}~\bibnamefont
  {Magesan}}, \bibinfo {author} {\bibfnamefont {J.~M.}\ \bibnamefont
  {Gambetta}},\ and\ \bibinfo {author} {\bibfnamefont {J.}~\bibnamefont
  {Emerson}},\ }\href {https://doi.org/10.1103/PhysRevA.85.042311} {\bibfield
  {journal} {\bibinfo  {journal} {Physical Review A - Atomic, Molecular, and
  Optical Physics}\ }\textbf {\bibinfo {volume} {85}},\ \bibinfo {pages}
  {042311} (\bibinfo {year} {2012})},\ \Eprint
  {https://arxiv.org/abs/1109.6887} {arXiv:1109.6887} \BibitemShut {NoStop}%
\bibitem [{\citenamefont {Dalibard}\ \emph {et~al.}(1992)\citenamefont
  {Dalibard}, \citenamefont {Castin},\ and\ \citenamefont
  {Molmer}}]{Dalibard1992}%
  \BibitemOpen
  \bibfield  {author} {\bibinfo {author} {\bibfnamefont {J.}~\bibnamefont
  {Dalibard}}, \bibinfo {author} {\bibfnamefont {Y.}~\bibnamefont {Castin}},\
  and\ \bibinfo {author} {\bibfnamefont {K.}~\bibnamefont {Molmer}},\
  }\href@noop {} {\bibfield  {journal} {\bibinfo  {journal} {Phys. Rev. Lett.}\
  }\textbf {\bibinfo {volume} {68}},\ \bibinfo {pages} {580} (\bibinfo {year}
  {1992})}\BibitemShut {NoStop}%
\bibitem [{\citenamefont {Carmichael}(1993)}]{Carmichael1993}%
  \BibitemOpen
  \bibfield  {author} {\bibinfo {author} {\bibfnamefont {H.~J.}\ \bibnamefont
  {Carmichael}},\ }\href@noop {} {\bibfield  {journal} {\bibinfo  {journal}
  {Lecture Notes in Physics, New Series m: Monographs Vol. m18, Springer,
  Berlin}\ } (\bibinfo {year} {1993})}\BibitemShut {NoStop}%
\bibitem [{\citenamefont {Bergmann}\ \emph {et~al.}(1998)\citenamefont
  {Bergmann}, \citenamefont {Theuer},\ and\ \citenamefont
  {Shore}}]{Bergmann1998}%
  \BibitemOpen
  \bibfield  {author} {\bibinfo {author} {\bibfnamefont {K.}~\bibnamefont
  {Bergmann}}, \bibinfo {author} {\bibfnamefont {H.}~\bibnamefont {Theuer}},\
  and\ \bibinfo {author} {\bibfnamefont {B.~W.}\ \bibnamefont {Shore}},\ }\href
  {https://doi.org/10.1103/revmodphys.70.1003} {\bibfield  {journal} {\bibinfo
  {journal} {Rev. Mod. Phys.}\ }\textbf {\bibinfo {volume} {70}},\ \bibinfo
  {pages} {1003} (\bibinfo {year} {1998})}\BibitemShut {NoStop}%
\bibitem [{\citenamefont {Vitanov}\ \emph {et~al.}(2001)\citenamefont
  {Vitanov}, \citenamefont {Halfmann}, \citenamefont {Shore},\ and\
  \citenamefont {Bergmann}}]{Vitanov.2001}%
  \BibitemOpen
  \bibfield  {author} {\bibinfo {author} {\bibfnamefont {N.~V.}\ \bibnamefont
  {Vitanov}}, \bibinfo {author} {\bibfnamefont {T.}~\bibnamefont {Halfmann}},
  \bibinfo {author} {\bibfnamefont {B.~W.}\ \bibnamefont {Shore}},\ and\
  \bibinfo {author} {\bibfnamefont {K.}~\bibnamefont {Bergmann}},\ }\href
  {https://doi.org/10.1146/annurev.physchem.52.1.763} {\bibfield  {journal}
  {\bibinfo  {journal} {Annual Review of Physical Chemistry}\ }\textbf
  {\bibinfo {volume} {52}},\ \bibinfo {pages} {763} (\bibinfo {year}
  {2001})}\BibitemShut {NoStop}%
\bibitem [{\citenamefont {Miao}\ \emph {et~al.}(2007)\citenamefont {Miao},
  \citenamefont {Wertz}, \citenamefont {Cohen},\ and\ \citenamefont
  {Metcalf}}]{Miao2007}%
  \BibitemOpen
  \bibfield  {author} {\bibinfo {author} {\bibfnamefont {X.}~\bibnamefont
  {Miao}}, \bibinfo {author} {\bibfnamefont {E.}~\bibnamefont {Wertz}},
  \bibinfo {author} {\bibfnamefont {M.~G.}\ \bibnamefont {Cohen}},\ and\
  \bibinfo {author} {\bibfnamefont {H.}~\bibnamefont {Metcalf}},\ }\href
  {https://doi.org/10.1103/PhysRevA.75.011402} {\bibfield  {journal} {\bibinfo
  {journal} {Physical Review Letters}\ }\textbf {\bibinfo {volume} {75}},\
  \bibinfo {pages} {011402} (\bibinfo {year} {2007})}\BibitemShut {NoStop}%
\bibitem [{\citenamefont {Guery-Odelin}\ \emph {et~al.}(2019)\citenamefont
  {Guery-Odelin}, \citenamefont {Ruschhaupt}, \citenamefont {Kiely},
  \citenamefont {Torrontegui}, \citenamefont {Mart{\'{i}}nez-Garaot},\ and\
  \citenamefont {Muga}}]{Guery-Odelin2019}%
  \BibitemOpen
  \bibfield  {author} {\bibinfo {author} {\bibfnamefont {D.}~\bibnamefont
  {Guery-Odelin}}, \bibinfo {author} {\bibfnamefont {A.}~\bibnamefont
  {Ruschhaupt}}, \bibinfo {author} {\bibfnamefont {A.}~\bibnamefont {Kiely}},
  \bibinfo {author} {\bibfnamefont {E.}~\bibnamefont {Torrontegui}}, \bibinfo
  {author} {\bibfnamefont {S.}~\bibnamefont {Mart{\'{i}}nez-Garaot}},\ and\
  \bibinfo {author} {\bibfnamefont {J.~G.}\ \bibnamefont {Muga}},\ }\href
  {https://doi.org/10.1103/RevModPhys.91.045001} {\bibfield  {journal}
  {\bibinfo  {journal} {Rev. Mod. Phys.}\ }\textbf {\bibinfo {volume} {91}},\
  \bibinfo {pages} {045001} (\bibinfo {year} {2019})},\ \Eprint
  {https://arxiv.org/abs/1904.08448} {arXiv:1904.08448} \BibitemShut {NoStop}%
\bibitem [{\citenamefont {Berry}(1984)}]{Berry.1984}%
  \BibitemOpen
  \bibfield  {author} {\bibinfo {author} {\bibfnamefont {M.~V.}\ \bibnamefont
  {Berry}},\ }\href {https://doi.org/10.1098/rspa.1984.0023} {\bibfield
  {journal} {\bibinfo  {journal} {Proceedings of the Royal Society of London.
  A. Mathematical and Physical Sciences}\ }\textbf {\bibinfo {volume} {392}},\
  \bibinfo {pages} {45} (\bibinfo {year} {1984})}\BibitemShut {NoStop}%
\bibitem [{\citenamefont {Zhu}\ and\ \citenamefont {Wang}(2002)}]{Zhu2002}%
  \BibitemOpen
  \bibfield  {author} {\bibinfo {author} {\bibfnamefont {S.-l.}\ \bibnamefont
  {Zhu}}\ and\ \bibinfo {author} {\bibfnamefont {Z.~D.}\ \bibnamefont {Wang}},\
  }\href {https://doi.org/10.1103/PhysRevLett.89.097902} {\bibfield  {journal}
  {\bibinfo  {journal} {Phys. Rev. Lett.}\ }\textbf {\bibinfo {volume} {89}},\
  \bibinfo {pages} {097902} (\bibinfo {year} {2002})}\BibitemShut {NoStop}%
\bibitem [{\citenamefont {Gustavson}(2000)}]{GustavsonThesis}%
  \BibitemOpen
  \bibfield  {author} {\bibinfo {author} {\bibfnamefont {T.~L.}\ \bibnamefont
  {Gustavson}},\ }\href@noop {} {Ph.D. thesis},\ \bibinfo  {school} {Standford
  University} (\bibinfo {year} {2000})\BibitemShut {NoStop}%
\bibitem [{\citenamefont {Lu}\ \emph {et~al.}(2005)\citenamefont {Lu},
  \citenamefont {Miao},\ and\ \citenamefont {Metcalf}}]{lu2005}%
  \BibitemOpen
  \bibfield  {author} {\bibinfo {author} {\bibfnamefont {T.}~\bibnamefont
  {Lu}}, \bibinfo {author} {\bibfnamefont {X.}~\bibnamefont {Miao}},\ and\
  \bibinfo {author} {\bibfnamefont {H.}~\bibnamefont {Metcalf}},\ }\href
  {https://doi.org/10.1103/PhysRevA.71.061405} {\bibfield  {journal} {\bibinfo
  {journal} {Phys. Rev. A}\ }\textbf {\bibinfo {volume} {71}},\ \bibinfo
  {pages} {061405} (\bibinfo {year} {2005})}\BibitemShut {NoStop}%
\bibitem [{\citenamefont {Sievers}\ \emph {et~al.}(2015)\citenamefont
  {Sievers}, \citenamefont {Kretzschmar}, \citenamefont {Fernandes},
  \citenamefont {Suchet}, \citenamefont {Rabinovic}, \citenamefont {Wu},
  \citenamefont {Parker}, \citenamefont {Khaykovich}, \citenamefont {Salomon},\
  and\ \citenamefont {Chevy}}]{Sievers2015}%
  \BibitemOpen
  \bibfield  {author} {\bibinfo {author} {\bibfnamefont {F.}~\bibnamefont
  {Sievers}}, \bibinfo {author} {\bibfnamefont {N.}~\bibnamefont
  {Kretzschmar}}, \bibinfo {author} {\bibfnamefont {D.}~\bibnamefont
  {Fernandes}}, \bibinfo {author} {\bibfnamefont {D.}~\bibnamefont {Suchet}},
  \bibinfo {author} {\bibfnamefont {M.}~\bibnamefont {Rabinovic}}, \bibinfo
  {author} {\bibfnamefont {S.}~\bibnamefont {Wu}}, \bibinfo {author}
  {\bibfnamefont {C.}~\bibnamefont {Parker}}, \bibinfo {author} {\bibfnamefont
  {L.}~\bibnamefont {Khaykovich}}, \bibinfo {author} {\bibfnamefont
  {C.}~\bibnamefont {Salomon}},\ and\ \bibinfo {author} {\bibfnamefont
  {F.}~\bibnamefont {Chevy}},\ }\href
  {https://doi.org/10.1103/PhysRevA.91.023426} {\bibfield  {journal} {\bibinfo
  {journal} {Physical Review A - Atomic, Molecular, and Optical Physics}\
  }\textbf {\bibinfo {volume} {91}},\ \bibinfo {pages} {023426} (\bibinfo
  {year} {2015})}\BibitemShut {NoStop}%
\bibitem [{\citenamefont {Genov}\ and\ \citenamefont
  {Vitanov}(2013)}]{Genov2013}%
  \BibitemOpen
  \bibfield  {author} {\bibinfo {author} {\bibfnamefont {G.~T.}\ \bibnamefont
  {Genov}}\ and\ \bibinfo {author} {\bibfnamefont {N.~V.}\ \bibnamefont
  {Vitanov}},\ }\href {https://doi.org/10.1103/PhysRevLett.110.133002}
  {\bibfield  {journal} {\bibinfo  {journal} {Phys. Rev. Lett.}\ }\textbf
  {\bibinfo {volume} {110}},\ \bibinfo {pages} {133002} (\bibinfo {year}
  {2013})}\BibitemShut {NoStop}%
\bibitem [{\citenamefont {Kang}\ \emph {et~al.}(2018)\citenamefont {Kang},
  \citenamefont {Chen}, \citenamefont {Shi}, \citenamefont {Huang},
  \citenamefont {Song},\ and\ \citenamefont {Xia}}]{Kang2018}%
  \BibitemOpen
  \bibfield  {author} {\bibinfo {author} {\bibfnamefont {Y.~H.}\ \bibnamefont
  {Kang}}, \bibinfo {author} {\bibfnamefont {Y.~H.}\ \bibnamefont {Chen}},
  \bibinfo {author} {\bibfnamefont {Z.~C.}\ \bibnamefont {Shi}}, \bibinfo
  {author} {\bibfnamefont {B.~H.}\ \bibnamefont {Huang}}, \bibinfo {author}
  {\bibfnamefont {J.}~\bibnamefont {Song}},\ and\ \bibinfo {author}
  {\bibfnamefont {Y.}~\bibnamefont {Xia}},\ }\href
  {https://doi.org/10.1103/PhysRevA.97.033407} {\bibfield  {journal} {\bibinfo
  {journal} {Phys. Rev. A}\ }\textbf {\bibinfo {volume} {97}},\ \bibinfo
  {pages} {033407} (\bibinfo {year} {2018})}\BibitemShut {NoStop}%
\bibitem [{\citenamefont {Jo}\ \emph {et~al.}(2019)\citenamefont {Jo},
  \citenamefont {Song},\ and\ \citenamefont {Ahn}}]{Jo2019}%
  \BibitemOpen
  \bibfield  {author} {\bibinfo {author} {\bibfnamefont {H.}~\bibnamefont
  {Jo}}, \bibinfo {author} {\bibfnamefont {Y.}~\bibnamefont {Song}},\ and\
  \bibinfo {author} {\bibfnamefont {J.}~\bibnamefont {Ahn}},\ }\href
  {https://doi.org/10.1364/oe.27.003944} {\bibfield  {journal} {\bibinfo
  {journal} {Opt. Express}\ }\textbf {\bibinfo {volume} {27}},\ \bibinfo
  {pages} {3944} (\bibinfo {year} {2019})}\BibitemShut {NoStop}%
\bibitem [{\citenamefont {Shore}(2014)}]{Shore2014}%
  \BibitemOpen
  \bibfield  {author} {\bibinfo {author} {\bibfnamefont {B.~W.}\ \bibnamefont
  {Shore}},\ }\href {https://doi.org/10.1080/09500340.2013.837205} {\bibfield
  {journal} {\bibinfo  {journal} {J. Mod. Opt.}\ }\textbf {\bibinfo {volume}
  {61}},\ \bibinfo {pages} {787} (\bibinfo {year} {2014})}\BibitemShut
  {NoStop}%
\bibitem [{\citenamefont {Muniz}\ \emph {et~al.}(2018)\citenamefont {Muniz},
  \citenamefont {Norcia}, \citenamefont {Cline},\ and\ \citenamefont
  {Thompson}}]{Muniz2018}%
  \BibitemOpen
  \bibfield  {author} {\bibinfo {author} {\bibfnamefont {J.~A.}\ \bibnamefont
  {Muniz}}, \bibinfo {author} {\bibfnamefont {M.~A.}\ \bibnamefont {Norcia}},
  \bibinfo {author} {\bibfnamefont {J.~R.~K.}\ \bibnamefont {Cline}},\ and\
  \bibinfo {author} {\bibfnamefont {J.~K.}\ \bibnamefont {Thompson}},\
  }\href@noop {} {\bibfield  {journal} {\bibinfo  {journal} {arXiv:1806.00838}\
  } (\bibinfo {year} {2018})}\BibitemShut {NoStop}%
\bibitem [{foo({\natexlab{a}})}]{foot:circular}%
  \BibitemOpen
  \bibinfo {note} {It is worth pointing out that the $\Delta m\neq 0$ spin
  leakage is naturally suppressed when Raman transitions are driven by beams
  with a same circular polarization~\cite{GustavsonThesis}. Here, circularly
  polarized nanosecond SDK can be implemented on a compact delay line. When
  combined with the balanced chirp-alternating scheme, coherent control of a
  specific Zeeman spinor matterwave component can be achieved with robustly
  cancelled dynamic phases. However, since both the 2-photon shift and the
  Raman Rabi frquency of the circular polarized SDK are highly $m-$dependent,
  it would be more difficult to achieve the parallel multi-Zeeman spinor
  control.}\BibitemShut {Stop}%
\bibitem [{\citenamefont {Bertoldi}\ \emph {et~al.}(2019)\citenamefont
  {Bertoldi}, \citenamefont {Minardi},\ and\ \citenamefont
  {Prevedelli}}]{Bertoldi2019}%
  \BibitemOpen
  \bibfield  {author} {\bibinfo {author} {\bibfnamefont {A.}~\bibnamefont
  {Bertoldi}}, \bibinfo {author} {\bibfnamefont {F.}~\bibnamefont {Minardi}},\
  and\ \bibinfo {author} {\bibfnamefont {M.}~\bibnamefont {Prevedelli}},\
  }\href {https://doi.org/10.1103/PhysRevA.99.033619} {\bibfield  {journal}
  {\bibinfo  {journal} {Phys. Rev. A}\ }\textbf {\bibinfo {volume} {99}},\
  \bibinfo {pages} {033619} (\bibinfo {year} {2019})},\ \Eprint
  {https://arxiv.org/abs/1812.11890} {arXiv:1812.11890} \BibitemShut {NoStop}%
\bibitem [{foo({\natexlab{b}})}]{foot:biasRotation}%
  \BibitemOpen
  \bibinfo {note} {In atom interferometry, the $R_{\varphi}(\theta)$ operations
  are usually achieved with simple Raman pulses with little resiliance to
  control errors. Methods for error-resliance realization of
  $R_{\varphi}(\theta)$ are discussed in {\it e.g.} ref.~\cite{Saywell2018}.
  Similarly, we find $R_{\varphi}(\theta)$ can be achieved with composite
  pulses. The results will be presented in a future publication}\BibitemShut
  {NoStop}%
\bibitem [{\citenamefont {Sidorenkov}\ \emph {et~al.}(2020)\citenamefont
  {Sidorenkov}, \citenamefont {Gautier}, \citenamefont {Altorio}, \citenamefont
  {Geiger},\ and\ \citenamefont {Landragin}}]{Sidorenkov2020}%
  \BibitemOpen
  \bibfield  {author} {\bibinfo {author} {\bibfnamefont {L.~A.}\ \bibnamefont
  {Sidorenkov}}, \bibinfo {author} {\bibfnamefont {R.}~\bibnamefont {Gautier}},
  \bibinfo {author} {\bibfnamefont {M.}~\bibnamefont {Altorio}}, \bibinfo
  {author} {\bibfnamefont {R.}~\bibnamefont {Geiger}},\ and\ \bibinfo {author}
  {\bibfnamefont {A.}~\bibnamefont {Landragin}},\ }\href
  {https://doi.org/10.1103/PhysRevLett.125.213201} {\bibfield  {journal}
  {\bibinfo  {journal} {Phys. Rev. Lett.}\ }\textbf {\bibinfo {volume} {125}},\
  \bibinfo {pages} {213201} (\bibinfo {year} {2020})},\ \Eprint
  {https://arxiv.org/abs/2006.08371} {arXiv:2006.08371} \BibitemShut {NoStop}%
\bibitem [{\citenamefont {Dubetsky}\ and\ \citenamefont
  {Kasevich}(2006)}]{Dubetsky2006}%
  \BibitemOpen
  \bibfield  {author} {\bibinfo {author} {\bibfnamefont {B.}~\bibnamefont
  {Dubetsky}}\ and\ \bibinfo {author} {\bibfnamefont {M.~A.}\ \bibnamefont
  {Kasevich}},\ }\href {https://doi.org/10.1103/PhysRevA.74.023615} {\bibfield
  {journal} {\bibinfo  {journal} {Phys. Rev. A - At. Mol. Opt. Phys.}\ }\textbf
  {\bibinfo {volume} {74}},\ \bibinfo {pages} {023615} (\bibinfo {year}
  {2006})},\ \Eprint {https://arxiv.org/abs/0604082} {arXiv:0604082 [physics]}
  \BibitemShut {NoStop}%
\bibitem [{\citenamefont {Petitjean}\ \emph {et~al.}(2007)\citenamefont
  {Petitjean}, \citenamefont {Bevilaqua}, \citenamefont {Heller},\ and\
  \citenamefont {Jacquod}}]{Petitjean2007}%
  \BibitemOpen
  \bibfield  {author} {\bibinfo {author} {\bibfnamefont {C.}~\bibnamefont
  {Petitjean}}, \bibinfo {author} {\bibfnamefont {D.~V.}\ \bibnamefont
  {Bevilaqua}}, \bibinfo {author} {\bibfnamefont {E.~J.}\ \bibnamefont
  {Heller}},\ and\ \bibinfo {author} {\bibfnamefont {P.}~\bibnamefont
  {Jacquod}},\ }\href {https://doi.org/10.1103/PhysRevLett.98.164101}
  {\bibfield  {journal} {\bibinfo  {journal} {Phys. Rev. Lett.}\ }\textbf
  {\bibinfo {volume} {98}},\ \bibinfo {pages} {164101} (\bibinfo {year}
  {2007})}\BibitemShut {NoStop}%
\bibitem [{\citenamefont {Su}\ \emph {et~al.}(2010)\citenamefont {Su},
  \citenamefont {Wu},\ and\ \citenamefont {Prentiss}}]{Su2010}%
  \BibitemOpen
  \bibfield  {author} {\bibinfo {author} {\bibfnamefont {E.}~\bibnamefont
  {Su}}, \bibinfo {author} {\bibfnamefont {S.}~\bibnamefont {Wu}},\ and\
  \bibinfo {author} {\bibfnamefont {M.}~\bibnamefont {Prentiss}},\ }\href
  {https://doi.org/10.1103/PhysRevA.81.043631} {\bibfield  {journal} {\bibinfo
  {journal} {Phys. Rev. A - At. Mol. Opt. Phys.}\ }\textbf {\bibinfo {volume}
  {81}},\ \bibinfo {pages} {043631} (\bibinfo {year} {2010})}\BibitemShut
  {NoStop}%
\bibitem [{foo({\natexlab{c}})}]{foot:AIc}%
  \BibitemOpen
  \bibinfo {note} {The projected interferometry performance is subtly enhanced
  by the time-reversal symmetry of the interferometry itself, introduced by the
  $R_2$ rotation. Practically, the symmetry can be broken by atomic motion
  during the long interrogation time $T$. In this case, the high-level
  performance can still be retained by avoding the narrow low-$\mathcal{F}$
  intensity regions in all the SDK sequences.}\BibitemShut {Stop}%
\bibitem [{\citenamefont {Gross}\ and\ \citenamefont
  {Bloch}(2017)}]{Gross.2017}%
  \BibitemOpen
  \bibfield  {author} {\bibinfo {author} {\bibfnamefont {C.}~\bibnamefont
  {Gross}}\ and\ \bibinfo {author} {\bibfnamefont {I.}~\bibnamefont {Bloch}},\
  }\href {https://doi.org/10.1126/science.aal3837} {\bibfield  {journal}
  {\bibinfo  {journal} {Science}\ }\textbf {\bibinfo {volume} {357}},\ \bibinfo
  {pages} {995} (\bibinfo {year} {2017})}\BibitemShut {NoStop}%
\bibitem [{\citenamefont {Blatt}\ and\ \citenamefont {Roos}(2012)}]{Blatt2012}%
  \BibitemOpen
  \bibfield  {author} {\bibinfo {author} {\bibfnamefont {R.}~\bibnamefont
  {Blatt}}\ and\ \bibinfo {author} {\bibfnamefont {C.~F.}\ \bibnamefont
  {Roos}},\ }\href {https://doi.org/10.1038/nphys2252} {\bibfield  {journal}
  {\bibinfo  {journal} {Nat. Phys.}\ }\textbf {\bibinfo {volume} {8}},\
  \bibinfo {pages} {277} (\bibinfo {year} {2012})}\BibitemShut {NoStop}%
\bibitem [{\citenamefont {Wang}\ \emph {et~al.}(2005)\citenamefont {Wang},
  \citenamefont {Anderson}, \citenamefont {Bright}, \citenamefont {Cornell},
  \citenamefont {Diot}, \citenamefont {Kishimoto}, \citenamefont {Prentiss},
  \citenamefont {Saravanan}, \citenamefont {Segal},\ and\ \citenamefont
  {Wu}}]{wang2005}%
  \BibitemOpen
  \bibfield  {author} {\bibinfo {author} {\bibfnamefont {Y.-J.}\ \bibnamefont
  {Wang}}, \bibinfo {author} {\bibfnamefont {D.}~\bibnamefont {Anderson}},
  \bibinfo {author} {\bibfnamefont {V.}~\bibnamefont {Bright}}, \bibinfo
  {author} {\bibfnamefont {E.}~\bibnamefont {Cornell}}, \bibinfo {author}
  {\bibfnamefont {Q.}~\bibnamefont {Diot}}, \bibinfo {author} {\bibfnamefont
  {T.}~\bibnamefont {Kishimoto}}, \bibinfo {author} {\bibfnamefont
  {M.}~\bibnamefont {Prentiss}}, \bibinfo {author} {\bibfnamefont
  {R.}~\bibnamefont {Saravanan}}, \bibinfo {author} {\bibfnamefont
  {S.}~\bibnamefont {Segal}},\ and\ \bibinfo {author} {\bibfnamefont
  {S.}~\bibnamefont {Wu}},\ }\href
  {https://doi.org/10.1103/PhysRevLett.94.090405} {\bibfield  {journal}
  {\bibinfo  {journal} {Phys. Rev. Lett.}\ }\textbf {\bibinfo {volume} {94}},\
  \bibinfo {pages} {090405} (\bibinfo {year} {2005})}\BibitemShut {NoStop}%
\bibitem [{\citenamefont {Hughes}\ \emph {et~al.}(2009)\citenamefont {Hughes},
  \citenamefont {Burke},\ and\ \citenamefont {Sackett}}]{Hughes2009}%
  \BibitemOpen
  \bibfield  {author} {\bibinfo {author} {\bibfnamefont {K.~J.}\ \bibnamefont
  {Hughes}}, \bibinfo {author} {\bibfnamefont {J.~H.}\ \bibnamefont {Burke}},\
  and\ \bibinfo {author} {\bibfnamefont {C.~A.}\ \bibnamefont {Sackett}},\
  }\href {https://doi.org/10.1103/PhysRevLett.102.150403} {\bibfield  {journal}
  {\bibinfo  {journal} {Phys. Rev. Lett.}\ }\textbf {\bibinfo {volume} {102}},\
  \bibinfo {pages} {150403} (\bibinfo {year} {2009})}\BibitemShut {NoStop}%
\bibitem [{\citenamefont {Debs}\ \emph {et~al.}(2011)\citenamefont {Debs},
  \citenamefont {Altin}, \citenamefont {Barter}, \citenamefont {D{\"{o}}ring},
  \citenamefont {Dennis}, \citenamefont {McDonald}, \citenamefont {Anderson},
  \citenamefont {Close},\ and\ \citenamefont {Robins}}]{Debs2011}%
  \BibitemOpen
  \bibfield  {author} {\bibinfo {author} {\bibfnamefont {J.~E.}\ \bibnamefont
  {Debs}}, \bibinfo {author} {\bibfnamefont {P.~A.}\ \bibnamefont {Altin}},
  \bibinfo {author} {\bibfnamefont {T.~H.}\ \bibnamefont {Barter}}, \bibinfo
  {author} {\bibfnamefont {D.}~\bibnamefont {D{\"{o}}ring}}, \bibinfo {author}
  {\bibfnamefont {G.~R.}\ \bibnamefont {Dennis}}, \bibinfo {author}
  {\bibfnamefont {G.}~\bibnamefont {McDonald}}, \bibinfo {author}
  {\bibfnamefont {R.~P.}\ \bibnamefont {Anderson}}, \bibinfo {author}
  {\bibfnamefont {J.~D.}\ \bibnamefont {Close}},\ and\ \bibinfo {author}
  {\bibfnamefont {N.~P.}\ \bibnamefont {Robins}},\ }\href
  {https://doi.org/10.1103/PhysRevA.84.033610} {\bibfield  {journal} {\bibinfo
  {journal} {Phys. Rev. A - At. Mol. Opt. Phys.}\ }\textbf {\bibinfo {volume}
  {84}},\ \bibinfo {pages} {033610} (\bibinfo {year} {2011})},\ \Eprint
  {https://arxiv.org/abs/1011.5804} {arXiv:1011.5804} \BibitemShut {NoStop}%
\bibitem [{\citenamefont {Abend}\ \emph {et~al.}(2016)\citenamefont {Abend},
  \citenamefont {Gebbe}, \citenamefont {Gersemann}, \citenamefont {Ahlers},
  \citenamefont {M{\"{u}}ntinga}, \citenamefont {Giese}, \citenamefont
  {Gaaloul}, \citenamefont {Schubert}, \citenamefont {L{\"{a}}mmerzahl},
  \citenamefont {Ertmer}, \citenamefont {Schleich},\ and\ \citenamefont
  {Rasel}}]{Abend2016}%
  \BibitemOpen
  \bibfield  {author} {\bibinfo {author} {\bibfnamefont {S.}~\bibnamefont
  {Abend}}, \bibinfo {author} {\bibfnamefont {M.}~\bibnamefont {Gebbe}},
  \bibinfo {author} {\bibfnamefont {M.}~\bibnamefont {Gersemann}}, \bibinfo
  {author} {\bibfnamefont {H.}~\bibnamefont {Ahlers}}, \bibinfo {author}
  {\bibfnamefont {H.}~\bibnamefont {M{\"{u}}ntinga}}, \bibinfo {author}
  {\bibfnamefont {E.}~\bibnamefont {Giese}}, \bibinfo {author} {\bibfnamefont
  {N.}~\bibnamefont {Gaaloul}}, \bibinfo {author} {\bibfnamefont
  {C.}~\bibnamefont {Schubert}}, \bibinfo {author} {\bibfnamefont
  {C.}~\bibnamefont {L{\"{a}}mmerzahl}}, \bibinfo {author} {\bibfnamefont
  {W.}~\bibnamefont {Ertmer}}, \bibinfo {author} {\bibfnamefont {W.~P.}\
  \bibnamefont {Schleich}},\ and\ \bibinfo {author} {\bibfnamefont {E.~M.}\
  \bibnamefont {Rasel}},\ }\href
  {https://doi.org/10.1103/PhysRevLett.117.203003} {\bibfield  {journal}
  {\bibinfo  {journal} {Phys. Rev. Lett.}\ }\textbf {\bibinfo {volume} {117}},\
  \bibinfo {pages} {203003} (\bibinfo {year} {2016})}\BibitemShut {NoStop}%
\bibitem [{\citenamefont {Macrae}\ \emph {et~al.}(2021)\citenamefont {Macrae},
  \citenamefont {Bongs},\ and\ \citenamefont {Holynski}}]{Macrae.2021}%
  \BibitemOpen
  \bibfield  {author} {\bibinfo {author} {\bibfnamefont {C.~D.}\ \bibnamefont
  {Macrae}}, \bibinfo {author} {\bibfnamefont {K.}~\bibnamefont {Bongs}},\ and\
  \bibinfo {author} {\bibfnamefont {M.}~\bibnamefont {Holynski}},\ }\href
  {https://doi.org/10.1364/ol.415963} {\bibfield  {journal} {\bibinfo
  {journal} {Optics letters}\ }\textbf {\bibinfo {volume} {46}},\ \bibinfo
  {pages} {1257} (\bibinfo {year} {2021})}\BibitemShut {NoStop}%
\bibitem [{\citenamefont {Wang}\ \emph {et~al.}(2022)\citenamefont {Wang},
  \citenamefont {He}, \citenamefont {Ji}, \citenamefont {Hu}, \citenamefont
  {Huang}, \citenamefont {Ma}, \citenamefont {Qiu}, \citenamefont {Zhao},\ and\
  \citenamefont {Wu}}]{Wang.2022}%
  \BibitemOpen
  \bibfield  {author} {\bibinfo {author} {\bibfnamefont {Y.}~\bibnamefont
  {Wang}}, \bibinfo {author} {\bibfnamefont {Y.}~\bibnamefont {He}}, \bibinfo
  {author} {\bibfnamefont {L.}~\bibnamefont {Ji}}, \bibinfo {author}
  {\bibfnamefont {J.}~\bibnamefont {Hu}}, \bibinfo {author} {\bibfnamefont
  {X.}~\bibnamefont {Huang}}, \bibinfo {author} {\bibfnamefont
  {Y.}~\bibnamefont {Ma}}, \bibinfo {author} {\bibfnamefont {L.}~\bibnamefont
  {Qiu}}, \bibinfo {author} {\bibfnamefont {K.}~\bibnamefont {Zhao}},\ and\
  \bibinfo {author} {\bibfnamefont {S.}~\bibnamefont {Wu}},\ }\href
  {https://doi.org/10.3788/COL202220.111401} {\bibfield  {journal} {\bibinfo
  {journal} {Chinese Opt. Lett.}\ }\textbf {\bibinfo {volume} {20}},\ \bibinfo
  {pages} {22} (\bibinfo {year} {2022})},\ \Eprint
  {https://arxiv.org/abs/2203.02915} {arXiv:2203.02915} \BibitemShut {NoStop}%
\bibitem [{\citenamefont {Jayich}\ \emph {et~al.}(2014)\citenamefont {Jayich},
  \citenamefont {Vutha}, \citenamefont {Hummon}, \citenamefont {Porto},\ and\
  \citenamefont {Campbell}}]{Jayich2014}%
  \BibitemOpen
  \bibfield  {author} {\bibinfo {author} {\bibfnamefont {A.~M.}\ \bibnamefont
  {Jayich}}, \bibinfo {author} {\bibfnamefont {A.~C.}\ \bibnamefont {Vutha}},
  \bibinfo {author} {\bibfnamefont {M.~T.}\ \bibnamefont {Hummon}}, \bibinfo
  {author} {\bibfnamefont {J.~V.}\ \bibnamefont {Porto}},\ and\ \bibinfo
  {author} {\bibfnamefont {W.~C.}\ \bibnamefont {Campbell}},\ }\href@noop {}
  {\bibfield  {journal} {\bibinfo  {journal} {Phys. Rev. A}\ }\textbf {\bibinfo
  {volume} {89}},\ \bibinfo {pages} {023425} (\bibinfo {year} {2014})},\
  \Eprint {https://arxiv.org/abs/arXiv:1312.4499v1} {arXiv:arXiv:1312.4499v1}
  \BibitemShut {NoStop}%
\bibitem [{\citenamefont {Liu}\ \emph {et~al.}(2022)\citenamefont {Liu},
  \citenamefont {Ma}, \citenamefont {Ji}, \citenamefont {Qiu}, \citenamefont
  {Ji}, \citenamefont {Tao},\ and\ \citenamefont {Wu}}]{Liu2022a}%
  \BibitemOpen
  \bibfield  {author} {\bibinfo {author} {\bibfnamefont {R.}~\bibnamefont
  {Liu}}, \bibinfo {author} {\bibfnamefont {Y.}~\bibnamefont {Ma}}, \bibinfo
  {author} {\bibfnamefont {L.}~\bibnamefont {Ji}}, \bibinfo {author}
  {\bibfnamefont {L.}~\bibnamefont {Qiu}}, \bibinfo {author} {\bibfnamefont
  {M.}~\bibnamefont {Ji}}, \bibinfo {author} {\bibfnamefont {Z.}~\bibnamefont
  {Tao}},\ and\ \bibinfo {author} {\bibfnamefont {S.}~\bibnamefont {Wu}},\
  }\href {https://doi.org/10.1364/oe.445719} {\bibfield  {journal} {\bibinfo
  {journal} {Opt. Express}\ }\textbf {\bibinfo {volume} {30}},\ \bibinfo
  {pages} {27780} (\bibinfo {year} {2022})},\ \Eprint
  {https://arxiv.org/abs/2110.15537} {arXiv:2110.15537} \BibitemShut {NoStop}%
\bibitem [{\citenamefont {Bruce}\ \emph {et~al.}(2017)\citenamefont {Bruce},
  \citenamefont {Haller}, \citenamefont {Peaudecerf}, \citenamefont {Cotta},
  \citenamefont {Andia}, \citenamefont {Wu}, \citenamefont {Johnson},
  \citenamefont {Lovett},\ and\ \citenamefont {Kuhr}}]{Bruce2017}%
  \BibitemOpen
  \bibfield  {author} {\bibinfo {author} {\bibfnamefont {G.~D.}\ \bibnamefont
  {Bruce}}, \bibinfo {author} {\bibfnamefont {E.}~\bibnamefont {Haller}},
  \bibinfo {author} {\bibfnamefont {B.}~\bibnamefont {Peaudecerf}}, \bibinfo
  {author} {\bibfnamefont {D.~A.}\ \bibnamefont {Cotta}}, \bibinfo {author}
  {\bibfnamefont {M.}~\bibnamefont {Andia}}, \bibinfo {author} {\bibfnamefont
  {S.}~\bibnamefont {Wu}}, \bibinfo {author} {\bibfnamefont {M.~Y.~H.}\
  \bibnamefont {Johnson}}, \bibinfo {author} {\bibfnamefont {B.~W.}\
  \bibnamefont {Lovett}},\ and\ \bibinfo {author} {\bibfnamefont
  {S.}~\bibnamefont {Kuhr}},\ }\href@noop {} {\bibfield  {journal} {\bibinfo
  {journal} {J. Phys. B: At. Mol. Opt. Phys}\ }\textbf {\bibinfo {volume}
  {50}},\ \bibinfo {pages} {095002} (\bibinfo {year} {2017})}\BibitemShut
  {NoStop}%
\bibitem [{\citenamefont {Dum}\ \emph {et~al.}(1992)\citenamefont {Dum},
  \citenamefont {Parkins}, \citenamefont {Zoller},\ and\ \citenamefont
  {Gardiner}}]{Dum1992}%
  \BibitemOpen
  \bibfield  {author} {\bibinfo {author} {\bibfnamefont {R.}~\bibnamefont
  {Dum}}, \bibinfo {author} {\bibfnamefont {A.~S.}\ \bibnamefont {Parkins}},
  \bibinfo {author} {\bibfnamefont {P.}~\bibnamefont {Zoller}},\ and\ \bibinfo
  {author} {\bibfnamefont {C.~W.}\ \bibnamefont {Gardiner}},\ }\href
  {https://journals-aps-org.proxy.scd.u-psud.fr/pra/pdf/10.1103/PhysRevA.46.4382}
  {\bibfield  {journal} {\bibinfo  {journal} {Phyiscal Rev. A}\ }\textbf
  {\bibinfo {volume} {46}},\ \bibinfo {pages} {4382} (\bibinfo {year}
  {1992})}\BibitemShut {NoStop}%
\end{thebibliography}%
\bibliographystyle{apsrev4-2}

\end{document}